\documentclass[a4paper,10pt]{article}

\def\nsection#1{\section{#1}\setcounter{equation}{0}}
\def\nsection#1{\section{#1}}

\def\nappendix#1{\vskip 1cm\noindent{\Large\bf Appendix
         #1}\def\thesection{#1} \setcounter{equation}{0}}

\usepackage[latin1]{inputenc}      
\usepackage[T1]{fontenc} 
\usepackage{amsmath} 
\usepackage{amssymb}
\usepackage{epsfig}
\usepackage{psfrag}
\usepackage{bm}

\usepackage[english]{babel} 
\newcommand{\tr}{{\rm tr}}
\newcommand{\Tr}{{\rm Tr}}
\newcommand{\dd}{{\rm d}}
\newcommand{\ii}{{\rm i}}
\newcommand{\ee}{{\rm e}}
\newcommand{\CA}{{\cal A}}
\newcommand{\CB}{{\cal B}}
\newcommand{\CC}{{\cal C}}
\newcommand{\CD}{{\cal D}}
\newcommand{\CH}{{\cal H}}
\newcommand{\CI}{{\cal I}}
\newcommand{\CJ}{{\cal J}}
\newcommand{\CN}{{\cal N}}
\newcommand{\CS}{{\cal S}}
\newcommand{\CT}{{\cal T}}
\newcommand{\CU}{{\cal U}}
\newcommand{\sfrac}[2]{\frac{_{#1}}{^{#2}}}

\newcommand{\bra}[1]{\left\langle #1 \right|}
\newcommand{\ket}[1]{\left| #1 \right\rangle}

\newcommand{\tket}[1]{| #1 \rangle}
\newcommand{\CF}{{\mathcal F}}
\newcommand{\qq}{\begin{equation}}
\newcommand{\qqq}{\end{equation}}

\begin{document}

\title{Nonequilibrium transport through quantum-wire junctions and
boundary defects for free massless bosonic fields} 
\author{Krzysztof Gaw\c{e}dzki\\
{\small{Laboratoire de Physique, C.N.R.S., ENS de Lyon,
Universit\'e de Lyon,}}\\\small{46 All\'ee d'Italie, 69364 Lyon, France}\\{}\\
Cl\'ement Tauber\\
{\small{Laboratoire de Physique, ENS de Lyon,
Universit\'e de Lyon,}}\\\small{46 All\'ee d'Italie, 69364 Lyon, France}}

\date{}
\maketitle

\begin{abstract}{\noindent We consider a model of quantum-wire junctions 
where the latter are described by conformal-invariant boundary 
conditions of the simplest type in the multicomponent compactified 
massless scalar free field theory representing the bosonized Luttinger 
liquids in the bulk of wires. The boundary conditions result in the 
scattering of charges across the junction with nontrivial reflection 
and transmission amplitudes. The equilibrium state of such a system, 
corresponding to inverse temperature $\beta$ and electric potential 
$V$, is explicitly constructed both for finite and for semi-infinite 
wires. In the latter case, a stationary nonequilibrium state describing
the wires kept at different temperatures and potentials may be 
also constructed following Ref.\,\cite{MSLL}. The main result of 
the present paper is the calculation of the full counting statistics
(FCS) of the charge and energy transfers through the junction in a
nonequilibrium situation. Explicit expressions are worked out for 
the generating function of FCS and its large-deviations asymptotics. 
For the purely transmitting case they coincide with those obtained 
in Refs.\,\cite{BD1,BD2}, but numerous cases of junctions with transmission 
and reflection are also covered. The large deviations rate function of FCS 
for charge and energy transfers is shown to satisfy the fluctuation 
relations of Refs.\,\cite{AGMT,BD3}. The expressions for FCS obtained here 
are compared with the Levitov-Lesovic formulae of Refs.\,\cite{LL,Klich}.}
\end{abstract}


\vskip 1cm

\nsection{Introduction}
\label{sec:intro}

The transport phenomena in quantum wires (carbon nanotubes, semiconducting, 
metallic and molecular nanowires, quantum Hall edges) and, in particular, 
across their junctions, have attracted a lot of interest in recent times, 
\,see e.g.\,\,\cite{Datta,BB}. To a good approximation, the charge carriers 
inside the wires may be described  by the Tomonaga-Luttinger model 
\cite{Voit,SassKram,G,Ishii}. In the low energy limit, such a model reduces  
to a relativistic 1+1 dimensional interacting fermionic field 
theory that can also be represented by free massless bosonic fields. 
The junction between the leads couples together the conformal 
field theories (CFTs) describing at low energies the bulk volumes of 
the wires. Specific features of the coupling depend on how the junction is 
realized. Various models that couple two or more wires locally at their 
connected extremities were considered in the literature, \,see 
e.g.\,\cite{FLS,NFLL,OCA0,OCA} where important results about transport 
properties of such models of wire-junctions were obtained. The low-energy 
long-distance effect of the interaction at the junction may be described 
with the use of boundary CFT, similarly as the effect of a magnetic 
impurity in the multi-channel Kondo problem \cite{Affleck}. Even if 
the coupling of the Luttinger liquid theories introduced by the junction 
breaks the conformal symmetry, the latter should be restored in the 
long-distance scaling limit. In the scaling limit, the effect of the 
junction will be represented, using the ``folding trick'' of 
ref.\,\cite{WA}, by a conformal boundary defect in the tensor product 
of the bulk CFTs of individual wires \cite{BdeBDO}. Such a boundary defect 
preserves half of the conformal symmetry of the bulk theory. Examples of 
conformal boundary defects that describe the renormalization group fixed 
points of Luttinger liquid theories with a coupling localized at the junction 
were discussed in \cite{FLS,NFLL,OCA0,OCA}. It was also realized that 
the boundary CFT description of the junction of wires gives via the 
Green-Kubo formalism a direct access to the low temperature electric 
conductance of junctions \cite{OCA,RHFCA,RHFOCA} that measure small 
currents induced by placing different wires in slightly different external 
electric potentials. Getting hold of the transport properties of 
the quantum-wire junctions beyond the linear response regime is more 
complicated, see \cite{FLS} for an early result using an exact 
integrability of a model of contact between two wires. The CFT approach 
seems also helpful here. It was shown in \cite{BD1,BD2,BD3,DHB} that 
for some boundary defects (those with pure transmission of charge or 
energy), not only the electric and thermal conductance but also the 
long-time asymptotics of the full counting statistics (FCS) of charge 
and energy transfers through the junction may be calculated for the wires 
initially equilibrated at different temperatures and different potentials. 
Moreover, steady nonequilibrium states obtained at long times from such 
initial conditions could be explicitly constructed. Physical restrictions 
for the applicability of the CFT approach in such a nonequilibrium 
situations were also discussed in some detail in those works, in particular 
in \cite{BD2}, see also \cite{BDLS,D,AJPP,BJP}. The incorporation 
of junctions corresponding to boundary defects with transmission and 
reflection into that approach poses more problems, although for a junction
of two CFTs a general scheme has been recently laid down in \cite{BDV}, 
together with some examples. 
\vskip 0.1cm

The present paper arose from an attempt to calculate the FCS for 
nonequilibrium charge and energy transfers for simple conformal boundary 
defects with transmission and reflection. We describe each of $\,N\,$ wires 
by a compactified free massless $1+1$-dimensional bosonic field, 
with the compactification 
radius related to the Luttinger model coupling constants that may be 
different for different wires. The product theory is a toroidal 
compactification of the massless $N$-component free field, \,i.e., \,on the 
classical level, its field takes values in the torus $\,U(1)^N$. \,In such 
a theory, we consider the simplest conformal boundary defects that restrict 
the boundary values of the field at the junction to a subgroup 
$\,\CB\subset U(1)^N\,$ isomorphic to the torus $\,U(1)^M\,$ with $M\leq N$. 
In the string-theory jargon, $\,\CB\,$ is called the D(irichlet)-brane 
\cite{Polchinski}. First, we study the wires of finite length $\,L\,$ 
with the reflecting boundary condition at their ends not connected to 
the junction. The overall $U(1)$-symmetry of the theory is imposed, 
leading to the conservation of the total electric charge. We show that 
the boundary defect gives rise to an $N\times N$ scattering matrix $\,S\,$ 
that relates linearly the left-moving and the right-moving components 
of the electric currents in various wires. The classical theory described 
above may be canonically quantized preserving the latter property. The exact 
solution for the quantum theory includes the formula for the partition 
function of the equilibrium state corresponding to inverse temperature 
$\beta$ and electric potential $V$ and for the equilibrium correlation 
functions of the chiral components of the electric currents. The thermodynamic 
limit $\,L\to\infty\,$ may then be performed giving rise to a free-field 
theory that was constructed directly for $\,L=\infty\,$ in \cite{MSLL}. 
In that limit, the equilibrium correlation functions involving 
only left-moving (or only right-moving) currents factorize into the product 
of contributions from the individual wires. This property was used in 
\cite{MSLL}, \,following the earlier work \cite{Mintchev}, to construct 
a nonequilibrium stationary state (NESS) where the correlation functions 
of left-moving currents factorize into the product of equilibrium 
contributions from individual wires, each corresponding to a different 
temperature and a different potential. The NESS correlation functions 
involving also the right-moving currents are reduced to those of the 
left-moving ones using the scattering relation between the chiral 
current components. Following the approach of \cite{BD1,BD2}, we 
show that such a state is obtained if one prepares disconnected wires 
each in the equilibrium state at different temperature and potential 
and then one connects the wires instantaneously and lets the initial 
state evolve for a long time \cite{Ruelle}. 
\vskip 0.1cm

The main aim of the present paper is the study of the FCS for 
charge and energy (heat) transfers through the junction modeled by 
the brane defect of the type described above. Similarly as in \cite{BD2}, 
the FCS is obtained from a two-time measurement protocol. First, the 
total charge and total energy is measured in each of the disconnected 
wires of finite length $\,L\,$ prepared in equilibria with different 
temperatures and potentials. Next the wires are instantaneously connected 
and evolve for time $\,t\,$ with the dynamics described by
the field theory with the brane defect.  After time $\,t$,
\,the wires are disconnected again and the second measurement 
of total charge and total energy in individual wires is performed. 
The FCS is encoded in the characteristic function of the probability 
distribution of the changes of total charge and total energy 
of individual wires. The above protocol is not practical 
for long wires as the total charge and and total energy of the
wires, unlike their change in time, behave extensively with $\,L$,
\,but a similar charge and energy transfer statistics should be 
obtainable from an indirect measurement protocol where one observes 
the evolution of gauges coupled appropriately to the wires and registering 
the flow of charge and energy through the junction, see \cite{LL,LLL}. 
In our model, we compute the generating function of FCS of charge 
transfers explicitly for any $\,L\,$ and $\,t\,$ and confirm that 
it takes for large $\,t\,$ the large-deviations exponential 
form that is independent of whether $\,L\,$ is sent to infinity 
first or, \,e.g., \,kept equal to $\,t/2$. \,The equality of the large
deviation forms for the two limiting procedures appears, however, 
to be less obvious than one could have expected. The choice 
$\,L=t/2\,$ leads to the simplest calculation of the large deviation 
rate function and was implicitly employed in \cite{BD1,BD2}, where 
it was argued that it reproduces correctly the large deviations of 
the FCS for the junction of semi-infinite wires. We also compute 
explicitly the generating function of the FCS for heat transfers 
for $\,L=t/2\,$ and its large deviations form. The case of general 
$\,L\,$ and $\,t\,$ could be also dealt with but the corresponding 
formulae are considerably heavier and we did not present them here. 
The generating function of the joint FCS of the charge and energy 
transfers for $\,L=t/2\,$ and its large deviations form were also 
obtained. To our knowledge, the calculations of FCS presented in 
this paper are the first ones obtained for junctions with transmission 
and reflection modeled by conformal boundary defects. It should be 
mentioned, however, that in a different physical setup, the FCS 
of charge transfers across an inhomogeneous Luttinger liquid conductor 
connected to two leads with distinct energy distributions was obtained 
by a ``nonequilibrium bosonization'' in \cite{GGM1,GGM2,NDBM}.  
\vskip 0.1cm

The present paper is organized as follows. In Sec.\,\ref{sec:QFT}, 
we briefly recall 
the description of relativistic free massless fermions and bosons on an 
interval. We discuss the correspondence between the two theories and 
how it extends to the case of the Luttinger model of interacting 
fermions. Sec.\,\ref{sec:bosjunc} describes in detail the model of a junction 
based on a toroidal compactification of the multi-component massless
bosonic free field with a boundary defect of the type mentioned above. 
We discuss first the classical theory on a space-interval of length 
$\,L\,$ and subsequently canonically quantize that theory in 
Sec.\,\ref{sec:Quant}. In particular, we show how the scattering matrix 
$\,S\,$ relating  the chiral components of the electric current arises 
from the brane describing the boundary defect. Sec.\,\ref{sec:equilib} 
constructs the equilibrium states of the quantized theory labeled by 
inverse temperature $\,\beta\,$ and electric potential $\,V$. In 
Sec.\,\ref{sec:funcint}, we discuss the Euclidean functional integral 
representation of the equilibrium state and in Sec.\,\ref{sec:clstr}, 
its dual closed-string representation resulting from the interchange 
of time and space in the functional integral. The closed-string picture 
is particularly convenient in the thermodynamic limit $\,L\to\infty\,$ 
of the equilibrium state that is analyzed in Sec.\,\ref{sec:thermlim}. 
\,Sec.\,\ref{sec:NESS} discusses the NESS of the junction of semi-infinite 
wires kept in different temperatures and different electric potentials. 
By considering the nonequilibrium state for close temperatures and 
potentials, we obtain as a byproduct the formulae for the electric and 
thermal conductance of the junction. The central Sec.\,\ref{sec:FCS} 
is devoted to the analysis of FCS for charge and heat transfers through
the junction. Subsecs.\,\ref{subsec:ch.transp} and \ref{subsec:exact.el} 
treat the charge transport, Subsec.\,\ref{subsec:th.trans} that of heat, 
and Subsec.\,\ref{subsec:ch.th.trans} the joint FCS for both. 
Sec.\,\ref{sec:compLL} compares the generating function of FCS for charge 
and heat transfers obtained in this paper with those given by the 
Levitov-Lesovik formulae for free fermions \cite{LL,LLL} and free bosons 
\cite{Klich}. In Sec.\,\ref{sec:exampl}, we specify our general formulae 
to few simplest cases of junctions of two and three wires. Finally, 
Sec.\,\ref{sec:concl} collects our conclusions and discusses the possible 
generalizations and open problems. Appendix {A} contains the 
calculations of the generating functional of FCS for charge transfers 
at general $\,t\,$ and $\,L$. \,Appendix {B} performs the computation 
of certain bosonic Fock space expectations that are needed to obtain 
the generating function of FCS for heat transfers through the junction. 
Appendix {C} calculates the quadratic contribution to the Levitov-Lesovik 
large-deviations rate function of charge transfers for free fermions. 
\vskip 0.1cm
     
Acknowledgements: The authors thank D. Bernard for discussions on
nonequilibrium CFT and J. Germoni for Ref.\,\cite{Sims}. A part of
the work of K.G. was done within the STOSYMAP project ANR-11-BS01-015-02.

\nsection{Field theory description of quantum wires}
\label{sec:QFT}
\setcounter{equation}{0}
\subsection{Classical fermions} 
\label{subsec:clferm}

Consider a fermionic 1+1-dimensional field theory describing noninteracting 
conduction electrons in a quantum wire of length $\,L$. To a good approximation 
such electrons have a linear dispersion relation around the Fermi surface.  
For simplicity, we shall ignore here the electron spin. The classical action 
functional of the anticommuting Fermi fields of such a theory has 
the form
\begin{equation}
S[\bar\psi,\psi]\,=\,\sfrac{2\ii}{\pi}\int\dd t
\int\limits_0^L\big[\bar\psi^\ell
\partial_-\psi^\ell\,+\,
\bar\psi^r\partial_+\psi^r\big]\dd x\,,
\label{Sf}
\end{equation}
where $\,\partial_\pm=\frac{1}{2}(\partial_t\pm\partial_x)$,
\,with the boundary conditions 
\begin{equation}
\psi^\ell(t,0)=\psi^r(t,0)\,,\qquad \psi^\ell(t,L)=
-\psi^r(t,L)\,.
\label{bdc}
\end{equation}
We use the Fermi velocity $\,v_F\,$ to express time in the
same units as length. 
The classical equations obtained by extremizing action (\ref{Sf}) are
\begin{equation}
\partial_-\psi^\ell=0=\partial_-\bar\psi^\ell\,,\qquad 
\partial_+\psi^r=0=\partial_+\bar\psi^r
\end{equation}
and their solutions take the form:
\begin{align}
&\psi^\ell(t,x)\,=\,\sqrt{\sfrac{\pi}{2L}}\,
\sum\limits_{p\in{\mathbb Z}+\sfrac{1}{2}}\hspace{-0.1cm}
c_{p}\,\ee^{-\frac{\pi\ii p(t+x)}{L}}\,=\,\psi^r(t,-x)\,,\label{psis}\\
&\bar\psi^\ell(t,x)\,=\,\sqrt{\sfrac{\pi}{2L}}\,
\sum\limits_{p\in{\mathbb Z}+\sfrac{1}{2}}\hspace{-0.1cm}
\bar c_{-p}\,\ee^{-\frac{\pi\ii p(t+x)}{L}}\,=\,\bar\psi^r(t,-x)\,.
\end{align}
The space of classical solutions comes equipped with the odd symplectic 
form
\begin{equation}
\Omega\,=\,\sfrac{\ii}{\pi}
\int\limits_0^L\big[
\delta\bar\psi^\ell\wedge\delta\psi^\ell+
\delta\bar\psi^r\wedge\delta\psi^r\big]\,\dd x\,=\,\ii
\sum\limits_{p\in{\mathbb Z}+\sfrac{1}{2}}\delta\bar c_{p}\wedge\delta c_{p}
\end{equation}
leading to the odd Poisson brackets
\begin{equation}
\{c_{p},c_{p'}\}\,=\,0\,=\,\{\bar c_{p},\bar c_{p'}\}\,,
\qquad\{c_{p},\bar c_{p'}\}\,
=\,-\ii\,\delta_{p,p'}\,.
\end{equation}
The $U(1)$ symmetry 
\begin{equation}
\psi^{\ell,r}\ \mapsto\ \ee^{-\ii\alpha}\psi^{\ell,r}\,,\qquad
\bar\psi^{\ell,r}\ \mapsto\ \ee^{\ii\alpha}\bar\psi^{\ell,r}
\end{equation}
corresponds to the Noether current
\begin{equation}
J^0=\sfrac{1}{\pi}(\bar\psi^\ell\psi^\ell+\bar\psi^r\psi^r)\,,\qquad
J^1=\sfrac{1}{\pi}(\bar\psi^r\psi^r-\bar\psi^\ell\psi^\ell)
\end{equation}
with the chiral components
\begin{equation}
J^\ell=\sfrac{1}{2}(J^0-J^1)=\sfrac{1}{\pi}\bar\psi^\ell\psi^\ell\,,
\qquad
J^r=\sfrac{1}{2}(J^0+J^1)=\sfrac{1}{\pi}\bar\psi^r\psi^r
\end{equation}
and the conserved charge
\qq
Q=\int\limits_0^LJ^0(t,x)\,\dd x
=\sum\limits_{p\in\frac{1}{2}+\mathbb Z}c_p^\dagger c_p\,.
\qqq 
The classical Hamiltonian is
\begin{equation}
H\,=\,\sfrac{i}{\pi}\int\limits_0^L
\big[\bar\psi^\ell\partial_x\psi^\ell-\bar\psi^r\partial_x\psi^r\big]\,
\dd x\,=\,\sfrac{\pi}{L}\sum
\limits_{p\in{\mathbb Z}+\sfrac{1}{2}}k\,\bar c_{p}c_{p}\,.
\end{equation}

\subsection{Quantum fermions} 
\label{subsec:qferm}

Quantized Fermi fields 
$\psi^\ell$ and $\psi^r$ are given by 
expressions (\ref{psis}) with operators $\,c_{p}\,$ and their 
adjoints $c_{p}^\dagger$ satisfying the canonical anticommutation relations
\begin{equation}
[c_{p},c_{p'}]_{_+}\,=\,0\,=\,[c_{p}^\dagger,c_{p'}^\dagger]_{_+}\,,\qquad
[c_{p},c_{p'}^\dagger]_{_+}\,=\,\delta_{p,p'}\,.
\end{equation}
They act in the fermionic Fock space $\CF_{\hspace{-0.05cm}f}$ 
built upon the normalized vacuum
state $\,|0\rangle_{\hspace{-0.05cm}_f}\,$ 
annihilated by $\,c_{p}\,$ and $\,c_{-p}^\dagger\,$ for $\,p>0\,$ (the 
annihilation operators of electrons and holes, respectively). 
\,Upon quantization, fields $\,\bar\psi^\ell\,$ and $\,\bar\psi^r\,$ become the 
hermitian adjoints of $\,\psi^\ell\,$ and $\,\psi^r$. 
\,The quantum $U(1)$ currents have the chiral components
\begin{equation}
J^\ell=\sfrac{1}{\pi}:\bar\psi^\ell\psi^\ell:\,,\qquad
J^r=\sfrac{1}{\pi}:\bar\psi^r\psi^r:
\end{equation}
and the conserved $U(1)$ (electric) charge 
is\footnote{Here and below, we measure 
the electric charge in the negative units $\,-e\,$ so that electron's
charge is $+1$.} 
\begin{equation}
Q=\int\limits_0^LJ^0\,\dd x=\int\limits_0^L(J^\ell+J^r)\,\dd x\,=\,
\sum\limits_{p\in{\mathbb Z}+\sfrac{1}{2}}:c_p^\dagger c_p:.
\end{equation}
The fermionic Wick ordering putting (electron and hole) creation operators 
$\,c_{p}\,$ and $\,c_{-p}^\dagger\,$ for $p<0$ to the left of
annihilators $\,c_{p}\,$ and $\,c_{-p}^\dagger\,$ for $p>0$, \,with
a minus sign whenever a pair is interchanged, assures that the vacuum 
$|0\rangle_{\hspace{-0.05cm}_f}$ has zero charge.
\,The quantum Hamiltonian is
\begin{equation}
H\,=\,\sfrac{\pi}{L}\Big(
\sum\limits_{p\in{\mathbb Z}+\sfrac{1}{2}}\hspace{-0.15cm}
p:c_p^\dagger c_p:-\,\sfrac{1}{24}\Big),
\label{Hf}
\end{equation}
where the constant contribution is that of the zeta-function regularized
zero-point energy 
\begin{equation}
\sum\limits_{p<0}p=-\sfrac{1}{2}\sum\limits_{n=1}^\infty n+\sum\limits_{n=1}^n n
=\sfrac{1}{2}\zeta(-1)=-\sfrac{1}{24}\,.
\end{equation}

\subsection{Classical bosons}
\label{subsec:clbos}
 
Consider now a bosonic 1+1-dimensional 
massless free field $\varphi(t,x)$ defined modulo $2\pi$ on the spacetime  
$\mathbb R\times[0,L]$, with the action functional
\begin{equation} \label{action_initial0}
S[\varphi]=\frac{_{r^2}}{^{4\pi}}\int
\dd t\int\limits_0^L\big[(\partial_t
\varphi)^2-(\partial_x\varphi)^2\big]\,\dd x\,.
\end{equation}
We shall impose on $\,\varphi(t,x)\,$ the Neumann boundary conditions
\begin{equation}
\partial_x\varphi(t,0)\,=\,0\,=\,\partial_x\varphi(t,L)\,.
\end{equation}
Such a scalar field will be viewed as having the range of its values 
compactified to the circle of radius $r$ with metric $r^2(d\varphi)^2$. 
The classical solutions extremizing action (\ref{action_initial0}) have 
the form
\begin{equation}
\varphi(t,x)=\varphi^\ell(t,x)+\varphi^r(t,x)
\end{equation}
with
\begin{equation}
\varphi^\ell(t,x)\,=\,\sfrac{1}{2}\varphi_0+\sfrac{\pi}{2L}\alpha_0(t+x)
+\ii\hspace{-0.1cm}\sum_{0\not=n\in{\mathbb Z}}\hspace{-0.1cm}
\sfrac{1}{2n}\,\alpha_{2n}\,
\ee^{-\frac{\pi\ii n(t+x)}{L}}=\varphi^r(t,-x)
\label{phiell}
\end{equation}
and $\,\bar\alpha_{2n}=\alpha_{-2n}$. \,The labeling of modes $\alpha_{2n}$
by even integers is for the later convenience. 
The symplectic form on the space of classical solutions is equal to 
\begin{equation}\label{Symp_final0}
  \Omega = \sfrac{r^2}{2} \delta \alpha_0 
\wedge \delta \varphi_0 -\sfrac{\ii\,r^2}{2} \sum_{n \neq 0} 
\sfrac{1}{2n} \delta \alpha_{2n}
\wedge \delta\alpha_{-2n}\,,
\end{equation}
leading to the Poisson brackets
\begin{equation}
\{\alpha_0,\varphi_0\}\,=\,-2r^{-2}\,,\qquad \{\alpha_{2n},\alpha_{2n'}\}\,=\,
-2n\ii r^{-2}\,\delta_{n+n',0}\,.
\end{equation}
The $U(1)$ symmetry 
\begin{equation}
\varphi\ \mapsto\ \varphi+\alpha
\end{equation}
corresponds to the Noether current
\begin{equation}
J^0\,=\,\sfrac{r^2}{2\pi}\partial_t\varphi\,,\qquad J^1\,=\,
-\sfrac{r^2}{2\pi}\partial_x\varphi
\end{equation}
with the chiral components
\begin{equation}
J^{\ell,r}(t,x)\,=\,\sfrac{r^2}{2\pi}\partial_\pm\varphi(t,x)\,=\,
\sfrac{r^2}{4L}\sum\limits_{n\in{\mathbb Z}}\alpha_{2n}\,
\ee^{-\frac{\pi\ii n(t\pm x)}{L}},
\label{leftrightc}
\end{equation}
where the upper sign pertains to the left-moving component depending
on $x^+=t+x$ and the lower one to the right-moving one depending
on $x^-=t-x$. \,The classical Hamiltonian takes
the form
\begin{equation}
H\,=\,\sfrac{r^2}{4\pi}\int\limits_0^L\big[(\partial_t\varphi)^2
+(\partial_x\varphi)^2\big]\dd x\,=\,\sfrac{\pi  r^2}{4L}
\sum\limits_{n\in{\mathbb Z}}\alpha_{2n}\alpha_{-2n}\,.
\label{Hb}
\end{equation}

\subsection{Quantum bosons} 
\label{subsec:qbos}

The space $\,\CH_0\,$ of quantum states 
corresponding to the zero modes $\,\varphi_0,\alpha_0\,$ 
may be represented as $\,L^2(U(1))$, \,with $\,\varphi_0\,$ viewed
as the angle in $\,U(1)$. $\,\alpha_0\,$ acts then as 
$\,-2\ii r^{-2}\frac{\partial}{\partial\varphi_0}\,$ assuring
the commutation relation $\,[\alpha_0,\varphi_0]=-2\ii r^{-2}$. 
\,An orthonormal basis of $\,\CH_0\,$ is composed of the states 
\begin{equation}
|k\rangle\,=\,\ee^{\ii k\varphi_0}
\end{equation}
with 
\begin{equation}
\alpha_0|k\rangle\,=\,2r^{-2}k\,|k\rangle\,.
\end{equation}
The excited modes $\,\alpha_{2n}=\alpha_{-2n}^\dagger$ with the commutation
relations 
\begin{equation}
[\alpha_{2n},\alpha_{2n'}]\,=\,2n\,r^{-2}\,\delta_{n+n',0}
\end{equation}
are represented in the bosonic Fock space $\,\CF_b\,$ built upon the vacuum
state $\,|0\rangle_b\,$ annihilated by $\,\alpha_{2n}\,$ with $n>0$. 
The total bosonic space of states is $\,\CH_b=\CH_0\otimes\CF_b$. 
\,We shall identify $\,\CH_0\,$ with its subspace 
$\,\CH_0\otimes|0\rangle_b\,$ and the state $\,|0\rangle\in\CH_0\,$ 
with the vacuum $\,|0\rangle\otimes|0\rangle_b$.
\,The chiral components $\,J^{\ell,r}\,$ of the quantum $U(1)$ current 
are given by the right hand side of Eq.\,(\ref{leftrightc}). The conserved 
$U(1)$ charge takes the form
\qq
Q\,=\,\int\limits_0^LJ^0\,\dd x\,=\,\sfrac{1}{2}r^2\alpha_0
\qqq
so that $\,Q|k\rangle=k|k\rangle\,$ and it acts trivially in $\,\CF_b$.
\,The quantum Hamiltonian requires a bosonic Wick reordering putting 
the creators $\alpha_{-2n}$ for $n>0$ to the left of annihilators 
$\alpha_{2n}$ in the classical expression. Explicitly,
\begin{equation}
H\,=\,\sfrac{\pi r^2}{4L}\alpha_0^2\,+\,\sfrac{\pi r^2}{2L}
\sum\limits_{n=1}^\infty\alpha_{-2n}\alpha_{2n}\,-\,\sfrac{\pi}{24L}\,,
\end{equation} 
where the constant term is the contribution of the zeta-function regularized 
zero-point energy
\qq
\sfrac{\pi r^2}{4L}\sum\limits_{n=1}^\infty 2nr^{-2}\,=\,\sfrac{\pi}{2L}\zeta(-1)
\,=\,-\sfrac{\pi}{24L}\,. 
\qqq

\subsection{Boson-fermion correspondence} 
\label{subsec:bfcorr}

In one space dimension there is an equivalence between quantum 
relativistic free fermions and free bosons that provides a powerful
tool for the analysis of such systems \cite{Senechal,Kane}, see
also \cite{Stone} for the historical account. 
In the context of the fermionic system described in 
Sec.\,\ref{subsec:qferm}, such an equivalence involves the free bosonic 
field of Sec.\,\ref{subsec:qbos}
with the compactification radius $\,r=\sqrt{2}\,$ and is realized by 
a unitary isomorphism $\,\CI:\CH_b\rightarrow\CH_{\hspace{-0.04cm}f}\,$ that
maps vacuum to vacuum, $\,\CI\,|0\rangle_b=|0\rangle_{\hspace{-0.04cm}f}$,
\,and intertwines the action of $U(1)$ currents and the 
Hamiltonians\footnote{We choose to represent free fermions by bosons
compactified on the radius $r=\sqrt{2}$ rather than on the more frequently
used dual radius $r=\frac{1}{\sqrt{2}}$ as better suited to 
the fermionic boundary conditions (\ref{bdc}).}.
\,In particular,
\begin{equation}
\CI\,\,\alpha_{2n}\,=\,\sum\limits_{p\in\frac{1}{2}+{\mathbb Z}}\hspace{-0.15cm}
:c_p^\dagger c_{p+n}:\,\CI\,.
\end{equation}
The Fermi fields are intertwined by $\,\CI\,$ with the bosonic vertex 
operators:
\begin{align}
\psi^\ell(t,x)\,\CI
\,=\,\CI\ \sqrt{\sfrac{\pi}{2L}}:\ee^{-2\ii\varphi^\ell(t,x)}:\,\ \equiv&\ \,\,
\CI\ \sqrt{\sfrac{\pi}{2L}}\,\ee^{\frac{\pi\ii}{2L}(t+x)}
\,\ee^{-\ii\varphi_0}\,\ee^{-\frac{\pi\ii}{L}
\alpha_0(t+x)}\cr
&\times\ \ee^{\,\sum\limits_{n<0}\frac{1}{n}\alpha_{2n}\ee^{-\frac{\pi\ii n(t+x)}{L}}}
\,\ee^{\,\sum\limits_{n>0}\frac{1}{n}\alpha_{2n}\ee^{-\frac{\pi\ii n(t+x)}{L}}},\cr\cr
\bar\psi^\ell(t,x)\,\CI
\,=\,\CI\ \sqrt{\sfrac{\pi}{2L}}:\ee^{2\ii\varphi^\ell(t,x)}:\,\ \equiv&\ \,\,
\CI\ \sqrt{\sfrac{\pi}{2L}}\,\ee^{\frac{\pi\ii}{2L}(t+x)}
\,\ee^{i\varphi_0}\,\ee^{\frac{\pi\ii}{L}
\alpha_0(t+x)}\cr
&\times\ \ee^{-\sum\limits_{n<0}\frac{1}{n}\alpha_{2n}\ee^{-\frac{\pi\ii n(t+x)}{L}}}
\,\ee^{-\sum\limits_{n>0}\frac{1}{n}\alpha_{2n}\ee^{-\frac{\pi\ii n(t+x)}{L}}}.
\end{align}

\subsection{Luttinger model} 
\label{subsec:Lutt}

The interaction of electrons
near the Fermi surface gives rise to the addition of a perturbation
to the free field Hamiltonian (\ref{Hf}) that in the leading order
takes the form of a combination of quartic terms in the free fermionic
fields:
\qq
H^{\rm int}\,=\,\sfrac{1}{2\pi^2}
\int\limits_0^L\big[2g_2(:\bar\psi^\ell\psi^\ell:)
(:\bar\psi^r\psi^r:)
+g_4\big((:\bar\psi^\ell\psi^\ell:)^2+(:\bar\psi^r\psi^r:)^2\big)\big]\,
\dd x\,+\,{\rm const.}\,,
\qqq
where an infinite constant is needed to make the operator 
well defined in the fermionic Fock space. Such a perturbation defines 
the Luttinger model of spinless electrons in one-dimensional crystal 
\cite{Voit}. The crucial fact that enables an exact solution of such a model
is that, under the bosonization map, the above perturbation becomes quadratic 
in the free bosonic field:
\begin{align}
H^{\rm int}\ \CI&=\,\CI\ 
\sfrac{1}{2\pi^2}\int\limits_0^L:\big[2g_2(\partial_+\varphi)
(\partial_-\varphi)\,+\,g_4\,((\partial_+\varphi)^2+(\partial_-\varphi)^2)
\big]:\,\dd x\,+\,{\rm const.}\cr
&=\,\CI\ \sfrac{1}{4\pi^2}\int\limits_0^L:\big[(g_4+g_2)(\partial_t\varphi)^2
+(g_4-g_2)(\partial_x\varphi)^2\big]:\,\dd x\,+\,{\rm const.}\,,
\end{align}
where on the bosonic side the Wick ordering takes care of the 
diverging part of the constant on the fermionic side.
The perturbed bosonic Hamiltonian has then the form
\qq
H^{\rm tot}=H+H^{\rm int}\,=\,\sfrac{1}{4\pi^2}\int\limits_0^L:\big[(2\pi+g_4+g_2)
(\partial_t\varphi)^2+(2\pi+g_4-g_2)(\partial_x\varphi)^2\big]:\,\dd x\,
+\,{\rm const.}
\qqq
in terms of the free field $\,\varphi(t,x)\,$ with the compactification
radius $\,r=\sqrt{2}$. \,$H^{\rm tot}\,$ corresponds to the classical Hamiltonian
\qq
H^{\rm tot}\,=\,\sfrac{1}{4}\int\limits_0^L\big[(2\pi+g_4+g_2)
\Pi^2+\sfrac{2\pi+g_4-g_2}{\pi^2}(\partial_x\varphi)^2\big]\,\dd x\,,
\qqq
where $\,\Pi(t,x)=\frac{1}{\pi}(\partial_t\varphi)(t,x)\,$ is the field 
canonically conjugate to $\,\varphi(t,x)$. \,The classical Lagrangian
related to the above classical Hamiltonian is obtained by the Legendre 
transform:
\begin{align}
L^{\rm tot}&=\,\sfrac{1}{2\pi}\int\limits_0^L\big[\sfrac{2\pi}{2\pi+g_4+g_2}
(\partial_t\varphi)^2-\sfrac{2\pi+g_4-g_2}{2\pi}(\partial_x\varphi)^2\big]\,
\dd x
&=\,\sfrac{r^2}{4\pi}\hspace{-0.15cm}\int\limits_0^{\alpha^{-1}L}
\hspace{-0.1cm}\big[
(\partial_t\varphi)^2-(\partial_{x'}\varphi)^2\big]\,
\dd x'
\end{align}
for $\,x=\alpha x'$, \,where
\qq
\sfrac{r^2}{2}\,\equiv\,K\,=\,\sqrt{\sfrac{2\pi+g_4-g_2}{2\pi+g_4+g_2}}\,,
\qquad\alpha\,\equiv\,\sfrac{v_{\rm ren}}{v_F}\,
=\,\sfrac{\sqrt{(2\pi+g_4)^2-g_2^2
}}{2\pi}\,.
\label{ralpha}
\qqq
Hence after the change of the spatial variable, \,Lagrangian 
$\,L^{\rm tot}\,$ becomes that of the free bosonic field compactified on 
the radius $\,r\,$ that is different from $\,r=\sqrt{2}\,$ if $\,g_2\not=0$. 
\,The factor $\,\alpha\,$ gives the multiplicative renormalization of 
the wave velocity $\,v_F\,$ due to the interactions (we assume that $\,|2\pi+g_4|>|g_2|$). The quantization of the free bosonic theory compactified
at radius $\,r\,$ discussed in Sec.\,\ref{subsec:qbos} provides the exact 
solution of the Luttinger model on the quantum level.

\nsection{Bosonic model of a junction of quantum wires}
\label{sec:bosjunc}
\setcounter{equation}{0}

In the spirit of the ``folding trick'' of \cite{WA,OA}, \,see Fig.\,1,
\,we shall model a junction of $N$ quantum wires 
by a compactified free field $\,\bm g(t,x)\,$ with $N$-components 
$\,g_i(t,x)=\ee^{\ii\varphi_i(t,x)}\in U(1)\,$ defined on the spacetime  
$\mathbb R\times[0,L]$, \,with the action functional
\begin{equation} \label{action_initial}
S[\bm g]=\sum\limits_{i=1}^N\frac{_{r_i^2}}{^{4\pi}}\int
\dd t\int\limits_0^{L}\big((\partial_t
\varphi_i)^2-(\partial_x\varphi_i)^2\big)\,\dd x
\end{equation}
and appropriate boundary conditions.
The compactification radii $\,r_i\,$ may be different
for different wires, corresponding to different quartic coupling constants 
$g_{2i}$ and $g_{4i}$ in the Luttinger models describing the electrons 
in the individual wires, see Eq.\,(\ref{ralpha}) of 
Sec.\,\ref{subsec:Lutt}. We shall impose the Neumann reflecting 
boundary conditions at the free ends of the wires:
\begin{equation}
 \partial_x\bm\varphi(t,L) = 0\,,
\label{bcatR/2}
\end{equation}
where $\,\bm\varphi\equiv(\varphi_i)$. \,Note that we use the rescaled spatial 
variables in the wires so that the lengths of the wires in physical variables 
are fixed to $\alpha_iL$. This will not matter much because the length $L$ 
will be ultimately sent to infinity.

\begin{figure}[th]
\leavevmode
\begin{center}
\vskip -0.2cm
\includegraphics[width=7.3cm,height=3.5cm]{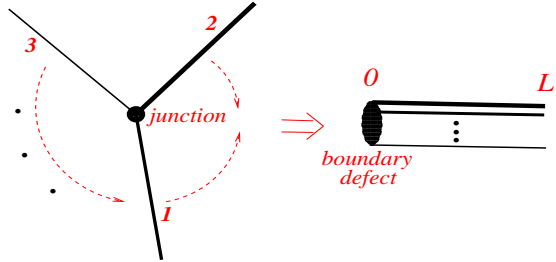}\\
\caption{Folding trick}
\end{center}
\end{figure}

\noindent The ``boundary defect'' representing in the folding trick 
the junction of wires at $\,x=0\,$ will be described by the boundary 
condition requiring that the $\,U(1)^N$-valued field 
$\,\bm g\,$ belongs to a ``brane'':
\begin{equation}
\bm g(t,0) \in\CB\equiv\kappa(U(1)^M)\subset U(1)^N\,,
\label{bcat0}
\end{equation}
where $\,\kappa:U(1)^M\rightarrow U(1)^N\,$ is a group homomorphism
\begin{equation}
 \big( e^{\ii \psi_m}\big)_{_{m=1}}^{^{M}}\,\mathop{\longmapsto}\limits^\kappa\  
\big( e^{\ii\sum_m\kappa_i^m \psi_m }\big)_{_{i=1}}^{^N}
\label{homnu}
\end{equation}
specified by integers $\kappa^m_i$. We shall assume that $\kappa$ is injective
so that $\kappa(U(1)^M)\cong U(1)^M$. \,As may be seen from the Smith normal 
form of matrix $\,\big(\kappa^m_n\big)$, such a property is assured if and 
only if the $M\times N$ matrix $\big(\kappa^m_n\big)$ has rank $M$ and 
the g.c.d. of its $M\times M$ minors is equal to 1, \,see Proposition 4.3
of \cite{Sims}. In particular, $M\leq N$ necessarily. \,Consider matrices
$\,T=(T^{mm'})\,$ and $\,P=(P_{ii'})\,$ defined by the relations
\begin{equation}
T^{mm'}=\sum_{i=1}^N r_i^2\kappa^m_i\kappa^{m'}_i\,,\qquad
P_{ii'}=\sum\limits_{m,m'=1}^M\kappa^m_i(T^{-1})_{mm'}\kappa^{m'}_{i'}r_{i'}^2\,.
\label{explicit_Pij}
\end{equation}
Matrix $\,P\,$ defines the projector on the subspace of $\,\mathbb R^N$ 
spanned 
by the vectors $\,\bm\kappa^m=(\kappa^m_1,\dots\kappa^m_N)\,$ that is orthogonal with 
respect to the scalar product
\begin{equation}\label{scalarR}
\bm a \cdot\bm b = \sum_i r_i^2 a_i b_i
\end{equation}
in $\mathbb R^N$. The boundary condition (\ref{bcat0}) implies that
\begin{equation}
\label{Dir}
P^\perp\partial_t\bm\varphi(t,0)=0\,,
\end{equation}
where $\,P^\perp\equiv I-P$.
\,The stationary points of the action
functional (\ref{action_initial}) satisfy, besides the imposed boundary
conditions, the equations
\begin{align}
&(\partial_t^2-\partial_x^2)\bm\varphi(t,x)=0\,,\label{clsW}\\
&P \,\partial_x\bm\varphi(t,0)=0\,.\label{clsN}
\end{align}
Note that relations (\ref{Dir}) and (\ref{clsN}) imply mixed 
Dirichlet-Neumann boundary conditions at $\,x=0\,$ for massless free fields
$\,\bm\varphi(t,x)$. \,The solutions of the classical equations decompose
in terms of the left- and right-movers:
\begin{equation} \label{solution+-}
\bm\varphi(t,x)=\bm\varphi^\ell(t+x)+\bm\varphi^r(t-x)
\end{equation}
with
\begin{equation}
\bm\varphi^{\ell,r}(t\pm x)\,=\,\bm\varphi^{\ell,r}_0+\sfrac{\pi}{2L}
\bm\alpha_0^{\ell,r}(t\pm x)
+\ii\sum\limits_{n\not=0}\frac{_1}{^n}\bm\alpha^{\ell,r}_n
\ee^{-\frac{\pi \ii n(t\pm x)}{2L}}\,,
\label{chiralf}
\end{equation}
where the upper sign relates to the $\bm\varphi^\ell$ and 
the lower one to $\bm\varphi^r$, \,and
\begin{equation}
\overline{\bm\alpha_{-n}^{\ell,r}}=\bm\alpha_n^{\ell,r}=(P-P^\perp)
\bm\alpha_n^{r,\ell}\,,
\qquad P\bm\alpha_{2n+1}^{\ell,r}=0\,,\qquad P^\perp\bm\alpha_{2n}^{\ell,r}=0
\label{arestr}
\end{equation}
with real $\,\bm\varphi_0=\bm\varphi_0^\ell+\bm\varphi_0^r\,$ such that 
$\,\ee^{\ii\bm\varphi_0}\in\CB$. \,In particular, $\,\bm\alpha_{2n+1}^r=
-\bm\alpha_{2n+1}^\ell\,$ and $\,\bm\alpha_{2n}^r=\bm\alpha_{2n}^\ell$. 
\,The space of classical solutions
comes equipped with the symplectic form
\begin{equation}\label{Symp_final}
  \Omega = \sfrac{1}{2}\delta\bm\alpha^\ell_0 
\cdot \wedge\delta\bm\varphi_0 -\sfrac{\ii}{2} \sum_{n \neq 0} 
\sfrac{1}{n} \delta\bm\alpha_{n}^\ell\cdot
\wedge\delta\bm\alpha_{-n}^\ell
\end{equation}
which determines the Poisson brackets of functionals on that space
that may be directly quantized.
\vskip 0.1cm

The particular case when $\,\kappa\,$ in (\ref{homnu}) is the identity mapping
of $U(1)^N$, \,corresponding to the ``space-filling'' brane $\CB_0=U(1)^M$,
\,describes the disconnected wires. In this case, $\,P=I\,$ (i.e. $P$ is the 
identity matrix) and field $\,\bm\varphi\,$
satisfies the Neumann boundary conditions both at $\,x=0\,$ and $\,x=L\,$
and only the even modes $\,\bm\alpha_{2n}^{\ell}=\bm\alpha_{2n}^r\,$ appear.
One obtains in this case the product of $N$ theories considered in
Sec.\,\ref{subsec:clbos}.

\nsection{Quantization}
\label{sec:Quant}
\setcounter{equation}{0}

\subsection{Space of states}
\label{subsec:Hspace}

The quantization of the bosonic theory of Sec.\,\ref{sec:bosjunc} is 
again straightforward but a little more involved than for the disconnected
wires. Let us first quantize the zero modes. According to the 
boundary conditions,
\begin{equation}
\bm\alpha^\ell_0 =\bm\alpha^r_0= \sum_m\beta_m\bm\kappa^m, 
\qquad \bm\varphi_0 = \sum_m \psi_m\bm\kappa^m\,,
\end{equation}
where $\,(\psi_m),\ m=1,\dots,M$, \,are angles parameterizing $\,U(1)^M$, 
\,so that
\begin{equation}
 \sfrac{1}{2}\delta\bm\alpha^\ell_0 \cdot \wedge \delta\bm\varphi_0 =
\sfrac{1}{2} \sum_{m,m'} T^{mm'}\delta\beta_m \wedge \delta
\psi_{m'}\,.
\end{equation}
The corresponding Poisson brackets are
\begin{equation}
 \big\{ \beta_m ,\,  \psi_{m'} \big\} \,=\, - 2
(T^{-1})_{mm'}
\end{equation}
leading to the commutators
\begin{equation}
 \big[  \beta_m ,\, \psi_{m'} \big] \,=\, - 2\ii 
(T^{-1})_{mm'}\,.
\end{equation}
Keeping in mind that the angular variables $\,\psi_m\,$ are
multivalued, the above commutators will be represented
in the Hilbert space $\,\CH_0=L^2((U(1)^M)\,$ of functions of $\,M\,$ 
angles $\,\psi_m$, \,square integrable in the Haar measure, by setting
\begin{equation}
 \beta_m = -2\ii  \sum_{m'} (T^{-1})_{mm'}
\dfrac{_{\partial}}{^{\partial\psi_{m'}}}\,.
\end{equation}
An orthonormal basis of $\,\CH_0\,$ is given by the states
\begin{equation}
\ket{k^1 \ldots k^{M}}\,=\,\exp \Big( \ii \sum_{m=1}^{M} k^m \psi_m
\Big) \qquad \text{for }\quad  k^m \in \mathbb Z
\end{equation}
such that
\begin{align}
 &\beta_m \ket{k^1 \ldots k^{M}} = 2
\sum_{m'}(T^{-1})_{mm'} k^{m'} \ket{k^1 \ldots
k^{M}}
\end{align}
and
\begin{equation} \label{actionalpha0}
\bm\alpha^\ell_0 \ket{k^1 \ldots k^{M}} = 2 \sum_{m,m'}\bm\kappa^m
(T^{-1})_{mm'} k^{m'} \ket{k^1 \ldots
k^{M}}.
\end{equation}
For the excited modes, it is convenient to introduce a basis 
$\,(\bm\Lambda_j)_{j=1}^N$ of vectors $\,\bm\Lambda_j\hspace{-0.06cm}
=(\Lambda_{j1},\dots,\Lambda_{jN})$
in $\,\mathbb R^N$ such that
\begin{equation}
\bm\Lambda_j\cdot\bm\Lambda_{j'}=\delta_{jj'},\qquad 
P\bm\Lambda_j=\bm\Lambda_j\quad\text{for}\quad
j\leq M\,,\qquad P^\perp\bm\Lambda_j=\bm\Lambda_j\quad\text{for}\quad j>M
\label{Lambdai}
\end{equation}
and the projected modes
\begin{equation}
\tilde\alpha^{\ell,r}_{nj}=\bm\Lambda_j\cdot\alpha_n^{\ell,r}\,.
\end{equation}
with the inverse formulae
\begin{equation}
\alpha^{\ell,r}_{ni}=\sum\limits_{j=1}^N\Lambda_{ji}\,
\tilde\alpha^{\ell,r}_{nj}\,.
\end{equation}
Note that relations (\ref{arestr}) imply that $\,\tilde\alpha^\ell_{(2n+1)\,j}
=0\,$ for $\,j\leq M\,$ and $\,\tilde\alpha^\ell_{(2n)\,j}=0\,$ for $\,j>M$.
\,The Poisson brackets of the non-zero operators $\,\tilde\alpha_{nj}\,$
take the form
\begin{equation}
\{\tilde\alpha^\ell_{nj},\tilde\alpha^\ell_{n'j'}\}
=-\ii n\,\delta_{jj'}\delta_{n+n',0}
\end{equation}
leading to the commutators
\begin{equation}
[\tilde\alpha^\ell_{nj},\tilde\alpha^\ell_{n'j'}]
=n\,\delta_{jj'}\delta_{n+n',0}\,.
\end{equation}
In the standard Fock space quantization, we take
\begin{equation}
 \mathcal F_e = \mathop{\otimes}_{j=1}^M \mathcal F_{ej}, \qquad 
\qquad \mathcal F_o =
\mathop{\otimes}_{j=M+1}^N \mathcal F_{oj}
\end{equation}
where $\mathcal F_{ej}$ and $\mathcal F_{oj}$ are 
generated by vectors
\begin{align}
&  \tilde \alpha_{-(2n_1) j}^\ell \ldots \tilde\alpha_{-(2n_l)j}^\ell
\ket{0}_{e} \qquad\qquad l \geq 0,
\qquad n_1 \geq \ldots \geq n_l \geq 1, \qquad j \in \lbrace 1,
\ldots, M \rbrace\,, \\
& \tilde\alpha_{-(2n_1+1) j}^\ell \ldots \tilde\alpha_{-(2n_{l}+1)j}^\ell
\ket{0}_{o} \qquad l \geq 0,
\qquad n_1 \geq \ldots \geq n_{l} \geq 0, \qquad  j\in \lbrace M+1,
\ldots, N \rbrace\ 
\end{align}
with the scalar product determined by the relations
\begin{align}
&\tilde \alpha_{(2n)\,j}^\ell \ket{0}_{e} = 0 \ \quad 
\text{for }\quad n > 0\,,\qquad
\tilde \alpha_{(2n+1)\,j}^\ell \ket{0}_{o} = 0 \ \quad 
\text{for }\quad n \geq 0\,,\\
&{}_e\langle 0 | 0 \rangle_e = {_o}{\langle} 0 | 0 \rangle_o = 1\,,
\qquad{(\tilde \alpha_{nj}^\ell)}^\dagger = \tilde \alpha_{(-n)j}^\ell\,. 
\end{align}
The total Hilbert space of states of the theory is 
\begin{equation}
 \mathcal H = \mathcal H_0 \otimes \mathcal F_e \otimes \mathcal F_o
\end{equation}
and in the following we identify
\begin{equation}
 \ket { k^1 \ldots k^{M} } \equiv \ket { k^1 \ldots k^{M} } \otimes
\ket{0}_{e} \otimes \ket{0}_{o}
\end{equation}

\subsection{Currents, charge and energy}
\label{subsec:curchen}

\noindent We shall be interested in the system that possesses global $U(1)$
symmetry acting on fields by $\,(g_i(t,x))\mapsto(ug_i(t,x))\,$ for 
$\,u\in U(1)$. \,Invariance of the theory requires that this action 
preserves the brane
$\,\CB=\kappa(U(M))\subset U(N)$. \,This holds if and only if the vector 
$\,\bm 1=(1,\dots,1)\,$ is in the image of projector $\,P$,
\,i.e. if $\,P\bm1=\bm1\,$ or 
\begin{equation}
\sum\limits_{i'}P_{ii'}=1\quad\ \text{for all}\quad i\,.
\label{symcond}
\end{equation}
The Noether (electric) current corresponding to the $U(1)$ symmetry has 
then the form
\begin{align}
J^0(t,x)=\sum\limits_{i=1}^NJ^0_i(t,x)\,,
\qquad
J^1(t,x)=\sum\limits_{i=1}^N J^1_i(t,x)
\end{align} 
in terms of the currents in individual wires with the left-right 
moving components
\begin{equation}
J_i^{\ell,r}(t,x)=\sfrac{1}{2}(J^0\mp J^1)(t,x)=\frac{_{r_i^2}}{^{2\pi}}
\sfrac{1}{2}(\partial_t\pm\partial_x) \varphi_i(t,x)
=\frac{_{r_i^2}}{^{4 L}} \sum\limits_{n\in{\mathbb Z}}
\alpha^{\ell,r}_{ni}\ee^{-\frac{\pi\ii n(t\pm x)}{2L}}\label{JLR}
\end{equation}
defining $J_i^\ell(t,x)$ and $J^r(t,x)$
as functions, respectively, of $t+x$ and $t-x$ for any real $t$ and $x$.
\,We shall use formulae (\ref{JLR}) also for quantum currents.
At $\,x=0$, \,the left and right currents are linearly related:
\begin{equation}\label{linkJLR}
J_i^r(t,0)\,=\,\sum_{i'}
S_{ii'}J_{i'}^\ell(t,0)\,,
\end{equation}
where
\begin{equation}
S_{ii'}\,=\,P_{i'i}-P^\perp_{i'i}
\label{Smatrix}
\end{equation}
according to \eqref{arestr} and the explicit expression
\eqref{explicit_Pij} for the matrix $P$.
The $N\times N$ \,''$S$-matrix'' $S=(S_{ii'})$ describes the flow of
the currents through the junction
of wires. It satisfies the relations
\begin{align}
&S_{i'i}=r_i^{-2}S_{ii'}r_{i'}^2\,,\qquad\sum_{i'} S_{ii'}S_{i'i"}
=\delta_{ii"}\,,\qquad\sum\limits_{i}S_{ii'}=1\,,\label{symcond1}
\end{align}
In other words, 
\begin{equation}
S_{ii'}=r_iO_{ii'}r_{i'}^{-1}
\label{O}
\end{equation}
where $O=(O_{ii'})$ is a symmetric orthogonal matrix such that
\begin{equation}
\sum\limits_ir_iO_{ii'}=r_{i'}\,.
\end{equation}
We shall use matrices $\,S\,$ and $\,O\,$ interchangeably.
For $\,N=2$, \,there are two possibilities:
\begin{equation}
O\,=\,\Big(\begin{matrix}1&0\cr 0&1\end{matrix}\Big)\qquad{\rm or}
\qquad O\,=\,\frac{1}{r_1^2+r_2^2}\Big(\begin{matrix}{r_1^2-r_2^2}
&{2r_1r_2}\cr {2r_1r_2}&
{r_2^2-r_1^2}\end{matrix}\Big).
\label{ON2}
\end{equation}
The first case corresponds to the identity
embedding $\,\kappa\,$ describing the disconnected wires whereas the
second one corresponds to the diagonal embedding of $\,U(1)\,$ into 
$\,U(1)^2\,$ that leads to a nontrivial junction. In the last case, 
the $r_1=r_2$ case corresponds to off-diagonal matrix $O=S$ with unit 
non-zero entries, i.e. to the pure transmission of currents through 
the junction, but for $\,r_1\not=r_2\,$ the currents are partly 
transmitted and partly reflected at the junction. 
\vskip 0.1cm
 
Eq. (\ref{linkJLR}) implies that the right currents are 
linear combinations of left currents if considered as functions of real 
$t$ and $x$:
\begin{align}\label{linkJLR1}
J_i^r(t,x) = \sum_{i'}
S_{ii'}J_{i'}^\ell(t,-x)\,.
\end{align}
At $\,x = L$, \,i.e. at the ends of the wires,
the left and right currents are equal:
\begin{equation}
 J_i^r(t,L) = J_i^\ell(t,L)
\end{equation}
which implies that
\begin{equation}
\label{linkJLR2}
 J_i^r(t,x) = J_i^\ell(t,-x+2L)
\end{equation}
if we treat the currents as functions of real $t$ and $x$. \,The
quantum currents satisfy the equal-time commutation relations
\begin{align}
[J^\ell_i(t,x),J^\ell_{i'}(t,y)]&=\sfrac{\ii r_i^2}{4\pi}
\sum_{n\in\mathbb Z}\big(P_{ii'}+(-1)^n(\delta_{ii'}-P_{ii'})\big)
\,\delta'(x-y+2nL)\cr
&=-[J^r_i(t,x),J^r_{i'}(t,y)]\,,\label{comrelcur}\\ \cr
[J^\ell_i(t,x),J^r_{i'}(t,y)]&=\sfrac{\ii r_i^2}{4\pi}
\sum_{n\in\mathbb Z}\big(P_{ii'}-(-1)^n(\delta_{ii'}-P_{ii'})\big)
\,\delta'(x+y+2nL)\,,
\label{comrelcur1}
\end{align}
In particular, the left-moving currents commute among themselves at equal 
times if their positions do not coincide modulo $2L$. Similarly
for the right-moving currents. The left-moving currents commute
with the right-moving ones at equal times if their positions are
not opposite modulo $2L$. Note that for $\,0<x,y\leq L$ the only terms 
that contribute to  (\ref{comrelcur}) and (\ref{comrelcur1}) have
$n=0$ or $n=1$, respectively, so that for such values of $x$ and $y$ 
the commutation relations of currents do not depend on the choice
of brane $\CB$. This permits to identify for different junctions 
the algebras of observables generated by currents 
$J^{\ell,r}_i(0,x)$ with $0<x\leq L$, i.e. localized away from 
the contact point. 
In particular, we may identify such observables for disconnected wires 
with those for connected wires, with the physical meaning that their 
measurement just before and just after establishing or breaking the 
connection between the wires should give the same result. \,Whatever 
the junction, the total charge
\begin{equation}
 Q(t) = \sum\limits_{i=1}^NQ_i(t)\,,
\end{equation}
where $Q_i(t)$ are the charges in the individual wires,
\begin{equation}
Q_i(t)=\int\limits_{0}^{L}J^0_i(t,x)\,\dd x\,,
\end{equation}
is conserved: 
\begin{align}
&\dfrac{\dd Q(t)}{\dd t}  = \sum_{i=1}^N 
\int\limits_{0}^{L} (\partial_t J_i^\ell(t,x) +
\partial_t J_i^r(t,x))\dd x\cr 
&= \sum_{i=1}^N \int\limits_{0}^{L} (\partial_x J_i^\ell(t,x) 
- \partial_x J_i^r(t,x))\dd x
= \sum_{i=1}^N \left[ J_i^\ell(t,x) - J_i^r(t,x)\right]_{x=0}^{x=L}\cr
&=2\sum_{i,i'=1}^N(\delta_{ii'}-P_{i'i})J^\ell_{i'}(t,0)=0
\end{align}
due to (\ref{symcond}).
In terms of the modes, 
\begin{equation}
 Q = \sum_{i} \frac{_{r_i^2}}{^{ 2}}\alpha_{0i}^\ell\,.
\end{equation}
Operator $\,Q\,$ acts only on $\,\mathcal H_0\,$:
\begin{equation}
Q \ket{ k^1 \ldots k^M}  = \sum_i \sum_{m,m'} r_i^2 \kappa_i^m (T^{-1})_{mm'}
k^{m'}
\ket{
k^1 \ldots k^M} = (\bm p , T^{-1} \bm k) \ket{
k^1 \ldots k^M},
\label{Qeigen}
\end{equation}
where $\,\bm p=(p^m) \in \mathbb R^M\,$ with
\begin{equation}\label{def_p_os}
 p^m = \sum_i r_i^2 \kappa_i^m=\bm1\cdot\bm\kappa^m\,,
\end{equation}
and 
\begin{equation}
 (\bm a,\bm b) = \sum_m a^m b^m\,,
\end{equation}
denotes the standard scalar product on $\mathbb R^M$,
to be distinguished from the one of (\ref{scalarR}) used in $\mathbb R^N$.
\,Note that the spectrum of $\,Q\,$ is composed of integers, as must be 
the case for the generator of a unitary action of $\,U(1)\,$ group. 
\,Indeed, since for each $1\leq i\leq N$,
\begin{equation}
\sum\limits_{m,m'=1}^Mp^m(T^{-1})_{mm'}\kappa^{m'}_i=(P\bm1)_i=1
\end{equation}
by (\ref{symcond}), the injectivity of the homomorphism (\ref{homnu}) implies
that the sums $\sum\limits_{m}p^m(T^{-1})_{mm'}$ are integers. This is not
the case for (non-conserved) charges in the individual wires
\begin{equation}
Q_i(t)=\int\limits_0^{L}J^0_i(t,x)\,\dd x=\sfrac{r_i^2}{2}\,
\alpha_{0i}^\ell-\sfrac{\ii r_i^2}{\pi}\sum\limits_{n}\sfrac{1}{2n+1}\,
\alpha^\ell_{(2n+1)i}\,\ee^{-\frac{\pi\ii(2n+1)t}{2L}}.
\end{equation}
\vskip 0.2cm

The energy of the bosonic system of Sec.\,\ref{sec:bosjunc} is given by
its classical Hamiltonian that may be expressed 
in terms of the left and right moving currents by the formula
\begin{equation}
 H(t) = \sum_{i=1}^N\sfrac{2\pi}{r_i^{2}}\int\limits_0^{L}\big((J_i^\ell(t,x))^2 +
(J_i^r(t,x))^2\big)\,\dd x\,.
\end{equation}
Its conservation, that holds independently of the condition (\ref{symcond}),
results from the identity
\begin{align}
 \dfrac{\dd H(t)}{\dd t} =  \sum_{i=1}^N 2\pi r_i^{-2}\left[
(J_i^\ell)^2(t,x) - (J_i^r)^2(t,x)\right]_{x=0}^{x=L}
\end{align}
whose right hand side vanishes because 
\begin{equation}
 (J_i^r)^2(t,L) = (J_i^\ell)^2(t,L)
\end{equation}
for all $1\leq i\leq N$, \,and because
\begin{align}
\sum_{i=1}^N r_i^{-2}(J_i^r)^2(t,0) = \sum_{i,i',i"=1}^N r_i^{-2}
S_{ii'}S_{ii"}J_{i'}^\ell(t,0)J_{i"}^\ell(t,0) =
\sum_{i=1}^N r_i^{-2}(J_i^\ell)^2(t,0)
\end{align}
in virtue of (\ref{symcond1}). 
\,The quantum Hamiltonian $\,H\,$ is given by the Wick reordered version 
of the classical expression:
\begin{align}
&H =  \sum_{i=1}^N \sfrac{2\pi}{r_i^{2}} \int\limits_{0}^{L}
(:J_i^\ell(t,x)^2:+:J_i^r(t,x)^2:)\,\dd x\,+\sfrac{\pi}{L}\sfrac{N-3M}{48}\cr
&= \sfrac{\pi}{4L}\,\bm\alpha^\ell_0\cdot\bm\alpha^\ell_0 +  
\sfrac{\pi}{2L} \sum_{i=1}^M \sum_{n>0} 
\tilde \alpha_{(-2n)i}^\ell \tilde \alpha_{(2n)i}^\ell
+  \sfrac{\pi}{2L} \sum_{i=M+1}^N 
\sum_{n\geq0} 
\tilde \alpha_{(-(2n+1))i}^\ell \tilde \alpha_{(2n+1)i}^\ell\,
\,+\sfrac{\pi}{L}\sfrac{N-3M}{48}\,,\qquad
\,,
\label{Hamilt}
\end{align}
where the last c-number term accounts for the $\zeta$-function regularized
zero-point energy of the excited modes:
\begin{equation}
\sfrac{\pi}{2L}\sfrac{M}{2}\sum_{n=1}^\infty 2n+
\sfrac{\pi}{2L}\sfrac{N-M}{2}\sum\limits_{n=0}^\infty
(2n+1)=\sfrac{\pi}{2L}\sfrac{3M-N}{2}\,\zeta(-1)=\sfrac{\pi}{L}
\sfrac{N-3M}{48}.
\end{equation}
The Hilbert space vectors
\begin{equation}
 \ket u = \tilde \alpha_{-(2n^{j_1}_1) j_1}^\ell \ldots
\tilde\alpha_{-(2n^{j_l}_l)j_l}^\ell
\tilde\alpha_{-(2n^{j'_1}_1+1)j'_1}^\ell \ldots
\tilde\alpha_{-(2n^{j'_{l'}}_{l'}+1)j'_{l'}}^\ell\ket{ k_1 \ldots k_{M} }
\end{equation}
with $j_1,\dots,j_l\leq M$ and $j'_1,\dots,j'_{l'}>M\,$ form a
basis of eigen-states of $H$ with
\begin{equation}
 H \ket u = \sfrac{\pi}{2L} \left(2\, (\bm k , T^{-1}
\bm k)  + n +\sfrac{N-3M}{24}\right)\ket{u}\,, 
\label{Heigen}
\end{equation}
where
\begin{equation}
 n = \sum_{k=1}^l 2 n^{j_k}_k + \sum_{k'=1}^{l'} (2
n^{j'_{k'}}_{k'} +1)\,.
\end{equation}
The energy density and the energy current in the wires correspond, 
respectively, to operators
$\,K^0_i(t,x)=T^r_i(t,x)+T^\ell_i(t,x)\,$ and $\,K^1_i(t,x)=T^r_i(t,x)
-T^\ell_i(t,x)$, 
\,where
\begin{align}
&T^\ell_i(t,x)=\sfrac{2\pi}{r_i^2}:J_i^\ell(t,x)^2:-\,\sfrac{\pi}{48L^2}P_{ii}\,+\,
\sfrac{\pi}{96L^2}P^\perp_{ii}\,,\label{KL}\\
&T^r_i(t,x)=\sfrac{2\pi}{r_i^2}:J_i^r(t,x)^2:-\,\sfrac{\pi}{48L^2}P_{ii}\,+\,
\sfrac{\pi}{96L^2}P^\perp_{ii}\,,
\label{KR}
\end{align} 
are the left-moving and right-moving
energy-momentum-tensor components. The constant terms are the zero-point 
energy contributions. Note that the above choice assures 
by virtue of relation $\,\sum\limits_{i=1}^NP_{ii}=M\,$ that
\begin{equation}
H=\sum\limits_{i=1}^NH_i(t)\,,
\end{equation}
where 
\begin{equation}
H_i(t)=\int\limits_0^{L}K_i^0(t,x)\,\dd x
\label{Hi}
\end{equation}
are the observables representing energy in individual wires.

\nsection{Equilibrium state.} 
\label{sec:equilib}
\setcounter{equation}{0}

\noindent The equilibrium state at inverse temperature
$\beta$ and (electric) potential $\,V\,$ is described by the density matrix
\begin{equation}
\rho_{\beta,V}=\frac{1}{Z_{\beta,V}}\,\ee^{-\beta(H-VQ)}\,,
\end{equation}
where $\,Z_{\beta,V}\,$ is the the partition function. Note that with our
conventions, positive $\,V\,$ plays the role of a positive chemical potential 
for electrons and of a negative one for holes.
$\,Z_{\beta,V}\,$ is easily calculable with the use of relations
(\ref{Heigen}) and (\ref{Qeigen}):
\begin{align}
&Z_{\beta,V}= \Tr_{\mathcal H} \Big( \ee^{-\beta (H-VQ)} \Big) \cr 
&=\ee^{-\frac{(N-3M) \pi
\beta}{48
L}} \bigg(\hspace{-0.05cm}\sum_{\bm k \in
\mathbb Z^M} \ee^{-\frac{\pi \beta }{L} (\bm k ,T^{-1} \bm k) 
+ \beta V(\bm p, T^{-1} \bm k)}\bigg)
\hspace{-0.05cm}\bigg(\sum_{n \geq 0}  p_e(n)\, 
\ee^{-\frac{ \pi \beta }{2L} n}
\bigg)^{\hspace{-0.1cm}M}\hspace{-0.1cm}\bigg(\sum_{n \geq 0}  p_o(n)\, \ee^{-
\frac{ \pi \beta }{2L} n}\bigg)^{\hspace{-0.1cm}N-M}\ \quad
\label{Zpart0}
\end{align}
with $p_{e}(n)$ ($p_o(n)$) standing for the number of partitions of $n$ 
into a sum of even (odd) numbers. The Poisson resummation formula
applied to the $\bm k$-sum and the standard relation of the 
generating function for partitions to the Dedekind function
$\,\eta(\tau)=\ee^{\frac{\pi i\tau}{12}}\prod\limits_{n=1}^\infty
\big(1-\ee^{2\pi\ii\tau n}\big)\,$ allow to rewrite (\ref{Zpart0}) as
\begin{equation}
Z_{\beta,V}
=\Big(\sfrac{L}{\beta}\Big)^{\hspace{-0.1cm}\frac{M}{2}}
\hspace{-0.1cm}\sqrt{\det(T)}\ \ee^{\frac{ L   \beta V^2}{4\pi}\sum_i
r_i^2}\bigg(
\sum_{\bm k
\in
\mathbb Z^{M}} \ee^{-\ii L V(\bm p,\bm k) -
\frac{\pi
L}{\beta}
(\bm k,
T\bm k) }\bigg)
\big[\eta( \frac{_{\ii
\beta}}{^{2L}})\big]^{\hspace{-0.05cm}N-2M}\big[\eta( \frac{_{\ii
\beta}}{^{4L}})\big]^{\hspace{-0.05cm}M-N},
\label{Zpart}
\end{equation}
or, with the use of the modular property of 
$\,\eta(\tau)=\frac{1}{\sqrt{-i\tau}}\eta(-\frac{1}{\tau})$, \,as 
\begin{equation}
Z_{\beta,V}
=2^{-N}\hspace{-0.05cm} 
\sqrt{\det(T)}\ \ee^{\frac{ L   \beta V^2}{4\pi}\sum_i
r_i^2}\bigg(\sum_{\bm k
\in
\mathbb Z^{M}} \ee^{-\ii L V(\bm p,\bm k) -
\frac{\pi
L}{\beta}
(\bm k, 
T\bm k) }\bigg)\big[\eta(\frac{_{2\ii
L}}{^{
\beta}})\big]^{\hspace{-0.05cm}N-2M}\big[\eta( \frac{_{4\ii L}}{^{
\beta}})\big]^{\hspace{-0.05cm}M-N}.
\label{Zbm}
\end{equation}
The equilibrium state $\,\omega^L_{\beta,V}$ expectations of the observable 
algebra generated by currents $J^{\ell,r}_i$ are defined by the formula
\begin{equation}
\omega^L_{\beta,V}(A)=\Tr\big(\rho_{\beta,V}A\big)\,.
\end{equation}
The superscript $L$ in $\,\omega^L_{\beta,V}$ stresses that the state pertains
to the junction of wires of length $L$. 
In forming observables, it is enough to consider only the currents 
$J^\ell(t,x)$ at fixed $t$ and real $x$. 
We shall decompose such currents into the contributions
from the zero modes and the excited modes:
\begin{equation}
J^\ell_i(t,x) = \sfrac{r_i^2}{4L}\alpha^\ell_{0i}+\hat J^\ell_i(t,x)\,,
\end{equation}
see (\ref{JLR}).
In the equilibrium state expectation of products of currents, 
the contributions from
the zero modes and from the excited modes factorize. In particular,
\begin{align}
\omega^L_{\beta,V}\Big(\prod\limits_{k=1}^K\sfrac{r_{i_k}^2}{2L}\,
\alpha_{0i_k}^\ell
\Big)&=\frac{1}{(2\beta V L)^{K}}\,\dfrac{
\prod\limits_{k=1}^K
\big(r_{i_k}^2\sum\limits_{m}\kappa^{m}_{i_k}\frac{\partial}{\partial p^m}\big)
\sum\limits_{\bm k\in\mathbb Z^M}\,
\ee^{-\frac{\pi \beta }{L} (\bm k ,T^{-1} \bm
k) + \beta V(\bm p,T^{-1} \bm
k)}}{\sum\limits_{\bm k\in\mathbb Z^M}\ee^{-\frac{ \pi \beta }{L} (\bm k ,
T^{-1} \bm k) + \beta V(\bm p,T^{-1} \bm k)}}\cr
&\hspace{-1.5cm}=\frac{1}{(2\beta V L)^{K}}\,\dfrac{
\prod\limits_{k=1}^K
\big(r_{i_k}^2\sum\limits_{m}\kappa^{m}_{i_k}\frac{\partial}{\partial p^m}\big)\,
\,\big(\ee^{\frac{ L  
\beta V^2}{4\pi}(\bm p,T^{-1} \bm p)}\hspace{-0.1cm}
\sum\limits_{\bm k \in \mathbb
Z^M} 
\ee^{-\ii L V(\bm p,\bm k) -
\frac{\pi
L}{\beta}
(\bm k,T\bm k)}\big)}
{\ee^{\frac{ L  
\beta V^2}{4\pi}(\bm p,T^{-1} \bm p)}
\sum\limits_{\bm k \in \mathbb Z^M} 
\ee^{-\frac{\ii L V}{2}(\bm p, \bm k) -
\frac{\pi L}{\beta}
(\bm k,T\bm k)}}\,,\quad
\label{omega0modes}
\end{align}
where the second equality results from the Poisson resummation.
On the other hand, the expectations of products of $\hat J^+_i$ are
calculated by the Wick rule with
\begin{align}
&\omega^L_{\beta,V}\big(\hat J^\ell_i(t,x)\big)=0\,,\label{1ptfn}\\
&\omega^L_{\beta,V}\big(\hat J^\ell_{i_1}(t,x_1)\,\hat J^\ell_{i_2}(t,x_2)\big)=
- \sfrac{1}{2} \left(\sfrac{r_{i_1}}{2\pi}\right)^2\Big(
P_{i_1i_2}\,f_e(x_1-x_2)+(\delta_{i_1i_2}-P_{i_1i_2})f_o(x_1-x_2)\Big),\quad
\label{2ptfn}
\end{align}
where
\begin{align}
f_{e}(x_1-x_2) &=\wp(x_1-x_2;2L,-\ii\beta) + C_e\cr
f_{o}(x_1-x_2) &=2 {\wp}(x_1-x_2;4L,-\ii\beta) -
{\wp}(x_1-x_2;2L,-\ii\beta)+C_{o} 
\end{align}
with the constants
\begin{align}
C_e=&\big(\sfrac{\pi}{2L}\big)^2 \Big(\sfrac{1}{3} - \sum\limits_{n \neq 0}
\sinh^{-2}(\sfrac{\pi n\beta}{2L})\Big),\\
C_{o}=&\big(\sfrac{\pi}{2L}\big)^2\Big(-\sfrac{1}{6} 
-\sfrac{1}{2}
\sum\limits_{n \neq0}
\sinh^{-2}(\sfrac{\pi n\beta}{4L}) + 
\sum\limits_{n \neq 0}
\sinh^{-2}(\sfrac{\pi n\beta}{2L})\Big).
\end{align}
Above, $\wp(z;\omega_1,\omega_2)=\wp(z;\omega_2,\omega_1)$ is 
the Weierstrass function of period
$\omega_1$ and $\omega_2$ \cite{WW}:
\begin{align}
\wp(z;\omega_1,\omega_2)&=\sfrac{1}{z^2}\,+\sum_{n^2+m^2 \ne 0} 
\Big( \sfrac{1}{(z+m\omega_1+n\omega_2)^2}- 
\sfrac{1}{\left(m\omega_1+n\omega_2\right)^2}\Big)\cr
&= (\sfrac{\pi}{\omega_2})^2\Big[-\sfrac{1}{3}+\sum\limits_{n}
\sin^{-2}\big(\sfrac{\pi(z-n\omega_1)}{\omega_2}\big)
-\sum\limits_{n\not=0}\sin^{-2}\big(\sfrac{\pi n\omega_1}
{\omega_2}\big)\Big]. 
\label{Weierstrass}
\end{align}
Note the singularity of the 2-point functions (\ref{2ptfn}) at
the insertion points coinciding modulo $2L$. For such points, the equal-time
commutators of currents have contact terms, see (\ref{comrelcur}).
\,For the 1-point function of the
left current, one obtains
\begin{equation} \label{1pt_init_develop}
 \omega^L_{\beta,V}\left( J_{i}^\ell(t,x)
\right)
= \sfrac{r_i^2 V}{4 \pi}  -
\sfrac{\ii r_i^2}{2\beta} \sum_m \kappa_{i}^m\,
\dfrac{\displaystyle \sum_{\bm k \in \mathbb Z^M}  k^{m} \ee^{-\ii L
 V(\bm p,\bm k) -
\frac{\pi
L}{\beta}
(\bm k,
T\bm k) }}{\displaystyle \sum_{\bm k \in \mathbb Z^M} \ee^{-\ii L
V(\bm p, \bm k) -
\frac{\pi L}{\beta}
(\bm k,T\bm k) }}\,,
\end{equation}
where we have used the relation
\begin{equation}
 \sfrac{r_i^2}{2L\beta V} \sum_{m}  \kappa_{i}^m
\sfrac{\partial\,}{\partial p^m}\big(\sfrac{ L  
\beta V^2}{4\pi}(\bm p,T^{-1} \bm p)\big) 
=  \sfrac{r_i^2 V}{4 \pi}
\sum_{m,m'} \kappa_{i}^m (T^{-1})_{mm'} p^{m'} 
 = \sfrac{r_i^2 V}{4 \pi} \sum_{j} P_{ij} 
 = \sfrac{r_i^2 V}{4 \pi}\,. 
\end{equation}
From (\ref{linkJLR2}) or (\ref{linkJLR1}), it follows
that $\,
 \omega^L_{\beta,V}\left( J_{i}^r(t,x)\right)=
 \omega^L_{\beta,V}\left( J_{i}^\ell(t,x)
\right)\,$
so that                                                                  
\begin{equation}
 \omega^L_{\beta,V}\left( J_{i}^0(t,x)\right)=
 2\,\omega^L_{\beta,V}\left( J_{i}^\ell(t,x)
\right)\,,\qquad 
\omega^L_{\beta,V}\left( J_{i}^1(t,x)\right)=0\,.
\end{equation}
Hence, in the equilibrium state, the mean charge density is constant 
in each wire, whereas the mean current vanishes. 
\vskip 0.1cm

The equilibrium state $\,\omega^L_{\beta,V}$ is invariant 
under the replacement $J^\ell\leftrightarrow J^r$, \,the property expressing 
its time-reversal invariance. For the energy-momentum tensor
components $T_i^{\ell,r}(t,x)$ defined by (\ref{KL}) and (\ref{KR}), \,we obtain:
\begin{align}
\omega^L_{\beta,V}\big(T_i^\ell(t,x)\big)&=\sfrac{\pi}{8L^2}r_i^2\,
\omega^L_{\beta,V}
\big((\alpha_{0i}^+)^2\big)\,
-\,\sfrac{1}{4\pi}\big(P_{ii}C_e+(1-P_{ii})C_o\big)\cr
&=\lim\limits_{\epsilon\to 0}\,\omega^L_{\beta,V}
\Big(\sfrac{2\pi}{r_i^2}\,J^\ell_i(t,x+\epsilon)\,
J^\ell_i(t,x)\,+\,\sfrac{1}{4\pi\epsilon^2}\Big)\cr
&=\lim\limits_{\epsilon\to 0}\,\omega^L_{\beta,V}
\Big(\sfrac{2\pi}{r_i^2}\,J^r_i(t,x+\epsilon)\,
J^r_i(t,x)\,+\,\sfrac{1}{4\pi\epsilon^2}\Big)\,
=\,\omega^L_{\beta,V}\big(T_i^r(t,x)\big),
\label{oOPE}
\end{align}
which is a consequence of the operator product 
expansion
\begin{align}
&\sfrac{2\pi}{r_i^2}\,J^\ell_i(t,x+\epsilon)\,J^\ell_i(t,x)=
-\sfrac{1}{4\pi\epsilon^2}+T_i^\ell(t,x)\,+\,\dots\label{OPEl}\\
&\sfrac{2\pi}{r_i^2}\,J^r_i(t,x+\epsilon)\,J^r_i(t,x)=
-\sfrac{1}{4\pi\epsilon^2}+T_i^r(t,x)\,+\,\dots\label{OPEr}
\end{align}
holding under the equilibrium expectations away from other insertions
points.

\nsection{Functional integral representation}
\label{sec:funcint}
\subsection{Case with $\,V=0$}
\label{subsec:V=0}
\setcounter{equation}{0}

\noindent For $\,V=0$, \,the partition function $\,Z_{\beta,0}\equiv Z_\beta\,$ 
and the expectations in the thermal equilibrium
states $\,\omega^L_{\beta,0}\equiv\omega^L_\beta\,$ may be represented 
by Euclidean functional integrals over a cylindrical open-string worldsheet, 
\,see Fig.\,2.

\begin{figure}[th]
\leavevmode
\begin{center}
\vskip -0.2cm
\includegraphics[width=6cm,height=2.8cm]{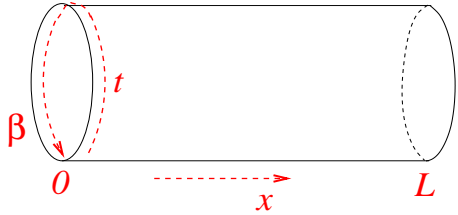}\\
\caption{Open string worldsheet}
\end{center}
\end{figure}

\noindent For the partition function,
\begin{equation}
Z_\beta=\int\ee^{-S_E[g]}\,\CD g
\label{fcin0}
\end{equation}
where the functional integral is over the maps 
$\,\bm g(t,x)=(\ee^{i\varphi_i(t,x)})\,$ from 
$\mathbb R\times[0,L]$ to $U(1)^N$ periodic in $t$ 
\begin{equation}
\bm g(t+\beta,x)=\bm g(t,x)
\end{equation}
with the boundary conditions
\begin{equation}
\bm g(t,0)\in\CB\,,\qquad P (\bm g^{-1}\partial_x\bm g)(t,0)=0\,,\qquad
(\bm g^{-1}\partial_x\bm g)(t,L)=0
\end{equation}
and the Euclidean action functional
\begin{equation}
S_E[\bm g]\,=\,\sfrac{1}{4\pi}\sum\limits_{i=1}^N\int\limits_0^\beta \dd t
\int\limits_0^L r_i^2\big((\partial_t\varphi_i)^2+(\partial_x\varphi_i)^2
\big)(t,x)\,\dd x\,\equiv\,S_E[\bm\varphi]\,.
\end{equation}
To give sense to the functional integrals,
\,one decomposes the multivalued fields 
$\,\varphi_i(t,x)\,$ into the linear part 
which winds in the time direction and the periodic part:
\begin{equation}
\varphi_i(t,x)=\sfrac{2\pi}{\beta}n_it+\tilde\varphi_i(t,x)\,,
\label{wdecom}
\end{equation}
where
\begin{equation}
n_i=\sum\limits_{m=1}^M\kappa^m_iq_m\,,\qquad\tilde\varphi_i(t,0)=\sum\limits_{m=1}^M
\kappa^m_i\tilde\psi_m(t)\,,\qquad P\,\partial_x\tilde{\bm\varphi}(t,0)=0
=\partial_x\tilde{\bm\varphi}(t,L)
\label{bdrc}
\end{equation}
with $\,q_m\in\mathbb Z$, $\,\tilde\psi_m\in\mathbb R$, \,and with the 
multivaluedness reduced to that of $\,\tilde\psi_m$ defined modulo 
$2\pi$. The Euclidean action functional decomposes accordingly:
\begin{equation}
S_E[\bm\varphi]=\sfrac{\pi L}{\beta}\,(\bm k,T\bm k)+S_E(P\tilde{\bm\varphi})+
S_E((I-P)\tilde{\bm\varphi})
\label{fcin}
\end{equation}
leading to the factorization of the functional integral
\begin{equation}
\int\ee^{-S_E[\bm g]}\,\CD\bm g=\sum\limits_{\bm k\in\mathbb Z^M}
\ee^{-\frac{\pi L}{\beta}\,
(\bm k,T\bm k)}\,\int\ee^{-S_E[P\tilde{\bm\varphi}]}\,\CD(P\tilde{\bm\varphi})
\,\int\ee^{-S_E[(I-P)\tilde{\bm\varphi}]}\,\CD((I-P)\tilde{\bm\varphi})\,.
\label{factoriz}
\end{equation}
The last factor is a standard Gaussian functional integral with the quadratic
form corresponding to the Laplacian with the periodic boundary conditions
in the $t$ direction and the mixed Dirichlet one at $x=0$ and the Neumann 
one at $x=L$ in the $x$ direction. Such Laplacian is strictly 
positive. Using the zeta-function regularization of such an 
infinite-dimensional Gaussian integral, one obtains:
\begin{equation} 
\int\ee^{-S_E(\ee^{\ii(I-P)\tilde{\bm\varphi}})}\,\CD((I-P)\tilde{\bm\varphi})
=\left|\eta\big(\frac{_{\ii \beta}}{^{2L}}\big)\right|^{N-M}
\left|\eta\big(\frac{_{\ii
\beta}}{^{4L}}\big)\right|^{-(N-M)}.
\label{Z2}
\end{equation}
In the first functional integral on the right hand side of
(\ref{factoriz}), we parameterize
\begin{equation}
P\tilde{\bm\varphi}=\sum\limits_{m=1}^M\tilde\psi_m\bm\kappa^m\,.
\end{equation}
$S_E[P\tilde{\bm\varphi}]$ becomes then a quadratic form in $(\tilde\psi_m)$
corresponding to the Laplacian with the       
periodic boundary conditions in the $t$ direction and the Neumann 
ones in the $x$ direction, with constant zero modes. 
The zero-mode integration may be turned to a one over $U(1)^M$
using collective coordinates and recalling that fields $\tilde\psi_m$
are determined modulo $2\pi$. Employing the zeta-function regularization
for the remaining Gaussian functional integral over the other modes, 
one obtains
\begin{equation}
\int\ee^{-S_E[P\tilde{\bm\varphi}]}\,\CD(P\tilde{\bm\varphi}) 
= \sqrt{\det T} \, \Big(\sfrac{L}{\beta}\Big)^{\hspace{-0.05cm}\frac{M}{2}}
\left|\eta\big(\frac{_{\ii\beta}}{^{2L}}\big) \right|^{-M}.
\label{Z1}
\end{equation}
Upon the substitution of (\ref{Z2}) and (\ref{Z1}) to (\ref{fcin}), 
\,the functional integral expression (\ref{fcin0}) for $\,Z_\beta\,$ reduces 
to (\ref{Zpart}) with $\,V=0$.  
\vskip 0.1cm

The expectations of products of equal-time currents in the thermal state 
$\,\omega^L_\beta\,$ are represented by the normalized functional integrals:
\begin{equation}
\omega^L_\beta\Big(\prod\limits_{k=1}^K J_{i_k}^\ell(t,x_k)
\prod\limits_{k'=1}^{K'}J_{i_{k'}}^r(t,y_{k'})\Big)
=\frac{1}{Z_\beta}\int\prod\limits_k j_{i_k}^\ell(0,x_k)\prod\limits_{k'}
j_{i_{k'}}^r(0,y_{k'})\,\,\ee^{-S_E(\bm g)}\,\CD\bm g\,,
\label{funcint}
\end{equation}
where on the right hand side
\begin{equation}
\hspace*{-0.6cm}j^\ell_i(t,x)
=\sfrac{r_i^2}{2\pi}\sfrac{1}{2}(\partial_x-i\partial_t)
\varphi_i(t,x)\,,\qquad
j^r_i(t,x)=-\sfrac{r_i^2}{2\pi}\sfrac{1}{2}(\partial_x+i\partial_t)
\varphi_i(t,x)\,,
\label{currf}
\end{equation}
are functionals of field $\,\bm\varphi(t,x)\,$ that
in terms of decomposition (\ref{wdecom}) take the form
\begin{equation}
j^\ell_i(t,x)=-i\sfrac{r_i^2n_i}{2\beta}+\sfrac{r_i^2}{2\pi}\sfrac{1}{2}
(\partial_x-i\partial_t)\tilde\varphi_i(t,x)\,,\quad\ 
j^r_i(t,x)=-i\sfrac{r_i^2n_i}{2\beta}-\sfrac{r_i^2}{2\pi}
\sfrac{1}{2}(\partial_x+i\partial_t)\tilde\varphi_i(t,x)\,.\ 
\end{equation}
The functional integral (\ref{funcint})
factorizes similarly as in (\ref{factoriz}), with terms
$-i\frac{r_i^2n_i}{2\beta}=-i\frac{r_i^2}{2\beta}\sum\limits_{m}\kappa^m_ik_m$ 
contributing to the factor with the sum over $\,\bm k\,$ and terms with 
derivatives of $\,\tilde{\bm\varphi}\,$ entering 
the factors involving the Gaussian 
integrals calculated by the Wick rule. The latter leads to combinations 
of products of derivatives of the Green functions of the Laplacians that 
reduce to expressions involving the Weierstrass functions. At the end,
one obtains the same formulae as the $\,V\to0\,$ limit of the ones worked
out before for the expectations of products of the left-moving currents 
resulting from applying the rule (\ref{linkJLR1}) to the right-moving 
currents.

\subsection{General case}
\label{subsec:gencas}

An imaginary potential $\,V\,$ may be included in the functional integral
approach by imposing the twisted-periodic boundary conditions in 
the time direction on the $U(1)^N$-valued fields $\,\bm g=(g_i)
=(\ee^{i\varphi_i})$:
\begin{equation}
g_i(t+\beta,x)=g_i(t,x)\,\ee^{-\beta V}\,.
\label{twistfi}
\end{equation}
The latter may be implemented in the functional integral by 
decomposing
\begin{equation}
\varphi_i(t,x)=(iV+\sfrac{2\pi}{\beta}n_i)t+\tilde\varphi_i(t,x)
\label{decommu}
\end{equation}
with $\tilde\varphi_i$ periodic in the time direction,
keeping the same boundary conditions in the $x$ direction that take again
the form (\ref{bdrc}). For real $\,V$, \,the above decomposition implies 
a complex shift of the functional integration contour over fields $\,\bm g$. 
\,Performing the functional integration the same way as before, one obtains 
the representation
\begin{align}
&Z_{\beta,V}=\int\ee^{-S_E[\bm g]}\,\CD\bm g\,,\label{Zbm1}\\
&\omega^L_{\beta,V}\Big(\prod\limits_{k=1}^K J_{i_k}^\ell(t,x_k)
\prod\limits_{k'=1}^{K'}J_{i_{k'}}^r(t,y_{k'})\Big)
=\frac{1}{Z_{\beta,V}}\int\prod\limits_k j_{i_k}^\ell(0,x_k)\prod\limits_{k'}
j_{i_{k'}}^r(0,y_{k'})\,\,\ee^{-S_E(\bm g)}\,\CD\bm g\,,
\label{corrbm}
\end{align}
where the currents are still given by Eq.\,(\ref{currf}) and 
the contour of functional integration depends on $\,V\,$ in the way
described by decomposition (\ref{decommu}).

\nsection{Closed-string picture}
\label{sec:clstr}
\setcounter{equation}{0}
\subsection{Classical description}
\label{subsec:cldescr}

\noindent A symmetric role of time and space in the functional
integration leads, upon reversing those roles, to a description of 
the equilibrium expectations in the closed-string picture, see Fig.\,3.

\begin{figure}[th]
\leavevmode
\begin{center}
\vskip -0.2cm
\includegraphics[width=6cm,height=2.8cm]{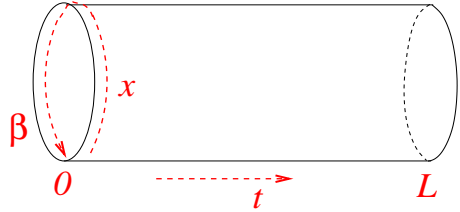}\\
\caption{Closed string worldsheet}
\end{center}
\end{figure}

\noindent In the latter, a collection of $N$ closed strings of length $\beta$, 
is described by fields $\,g_i(t,x)=\ee^{\ii\varphi_i(x,t)}\,$ defined for 
real $\,t\,$ and $\,x\,$ and twisted-periodic in the $\,x\,$ direction: 
\begin{equation}
g_i(t,x + \beta) = g_i(t,x)\,\ee^{-\beta V}\,,
\end{equation}
where $\,V\,$ is taken imaginary, compare to (\ref{twistfi}). On the classical
level and for Minkowski time, such fields are governed by 
the action functional
\begin{equation}
 S[\bm g]=\frac{_{1}}{^{4\pi}}\sum\limits_{i=1}^N\int
\dd t\int\limits_0^{\beta}\,r_i^2\big((\partial_t
\varphi_i)^2-(\partial_x\varphi_i)^2\big)\,\dd x\,.
\end{equation}
The twist in the periodicity condition may be absorbed by
setting
\begin{equation}
\varphi_i(t,x) \equiv \hat\varphi_i(t,x) + \ii V x\,,
\end{equation}
where
\begin{equation}\label{cs_period}
 \hat \varphi_i (t,x+\beta) = \hat \varphi_i (t,x) + 2 \pi m_i,
\qquad m_i \in \mathbb Z\,.
\end{equation}
The classical solutions have the form
\begin{equation}
\hat \varphi_i(t,x)=\hat \varphi_i^\ell(t+x)+\hat \varphi_i^r(t-x)\,,
\end{equation}
where
\begin{equation}
\hat{\bm\varphi}^{\ell,r}(t\pm x)\,=\,\hat{\bm\varphi}^{\ell,r}_0
+\frac{_{\sqrt{2}\,\pi}}{^{\beta}}\bm\alpha_0^{\ell,r}(t\pm x)
+\frac{_{\ii}}{^{\sqrt{2}\, }}\sum\limits_{n\not=0}\frac{_1}{^n}
\bm\alpha^{\ell,r}_n\ee^{-\frac{2\pi \ii n(t\pm x)}{\beta}}
\end{equation}
for $\,\overline{\bm\alpha^{\ell,r}_n}=\bm\alpha^{\ell,r}_{-n}$,
$\,\,\hat \varphi_0^\ell + \hat \varphi_0^r\equiv\hat \varphi_0 \in (\mathbb R /
2\pi \mathbb Z)^N$,
$\,\frac{1}{\sqrt{2}}(\bm\alpha_0^\ell + \bm\alpha_0^r)\equiv\bm p_0 
\in {\mathbb R}^N$, 
\,and
\begin{equation} 
\sfrac{1}{\sqrt{2}}(\bm\alpha_0^\ell - \bm\alpha_0^r) = \bm m
\end{equation}
where $\bm m$ is the vector of $N$ winding numbers $m_i\in\mathbb Z$. 
The symplectic form on the space of classical solutions is
\begin{align}
 \Omega\,=\,& \bm\delta p_{0}\cdot\wedge
\delta
\hat{\bm\varphi}_{0} - \frac{_\ii}{^2}
\sum_{n \neq 0} \frac{_1}{^n} \delta\bm\alpha_{n}^\ell
\cdot \wedge\bm\alpha_{(-n)}^\ell - \frac{_\ii}{^2} \sum_{n \neq 0}
\frac{_1}{^n}\delta\bm\alpha_{n}^r\cdot \wedge\bm\alpha_{(-n)}^r\,. 
\end{align}

\subsection{Quantization}
\label{subsec:qclstr}

The Poisson brackets obtained from $\Omega$ lead to the following 
canonical commutators:
\begin{equation}
\big[\varphi_{0i},p_{0j}\big]= \ii r_i^{-2}
\delta_{ij}\,,\quad
\big[\alpha_{ni}^\ell,\alpha_{n'j}^\ell\big]= 
r_i^{-2}\delta_{ij}\,n\,\delta_{n+n',0}\,,\quad
\big[\alpha_{ni}^r,\alpha_{n'j}^r\big]=
r_i^{-2}\delta_{ij}\,n\,\delta_{n+n',0}\,.
\end{equation}
For fixed winding numbers, the zero modes will be represented in 
the Hilbert space $\,L^2\big(U(1)^N\big)\,$ with an orthonormal-basis vectors
\begin{equation}
 \ket{\bm k} \equiv \ket{k^1 \ldots k^{N}}=\, 
\exp \Big( \ii
\sum_{i=1}^{N} k^i \varphi_{0i}
\Big) \qquad \text{for }\  k^i \in \mathbb Z
\end{equation}                                                                 
such that
\begin{equation}
r_i^2 p_{0i}  \ket{\bm k} \equiv - \ii 
\,\sfrac{\partial\,\,}{\partial
\varphi_{0i}}\,\exp \Big( \ii \sum_{i'=1}^{N} k^{i'} \varphi_{0i'}
\Big) = \,k^i \ket{\bm k}.
\end{equation}
The Hilbert space of states for the zero modes is 
a direct sum of an infinite number of copies of $\,L^2\big(U(1)^N\big)$, 
\,one for each winding vector,
\begin{equation}
 \mathcal H_0\,=\mathop{\oplus}\limits_{\bm m\in\mathbb Z^N}L^2
\big(U(1)^N\big)\,,
\end{equation}
with an orthonormal-basis vectors $\ket {\bm k, \bm m}$. 
The non-zero modes are represented in the tensor product of two 
standard Fock spaces $\CF^{\ell,r}$ generated by applying 
products of the $\alpha_{ni}^{\ell,r}$ with negative $n$ to
the normalized vectors $\tket{0}^{\ell,r}$, annihilated by
$\alpha_{ni}^{\ell,r}$ with positive $n$. The scalar products are defined 
by demanding that $(\alpha_{ni}^{\ell,r})^\dagger 
= \alpha_{(-n)i}^{\ell,r}$. The Hilbert 
space of the full theory is
\begin{equation}
 \mathcal H = \mathcal H_0 \otimes \CF^\ell \otimes \CF^r
\end{equation}
and we identify $\ket {\bm k, \bm m} \equiv \ket {\bm k, \bm m}
\otimes \ket{0}^\ell \otimes \ket{0}^r$.

\subsection{Current, energy and (magnetic) charge}
\label{subsec:curenmch}

As before, we define the left and right current for the closed string
\begin{equation}\label{current_cs}
\CJ_i^{\ell,r}(t,x) = \frac{_{r_i^2}}{^{2\pi}} \sfrac{1}{2}(\partial_t\pm 
\partial_x)\varphi_i(t,x) = 
\sfrac{r_i^2}{\sqrt{2}\,\beta} \sum_n \alpha_{ni}^{\ell,r}\,
\ee^{-\frac{2\pi \ii n(t\pm x)}{\beta}} \pm \sfrac{\ii r_i^2 V}{4 \pi}\,.
\end{equation}
The classical Hamiltonian of the system is
\begin{equation}
 H = \sfrac{1}{4\pi} \sum_i \int\limits_0^\beta r_i^2 \Big( (\partial_t \hat
\varphi_i)^2
+  (\partial_x \hat \varphi_i)^2 + 2 \ii V \partial_x \hat \varphi_i - V^2
\Big)\,\dd x\,.
\end{equation}
Once quantized, its $\,V$-independent part becomes the 
standard Hamiltonian of $N$ closed strings
\begin{equation}
\sfrac{2\pi}{\beta} \sum_i \big( L^\ell_{0i} + 
L_{0i}^r - \sfrac{1}{12} \big)\,\equiv\, H_{\rm cs}\,,
\end{equation}
where 
\begin{equation}
L^{\ell,r}_{0i}=\sfrac{r_i^2}{2}\sum\limits_n:\alpha^{\ell,r}_{ni}
\alpha^{\ell,r}_{-ni}:
\end{equation}
and the $-\frac{1}{12}$ term comes from the zero-point energy.
In the action on $\,\CH_0\,$ vectors,
\begin{equation}
H_{\rm cs}\ket{\bm k,\bm m}=\sfrac{\pi}{\beta}\Big(\sum_i
(r_i^{-2}(k^i)^2 + r_i^2 m_i^2)\,-\sfrac{N}{6}\Big)
\ket { \bm k, \bm m}\,.
\label{spec}
\end{equation}
The action of excited mode operators $\alpha_{(-n)i}^{\ell,r}$ for positive 
$n$ raises the eigenvalue of
$H_{\rm cs}$ by $\frac{2\pi n}{\beta}$. \,The part of the Hamiltonian linear in 
$\,V\,$ is equal to $\,\ii V\, Q^m_{cs}$, \,where 
\begin{equation}
 Q^{\rm m}_{\rm cs} = \sum_i \int_{0}^\beta\big(\hat{ \mathcal J}_i^\ell(x,t) - \hat{
\mathcal J}_i^r(x,t)\big)\,\dd x = 
\sfrac{1}{\sqrt 2} \sum_i r_i^2 (\alpha_{0i}^\ell -\alpha_{0i}^r)\,.
\end{equation}
is the total magnetic charge of the closed (untwisted) strings.
It acts only on  $\mathcal H_0$:
\begin{equation}
Q_{\rm cs}^{\rm m}\,\ket{\bm k, \bm m} = 
\sum_i r_i^2 m_i \ket{\bm k, \bm m}.
\label{spec1}
\end{equation}
Finally, the part of the Hamiltonian quadratic in $\,V\,$ is an additive 
constant, so that the full quantum Hamiltonian of the closed-string system 
becomes 
\begin{equation}\label{H_cs_mu}
 H\,\equiv\, H_{\rm cs} + \ii V\,Q^{\rm m}_{\rm cs} -\sfrac{ V^2\beta}{4 \pi} 
\sum_i r_i^2.
\end{equation}

\subsection{Boundary states}
\label{subsec:bdstates}

In the closed-string pictures, the boundary conditions in the space
direction, which in that picture becomes the time direction, are represented 
by the boundary states in the (completion of) the closed-string space 
of states \cite{Polchinski,Gaberdiel}.
The boundary state that corresponds to Neumann boundary condition
for all field component is 
\begin{equation}
\ket{\ket{\CN}\hspace{-0.04cm}} = A_\CN\sum_{\bm m \in
\mathbb Z^N}\ee^{-\sum\limits_{n=1}^\infty\frac{1}{n}\bm\alpha^\ell_{-n}\cdot
\hspace{0.03cm}\bm\alpha_{-n}^r}\,\ket{\bm 0, \bm m}\,,
\label{bdNeum}
\end{equation}
where $A_N$ is a suitable normalization constant.
This boundary state satisfies the relation
\begin{equation}
 \partial_t \varphi_i(0,x) \ket{\ket{\CN}\hspace{-0.035cm}} = 0\,
\end{equation}
whose excited-mode part implies that 
\begin{equation}
(\alpha_{ni}^\ell+\alpha_{(-n)i}^r)\ket{\ket{\CN}\hspace{-0.035cm}}=0
\end{equation} 
determining the form of the Ishibashi-type dependence of 
$\ket{\ket{\CN}\hspace{-0.035cm}}$
on those modes.
For the mixed Dirichlet-Neumann boundary
condition describing the junction of wires, the boundary state
has a more complicated form
\begin{align} \label{BS_cs_mu}
&\hspace*{-0.2cm}\ket{\ket{\CB}\hspace{-0.035cm}} = A_{DN}\,(2\pi)^{M-N}
\sqrt{\det T}\cr
&\hspace*{-0.2cm}
\times\hspace{-0.1cm}\sum_{\substack{\bm k
\in \mathbb Z^N \\ \sum_i k^i \kappa_i^m =0}} \ \sum_{\substack{\bm m \in
\mathbb Z^N \\ m_i = \sum_i \kappa_i^m s_m}}
\ee^{-\sum\limits_{j=1}^M\sum\limits_{n=1}^\infty\frac{1}{n}
\tilde\alpha_{(-n)j}^\ell
\tilde\alpha_{(-n)j}^r\,+\sum\limits_{j=M+1}^N\sum\limits_{n=1}^\infty\frac{1}{n}
\tilde\alpha_{(-n)j}^\ell\tilde\alpha_{(-n)j}^r}\,
\ket{\bm k, \bm m},\qquad 
\end{align}
where $s_m$ run through integers and 
$\,\tilde\alpha_{nj}^{\ell,r}=\bm\Lambda_j\cdot\bm\alpha^{\ell,r}_n$, 
\,see (\ref{Lambdai}). \,One has
\begin{equation}\label{mix_bd_st}
 P\hspace{0.03cm}\partial_t\bm\varphi(0,x) \ket{\ket{\CB}\hspace{-0.035cm}} 
= 0, \qquad (1-P)\hspace{0.02cm}\bm\varphi(0,x)
\ket{\ket{\CB}\hspace{-0.035cm}} = 0\,,
\end{equation}
where field $\,\bm\varphi\,$ may be equivalently replaced by 
$\,\hat{\bm\varphi}$.
\,The excited-mode part of these conditions implies that 
\begin{equation}
(\tilde\alpha_{nj}^\ell+\tilde\alpha_{(-n)j}^r)\ket{\ket{\CB}\hspace{-0.035cm}}=0
\quad\text{for}\quad j\leq M\,,\qquad
(\tilde\alpha_{nj}^\ell-\tilde\alpha_{(-n)j}^r)\ket{\ket{\CB}\hspace{-0.035cm}}=0
\quad\text{for}\quad j>M\,,
\label{BS_modes}
\end{equation}
fixing the form of the Ishibashi building-blocks of $\,\ket{\ket{\CB}
\hspace{-0.03cm}}$.
The zero-mode part of the first of relations (\ref{mix_bd_st}) assures that
$\sum\limits_i k^i\kappa^m_i=0$, whereas the zero-mode part of the second 
relations implies that $(1-P)\bm m=0$ which is solved by 
$m_i=\sum\limits_m\kappa^m_is_m$ for integer $s_m$.
The sum
\begin{equation}
(2\pi)^{M-N}\,\sqrt{\det T}\sum_{\substack{\bm k
\in \mathbb Z^N \\ \sum_i k^i \kappa_i^m =0}}\ket{\bm k} 
\label{sum}
\end{equation}
represents the delta-function supported by the brane $\CB=\kappa(U(1)^M)\subset
U(1)^N$ defined by the integral 
\begin{equation}
 \delta_{B}(\varphi_0) \equiv  \int \prod_i \delta \big( \varphi_{0i} - 
\sum_m \kappa_i^m \psi_m)\,\sqrt{\det T} \, \prod_m \dd \psi_m
\end{equation}
over $\,U(1)^M$ of \,the $\,2\pi-$periodic $\,N$-dimensional
$\,\delta$-function. Indeed,
\begin{align}
 \delta_{B}(\varphi_0) & = \sqrt{\det T} \,  \int \frac{_1}{^{(2 \pi)^N}}
\sum_{\bm k \in \mathbb Z^N} \ee^{i\sum_i k^i (\varphi_{0i} - \sum_m \kappa_i^m
\psi_m)} \prod_m \dd \psi_m \cr
& = (2\pi)^{M-N} \sqrt{\det T}\, \sum_{\bm k \in \mathbb Z^N} \ee^{i\sum_i
k^i \varphi_{0i}} \delta_{\sum_i k^i \kappa_i^m,0}
\end{align}
which reproduces (\ref{sum}).

\subsection{Partition function}
\label{subsec:partfct}

In the closed-string picture, the partition function $Z_{\beta,V}$ is
represented by the matrix element of the Euclidean evolution operator
$\ee^{-LH}$ between the boundary states. A direct calculation
gives:
\begin{align}
Z_{\beta,V} & = \bra{\hspace{-0.035cm}\bra{\CN}} \ee^{-LH} 
\ket{\ket{\CB}\hspace{-0.035cm}} \cr
&=A_NA_B(2\pi)^{M-N} \sqrt{\det T} \,\left[\eta
\big( \frac{_{2\ii L}}{^{\beta}} \big)\right]^{N-2M} \left[\eta 
\big( \frac{_{4\ii L}}{^{\beta}} \big)\right]^{M-N}\,\ee^{\sfrac{N V^2 \beta L}{4 \pi}\sum_i
r_i^2}\cr
&\times \sum_{\substack{\bm m \in
\mathbb Z^N \\ m_i = \sum \kappa_i^m s_m}} \ee^{-
\frac{\pi L }{\beta} \sum (r_i m_i)^2 - \ii L V \sum r_i^2 m_i}\,. 
\end{align}
This coincides with expression \eqref{Zbm} upon relabeling 
$\bm s=\bm k$ and recalling the definition \eqref{def_p_os}
of vector $\,\bm p$, \,provided that 
\begin{equation}
A_NA_B(2\pi)^{M-N}=\,2^{-N}\,.
\end{equation} 
The latter identity is assured if we take $A_N=A^N$ and $A_B=A^MB^{N-M}$
for $A=\frac{1}{\sqrt{2}}$ and $B=\sqrt{2}\pi$.

\subsection{Expectations}
\label{subsec:expect}

\noindent The expectation values of products of currents in equilibrium state
$\,\omega^L_{\beta,V}$ take in the open string picture the form of
the matrix elements between the boundary states of the time ordered products 
of Euclidean versions of currents $\CJ^{\ell,r}$:
\begin{align}
\label{csexpec}
\omega^L_{\beta,V}\Big(\prod\limits_{k=1}^K J_{i_k}^\ell(0,x_k)
\prod\limits_{k'=1}^{K'}J_{i_{k'}}^r(0,y_{k'})\Big)=&\,\frac{(-\ii)^K\ii^{K'}}
{\bra{\hspace{-0.035cm}\bra{\CN}} \ee^{-LH} \ket{\ket{\CB}\hspace{-0.035cm}}}\cr
&\hspace{-2.5cm}\times\ \bra{\hspace{-0.035cm}\bra{\CN}} \ee^{-LH}\,\CT
\prod\limits_{k=1}^K \CJ_{i_k}^\ell(-\ii x_k,0)
\prod\limits_{k'=1}^{K'}\CJ_{i_{k'}}^r(-\ii y_{k'},0) \ket{\ket{\CB}
\hspace{-0.035cm}},
\end{align}
where the Euclidean time ordering puts the operators at bigger $\,x_k\,$ or 
$\,y_{k'}\,$ to the left. The powers of $\,-\ii\,$ and $\,\ii\,$ represent 
the derivatives of the Euclidean conformal change of variables, 
$\,x+\ii t\mapsto t-\ii x\,$ and $\,y-\ii t\mapsto t+\ii y$, \,respectively, 
\,that reverses the roles of time and space.
\vskip 0.1 cm

The proof of \eqref{csexpec} in done in few steps. First, consider only the
left currents. As in the initial picture, \,we distinguish the constant part
from the excited terms,
\begin{equation}
 \CJ_i^\ell(t,x) =  
\sfrac{\ii r_i^2 V}{4 \pi} +
  \sfrac{r_i^2}{\sqrt{2}\,\beta} \alpha_{0i}^\ell + \hat \CJ_i^\ell(t,x) \,, 
\end{equation}
see \eqref{current_cs}. This decomposition factorizes 
in the expectation values.
For the constant terms, we get by direct calculation:
\begin{align}\label{cs_zeromodes}
& (-\ii)^K Z_{\beta,V}^{-1} \bra{\hspace{-0.03cm}\bra{\CN}}  
\ee^{-LH} \prod_{k=1}^K
\Big( \sfrac{\ii r_{i_k}^2 V}{4 \pi} +  \sfrac{r_{i_k}^2}{\sqrt{2}\,\beta}
\alpha_{0i_k}^\ell\Big)\ket{\ket{\CB}\hspace{-0.03cm}} \cr
& = \sfrac{1}{(2\beta V L)^K} \prod_{k=1}^K 
r_{i_k}^2\sum\limits_{m}\kappa^{m}_{i_k}\Big(\dfrac{\frac{\partial}{\partial
p^m}\ee^{\frac{ L  
\beta V^2}{4\pi}\bm p\cdot T^{-1} \bm p} }{\ee^{\frac{ L  
\beta V^2}{4\pi}\bm p\cdot T^{-1} \bm p}} + \,\dfrac{
\frac{\partial}{\partial p^m} \Big(
\sum\limits_{\bm k \in \mathbb
Z^M} 
\ee^{-\ii L V\bm p  \cdot \bm k -
\frac{\pi L}{\beta}
\bm k  \cdot
T\bm k }\Big)}
{
\sum\limits_{\bm k \in \mathbb Z^M} 
\ee^{-\ii L V\bm p  \cdot \bm k -
\frac{\pi L}{\beta}
\bm k  \cdot
T\bm k }}\Big) \cr
& = \omega^L_{\beta,V}\Big( \prod_{k=1}^K \big(J^\ell_{i_k}(t,x) - \hat
J^\ell_{i_k}(t,x)\big)\Big) - \sum_{\lbrace\ldots( k_p,l_p)\ldots \rbrace} \prod_p
\sfrac{r^2_{i_{k_p}}}{8\pi L \beta} P_{i_{k_p} i_{l_p}} \,,
\end{align}
which almost reproduces the zero mode part 
expectation value \eqref{omega0modes}
of the initial calculation but with one extra term, where $\lbrace\ldots(
k_p,l_p)\ldots \rbrace$ runs through all possible pairing of $\{1,\ldots,K\}$,
as in the Wick theorem. The presence of this term can be seen by induction 
on $K$. On the other hand, the expectations of products of $\hat \CJ^\ell_i$ 
are calculated by the Wick rule with
\begin{align}
&-\ii Z_{\beta,V}^{-1} \bra{\hspace{-0.03cm}\bra{\CN}}  \ee^{-LH}\hat
\CJ^\ell_i(-\ii x,0)\ket{\ket{\CB}\hspace{-0.03cm}}=0\,,\\
&(-\ii)^2 Z_{\beta,V}^{-1} \bra{\hspace{-0.03cm}\bra{\CN}}  \ee^{-LH}\hat
\CJ^\ell_{i_1}(-\ii x_1,0) \hat \CJ^\ell_{i_2}(-\ii x_2,0)\ket{\ket{\CB}
\hspace{-0.03cm}}\\ 
&  \hspace{2cm} = - \sfrac{1}{2} \left(\sfrac{r_{i_1}}{2\pi}\right)^2\Big(
P_{i_1i_2}\,(\wp(x_1-x_2;2L,-\ii\beta) + C_e^\prime) \cr
& \hspace{2.4cm} +(\delta_{i_1i_2}-P_{i_1i_2})(2 {\wp}(x_1-x_2;4L,-\ii\beta) -
{\wp}(x_1-x_2;2L,-\ii\beta)+C_{o}^\prime)\Big),\quad \label{cs_2pts}\quad
\end{align}
where we get expressions with the Weierstrass function similar to 
\eqref{2ptfn} but with different constants
\begin{align}
C_e^\prime =&\left(\sfrac{\pi}{-\ii\beta}\right)^2 \Big(\sfrac{1}{3} - \sum_{p
\neq 0} \big(
\sinh(\pi \sfrac{2Lp}{\beta}) \big)^{-2} \Big),\cr
C_{o}^\prime =&\left( \sfrac{\pi}{-\ii \beta} \right)^2 \Big( \sfrac{1}{3} -
2\sum_{p \neq
0}\big[\sinh\big(\pi
\sfrac{4L p}{\beta}\big)\big]^{-2} + \sum_{p \neq 0}\big[\sinh\big(\pi
\sfrac{2L p}{\beta}\big)\big]^{-2} \Big).
\end{align}
The theory of Weierstrass function of periods $\omega_1$ and $\omega_2$ 
\cite{WW} provides the identity 
\begin{equation}\label{beautiful}
\sfrac{ \omega_1}{ \omega_2}\Big( \sfrac{1}{3} +
 \sum_{n \neq 0}
\sin^{-2}\big(\sfrac{\pi n\omega_1}{\omega_2}\big)
\Big)-
 \sfrac{ \omega_2}{ \omega_1} \Big( \sfrac{1}{3} +
 \sum_{n \neq 0}
\sin^{-2} \big(\sfrac{\pi n\omega_2}{\omega_1}\big)
\Big)  = \pm\sfrac{\ii}{\pi}\,,
\end{equation}
where the sign on the right hand side is that of the imaginary part of
$\omega_1/\omega_2$. This leads to the relations
\begin{align}
C_o^\prime = C_o\,,\qquad C_e^\prime = C_e  - \sfrac{\pi}{L \beta} \,.
\end{align}
The contribution from the last term will cancel exactly the last 
contribution appearing in \eqref{cs_zeromodes} establishing
identity \eqref{csexpec} for any product of left currents. Finally, 
the closed-string expectation value of a general product of left and 
right current will be a combination of factors
corresponding to the decomposition \eqref{current_cs}. By direct calculation,
the constant part of the right currents can be expressed via the $S$ matrix in
terms of the one of the left currents with the use of \eqref{symcond1} and the
fact that $S$ also preserves vectors $\,\bm\kappa^m$. \,In 
the computation of the
excited part, the $S$ matrix appears naturally upon noticing that in 
the proper basis defined in \eqref{Lambdai}, it becomes
\begin{equation}
\tilde S = \text{diag}(\underbrace{1,\ldots,1}_{M},-1,\ldots,-1)
\end{equation}
which is precisely how the excited modes $\tilde \alpha_{ni}^\ell$ and $\tilde
\alpha_{ni}^r$ are related when they act on $\ket{\ket{\CB}\hspace{-0.03cm}}$, 
see \eqref{BS_modes}. Finally, under the closed-string 
expectation every right 
current is related to the left one by the $S$ matrix, exactly as in the 
initial picture \eqref{linkJLR1}. This proves identity \eqref{csexpec} 
in the general case.

\nsection{Thermodynamic limit}
\label{sec:thermlim}
\setcounter{equation}{0}

In the thermodynamic limit $L\to\infty$ the wires become infinitely long.
The partition function $Z_{\beta,V}$ diverges in that situation but
the free energy per unit length has a limit:
\begin{equation}
f^L_{\beta,V}\,=\,-\sfrac{1}{L\beta}\ln{Z_{\beta,V}}\ \,
\mathop{\longrightarrow}\limits_{L\to\infty}\ \,
-\sfrac{V^2}{4\pi}\sum\limits_{i=1}^Nr_i^2\,
-\sfrac{\pi N}{6\beta^2}\ \equiv\ f_{\beta,V}\,,
\label{freeen}
\end{equation}
as easily follows from its form (\ref{Zbm}).
The equilibrium state expectation values of the products of currents also
possess the $L\to\infty$ limit. In particular, it follows from
(\ref{omega0modes}) that
\begin{equation}
\lim\limits_{L\to\infty}\,
\omega^L_{\beta,V}\Big(\prod\limits_{k=1}^K\sfrac{r_{i_k}^2}{4L}\,
\alpha_{0i_k}^\ell\Big)\ =\ \prod\limits_{k=1}^K
\sfrac{r_{i_k}^2V}{4\pi}
\label{1stlimit}
\end{equation}
for real $\,V\,$ and relations (\ref{1ptfn}) and (\ref{2ptfn}) imply that
\begin{align}
&\lim\limits_{L\to\infty}\,\omega^L_{\beta,V}\big(\hat J^\ell_{i}(t,x)\big)\ 
=\ 0\,,\\
&\lim\limits_{L\to\infty}\,\omega^L_{\beta,V}\big(\hat J^\ell_{i_1}(t,x_1)\,
\hat J^\ell_{i_2}(t,x_2)\big)
\ =\ -\,\delta_{i_1i_2}\,\sfrac{r_{i_1}^2}{8\beta^2}
\,\sinh^{-2}\left(\sfrac{\pi(x_1-x_2)}{\beta}\right),
\label{2ndlimit}
\end{align}
as both $\,f_e(x)\,$ and $\,f_o(x)\,$ tend to
$\,\big(\frac{\pi}{\beta}\big)^2\,\sinh^{-2}\big(\frac{\pi x}{\beta}\big)\,$
when $\,L\to\infty$. The latter property follows
from (\ref{Weierstrass}) and the identity (\ref{beautiful}).
Eqs.\,(\ref{1stlimit}), (\ref{1ptfn}), (\ref{2ndlimit})
and the Wick rule, as well as the relation (\ref{linkJLR1}), determine
the $L\to\infty$ limit $\,\omega_{\beta,V}$ of the states 
$\,\omega^L_{\beta,V}$. Unlike for finite $L$, that limit
is not represented by a trace with a density matrix  
(for $L=\infty$, the Hamiltonian has a continuous spectrum and the operator 
$\ee^{-\beta(H-V Q)}$ is not traceclass). In particular, one obtains:
\begin{align}
&\omega_{\beta,V}\big(J^\ell_i(t,x)\big)\,=\,
\sfrac{r_{i}^2V}{4\pi}\,,\label{lim1pt}\\
&\omega_{\beta,V}\big(J^\ell_{i_1}(t,x_1)\,
J^\ell_{i_2}(t,x_2)\big)\,=\,\sfrac{r_{i_1}^2r_{i_2}^2V^2}{16\pi^2}\,
-\,\delta_{i_1i_2}\,\sfrac{r_{i_1}^2}{8\beta^2}
\,\sinh^{-2}\left(\sfrac{\pi(x_1-x_2)}{\beta}\right).\label{lim2pt}
\end{align}
The operator product expressions (\ref{oOPE}) and the limit (\ref{2ndlimit})
(that is uniform in small $|x_1-x_2|$) imply that
\begin{align}
\omega_{\beta,V}\big(T^\ell_i(t,x)\big)&=\sfrac{r_i^2V^2}{8\pi}\,+\,
\lim\limits_{\epsilon\to0}\,\Big(-\sfrac{\pi}{4\beta^2}\sinh^{-2}
\big(\sfrac{\pi\epsilon}{\beta}\big)+\sfrac{1}{4\pi\epsilon^2}\Big)=
\sfrac{r_i^2V^2}{8\pi}+\sfrac{\pi}{12\beta^2}\cr
&=\,\omega_{\beta,V}\big(T^r_i(t,x)\big).
\end{align}
In particular, the mean energy density in the equilibrium state is constant 
in each semi-infinite wire (but differs from one wire to another) 
and the mean energy current vanishes. 
\vskip 0.1 cm

The $L=\infty$ state is easy
to represent in the closed-string picture: by examining the right hand side
of (\ref{csexpec}), one infers that the 
boundary state $\ket{\ket{\CN}}$ of (\ref{bdNeum}) is projected 
when $L\to\infty$ to the 
closed-string vacuum
$\ket{\bm 0,\bm 0}\,$
so that 
\begin{align}
\omega_{\beta,V}\Big(\prod\limits_{k=1}^K 
J_{i_k}^\ell(0,x_k)
\prod\limits_{k'=1}^{K'}J_{i_{k'}}^r(0,y_{k'})\Big)\ 
=\ &\frac{(-\ii)^K\ii^{K'}}
{\langle\bm 0,\bm 0\ket{\ket{\CB}\hspace{-0.03cm}}}\cr
&\hspace{-2cm}\times\ \bra{\bm 0,\bm 0}\,\CT
\prod\limits_k \CJ_{i_k}^\ell(-i x_k,0)
\prod\limits_{k'}\CJ_{i_{k'}}^r(-iy_{k'},0) \ket{\ket{\CB}\hspace{-0.03cm}}
\label{csexpecinfty}
\end{align}
for $x_k,y_{k'}>0$,
\vskip 0.1 cm

One of the crucial observations that follows from (\ref{1stlimit})
and (\ref{2ndlimit}) is that, when restricted to the products of 
left-moving currents $\,J^\ell_i(0,x)\,$ with $\,x>0$, \,the limiting 
$\,L=\infty\,$ equilibrium expectations do not depend on the choice 
of the brane $\,\CB\,$ describing the contact of wires. In particular, 
such expectations are the same as for the space-filling brane $\,\CB_0$ with 
$\,S_{ii'}=\delta_{ii'}\,$ corresponding to the disconnected wires for which 
$\,J^r_i(t,x)=J_i^\ell(t,-x)\,$ and $\,\ket{\ket{\CB_0}\hspace{-0.03cm}}=
\ket{\ket{\CN}\hspace{-0.03cm}}$. \,The physical reason for this behavior
of the expectations of left-moving currents is that the latter
did not have contact with the junction up to time zero. 
The above observation is essential for the construction 
of nonequilibrium stationary state where the individual wires are kept
at different temperatures and at different potentials. For the 
disconnected wires, one has the obvious factorization:
\begin{equation}
\omega^L_{\beta,V}\Big(\prod\limits_{k=1}^KJ^\ell_{i_k}(t,x_k)\Big)\,=\,
\prod\limits_{i=1}^N\,\omega^L_{\beta,V}\Big(\prod\limits_{\substack{k\\i_k=i}}
J^\ell_{i}(t,x_k)\Big)\,.
\end{equation}     
Hence the same formula holds in the $L\to\infty$ limit of the equilibrium
state for any brane $\CB=\kappa(U(1)^M)$. 
\vskip 0.1 cm

For the disconnected wires, \,one also has the relation:
\begin{equation}
\omega^L_{\beta,V}\Big(\prod\limits_{{k'}=1}^{K'}
J^r_{i_{k'}}(t,y_{k'})\Big)\,=\,
\prod\limits_{i={k'}}^N\,\omega^L_{\beta,V}\Big(
\prod\limits_{\substack{k'\\i_{k'}=i}}
J^r_{i}(t,y_{k'})\Big)\,=\,\,\prod\limits_{i=1}^N\,\omega^L_{\beta,V}
\Big(\prod\limits_{\substack{k'\\i_{k'}=i}}J^\ell_{i}(t,-y_{k'})\Big).
\end{equation}     
It is easy to check using (\ref{linkJLR1}) and (\ref{symcond1})
that the latter factorization holds in the limit $L\to\infty$
also for other branes $\,\CB=\kappa(U(1)^M)$.

\nsection{Nonequilibrium stationary state}
\label{sec:NESS}
\setcounter{equation}{0}

Following \cite{MSLL,BD2}, see also \cite{Ruelle}, we shall consider 
a nonequilibrium stationary state (NESS) $\,\omega_{\rm neq}$ describing 
the situation when different 
semi-infinite wires are kept at different temperatures and different 
potentials. State $\,\omega_{\rm neq}$ may be obtained
by the following limiting procedure. For each disconnected
semi-infinite wire, one considers the algebra $\,\CA_i$ generated by products 
of currents $\,J^{\ell,r}_i(0,x)\,$ for $x>0$, together with a state 
$\,\omega^i_{\beta_i,V_i}$ given by the restriction to $\,\CA_i$ of 
the $L=\infty$ equilibrium state $\,\omega_{\beta_i,V_i}$ for the 
space-filling brane $\,\CB_0$. The product state 
$\,\omega_{\rm in}\equiv\mathop{\otimes}\limits_{i=1}^N
\omega^i_{\beta_,,V_i}$ on algebra $\,\CA=\mathop{\otimes_i}\CA_i$
describes the disconnected wires with each prepared in its 
own equilibrium state 
$\,\omega^i_{\beta_i,V_i}$. \,As in Sec.\,\ref{sec:Quant}, algebra 
$\,\CA\,$ may be identified with the 
one generated by currents $J^{\ell,r}_i(0,x)$ 
with $x>0$ and $\,1\leq i\leq N\,$ for the connected wires. 
Let $\,\CU_t\,$ for $t>0$ describe the forward in time Heisenberg-picture 
evolution of the currents $J^{\ell,r}_i(0,x)$ with $x>0$ in the presence 
of brane $\,\CB\,$: 
\begin{align}
&\CU_tJ_i^\ell(0,x)=J^\ell_i(t,x)=J^\ell_i(0,x+t)=J_i^\ell(0,t+x)\,,\label{Utl}\\
&\CU_tJ_i^r(0,x)=J^r_i(t,x)=J^r_i(0,x-t))=\begin{cases}\,J_i^r(0,x-t)
\hspace{1.34cm}{\rm for}
\quad t\leq x\cr
\,\sum\limits_{i'}S_{ii'}J^\ell_{i'}(0,t-x)\hspace{0.38cm}{\rm for}
\quad x\leq t\label{Utr}
\end{cases}
\end{align}
Then for $A\in\CA$,
\begin{equation}
\omega_{\rm neq}(A)\,=\,\lim\limits_{t\to\infty}\,\omega_{\rm in}
(\CU_tA)\,.
\end{equation}
In order to prove the above relation consider the backward in time 
Heisenberg evolution for decoupled wires, i.e. in the presence of brane 
$\CB_0$:
\begin{align}
&\hspace*{1cm}\CU^0_{-t}J^\ell_i(0,x)=J^\ell_i(-t,x)=J^\ell_i(0,x-t)
=\begin{cases}\,J_i^\ell(0,-t+x)\hspace{0.93cm}{\rm for}
\quad t\leq x\,,\cr
\,J^r_{i}(0,t-x)\hspace{1.18cm}{\rm for}\quad x\leq t\,,\end{cases}
\label{U0tl}\\
&\hspace*{1cm}\CU^0_{-t}J_i^r(0,x)=J^r_i(-t,x)=J^r_i(0,x+t)=J_i^r(0,t+x)\label{U0tr}
\end{align}
for $\,x,t>0$.
\,Such a decoupled evolution preserves the product state 
$\,\omega_{\rm in}\,$ so that
\begin{equation}
\lim\limits_{t\to\infty}\,\omega_{\rm in}(\CU_tA)\,=\,\omega_{\rm in}
(\CS A)\,,
\end{equation}
where
\begin{equation}
\CS=\lim\limits_{t\to\infty}\,\CU^0_{-t}\CU_t
\end{equation}
is the scattering operator in the action on algebra $\,\CA$. \,The explicit 
form of the latter in the action on the chiral currents follows
from equations (\ref{Utl}), (\ref{Utr}), (\ref{U0tl}) and (\ref{U0tr}):
\begin{align}
&\CS J^\ell_i(0,x)\,=\,J^\ell_i(0,x)\,,\\
&\CS J^r_i(0,x)\,=\,\sum\limits_{i'}S_{ii'}J^r_{i'}(0,x)
\end{align}
for $\,x>0$. \,Note that the nonequilibrium state $\omega_{\rm neq}$ 
is preserved by the Heisenberg evolution $\,\CU_t\,$ \,so that its 
stationarity follows. Hence the explicit formula: 
\begin{align}
&\omega_{\rm neq}\bigg(\Big(\prod\limits_{k=1}^KJ^\ell_{i_k}(t,x_k)\Big)
\Big(\prod\limits_{k'=1}^{K'}J^r_{i_{k'}}(t,y_{k'})\Big)\bigg)\cr
&=\,\prod\limits_{k'=1}^{K'}\sum\limits_{i'_{k'}}
S_{i_{k'}i'_{k'}}\,\prod\limits_{i=1}^N\,\omega^i_{\beta_i,V_i}
\bigg(\Big(\prod\limits_{\substack{k\\i_k=i}}
J^\ell_{i}(t,x_k)\Big)\Big(\prod\limits_{\substack{k'\\i'_{k'}=i}}
J^r_{i}(t,y_{k'})\Big)\bigg),
\label{omeganeq}
\end{align}
where on the left hand side the currents correspond to connected wires
and on the right hand side to disconnected ones and the values of $t$, 
$x_k$ and $x_{k'}$ may be taken arbitrary (with noncoincident 
$x_1,\dots,x_K,-y_1,\dots,-y_{K'}$ to avoid singularities). In particular,
\begin{equation}
\omega_{\rm neq}\bigg(\prod\limits_{k=1}^KJ^\ell_{i_k}(t,x_k)\bigg)
=\,\omega_{\rm in}\bigg(\prod\limits_{k=1}^KJ^\ell_{i_k}(t,x_k)\bigg)
\label{omomin}
\end{equation} 
so that the difference between $\omega_{\rm neq}$ and $\omega_{\rm in}$,
due to the junction between wires, arises only in the presence of 
right-moving currents. The left-moving currents do not feel the influence
of the junction. 
It should be stressed that the dynamics considered
above both in the presence of the junction and for decoupled wires  
is generated by the Hamiltonians that do not include the electric potentials 
in the bulk of the wires. Those play the role only in the preparation
of the initial product state and may be applied far away from the 
junction. That the ballistic evolution of chiral currents persists 
for long times in the bulk of the wires in such a nonequilibrium situation
should be assured by the integrable nature of the Luttinger liquids,
see the discussion at the end of \cite{BDV}. 
\vskip 0.1cm

Specifying Eq.\,(\ref{omeganeq}) to the 1-point expectations, one obtains:
\begin{equation}
\omega_{\rm neq}\big(J^\ell_i(t,x)\big)\,=\,\sfrac{r_i^2V_i}{4\pi}\,,\qquad
\omega_{\rm neq}\big(J^r_i(t,x)\big)\,=\,\sum\limits_{i'=1}^N
S_{ii'}\sfrac{r_{i'}^2V_{i'}}{4\pi}
\end{equation}
so that the mean charge density and mean current in the wires are
\begin{equation}
\omega_{\rm neq}\big(J^0_i(t,x)\big)\,=\,\sum\limits_{i'=1}^N
(S_{ii'}+\delta_{ii'})\sfrac{r_{i'}^2V_{i'}}{4\pi}\,,\qquad
\omega_{\rm neq}\big(J^1_i(t,x)\big)\,=\,\sum\limits_{i'=1}^N
(S_{ii'}-\delta_{ii'})\sfrac{r_{i'}^2V_{i'}}{4\pi}\,,
\label{noneq.J1}
\end{equation}
respectively. They are constant in each wire and the mean current 
does not vanish, in general, at difference with the equilibrium state. 
The electric conductance tensor of the junction (in the units
$\,e^2/\hbar$) \,is 
\begin{equation}
{}^{\rm el}G_{ii'} \equiv \left.\sfrac{\partial\,}{\partial V_{i'}}\,
\omega_{\rm neq}(J^1_i(0,x))\right|_{\substack{\beta_j = \beta \\ V_j = V}} =
\sfrac{1}{4\pi} (S_{ii'} - \delta_{ii'})r_{i'}^2\,.
\end{equation}
This agrees with the calculation of \cite{RHFCA,RHFOCA} based on the 
combination of the Green-Kubo formula with the conformal field theory 
representation of the equilibrium state. Note that the conductance 
vanishes for the decoupled wires.
The nonequilibrium current 2-point functions are given by
\begin{align}
&\omega_{\rm neq}\big(J^\ell_{i_1}(t,x_1)\,J^\ell_{i_2}(t,x_2)\big)\,=\,
\sfrac{r_{i_1}^2r_{i_2}^2V_{i_1}V_{i_2}}{16\pi^2}\,
-\,\delta_{i_1i_2}\,\sfrac{r_{i_1}^2}{8\beta_{i_1}^2}
\,\sinh^{-2}\left(\sfrac{\pi(x_1-x_2)}{\beta_{i_1}}\right),\\
&\omega_{\rm neq}\big(J^\ell_{i_1}(t,x_1)\,J^r_{i_2}(t,x_2)\big)\,=\,
\sum\limits_{i=1}^NS_{i_2i}\sfrac{r_{i_1}^2r_{i}^2V_{i_1}V_{i}}{16\pi^2}\,
-\,S_{i_2i_1}\,\sfrac{r_{i_1}^2}{8\beta_{i_1}^2}
\,\sinh^{-2}\left(\sfrac{\pi(x_1+x_2)}{\beta_{i_1}}\right),\\
&\omega_{\rm neq}\big(J^r_{i_1}(t,x_1)\,J^r_{i_2}(t,x_2)\big)\,=
\hspace{-0.05cm}\sum\limits_{i,i'=1}^N\hspace{-0.15cm}S_{i_1i}
S_{i_2i'}\sfrac{r_{i}^2r_{i'}^2V_{i}V_{i'}}{16\pi^2}
-\hspace{-0.1cm}\sum\limits_{i=1}^NS_{i_1i}S_{i_2i}\,
\sfrac{r_{i}^2}{8\beta_{i}^2}\,
\sinh^{-2}\left(\sfrac{\pi(x_1-x_2)}{\beta_{i}}\right).
\end{align}
Note that the nonequilibrium states $\,\omega_{\rm neq}$ with coupled wires 
break the time reversal symmetry $J^\ell\leftrightarrow J^r$. 
\vskip 0.1 cm

For the expectation value of the energy-momentum components, we obtain
from the operator product expansions (\ref{OPEl}) and (\ref{OPEr})
\begin{align}
\omega_{\rm neq}\big(T^\ell_i(t,x)\big)\,&=\,\sfrac{r_i^2V_i^2}{8\pi}\,+\,
\lim\limits_{\epsilon\to0}\Big(-\sfrac{\pi}{4\beta_i^2}\sinh^{-2}\Big(\sfrac{\pi\epsilon}{\beta_i}\Big)+\sfrac{1}{4\pi\epsilon^2}\Big)\,
=\,\sfrac{r_i^2V_i^2}{8\pi}\,+\,\sfrac{\pi}{12\beta_i^2}\,,\\
\omega_{\rm neq}\big(T^r_i(t,x)\big)\,&=\,\sfrac{1}{8\pi r_i^2}
\Big(\sum\limits_{i'=1}^NS_{ii'}
r_{i'}^2V_{i'}\Big)^{\hspace{-0.08cm}2}\,+\,
\lim\limits_{\epsilon\to0}\Big(-\sum\limits_{i'=1}^N(S_{ii'})^2
\sfrac{\pi r_{i'}^2}{4r_i^2\beta_{i'}^2}\sinh^{-2}\Big(
\sfrac{\pi\epsilon}{\beta_{i'}}\Big)+\sfrac{1}{4\pi\epsilon^2}\Big)\cr
&\,=\,\sfrac{1}{8\pi r_i^2}
\Big(\sum\limits_{i'=1}^NS_{ii'}
r_{i'}^2V_{i'}\Big)^{\hspace{-0.08cm}2}\,+\,\sfrac{\pi}{12r_i^2}
\sum\limits_{i'=1}^N\Big(S_{ii'}\sfrac{r_{i'}}{\beta_{i'}}\Big)^{\hspace{-0.08cm}2}
\end{align}
so that the mean energy density and current are, respectively,
\begin{equation}
\omega_{\rm neq}\big(K^{0,1}_i(t,x)\big)\,=\,\sfrac{1}{8\pi r_i^2}
\Big(\sum\limits_{i'}^NS_{ii'}
r_{i'}^2V_{i'}\Big)^{\hspace{-0.08cm}2}\,
\pm\,\sfrac{r_i^2V_i^2}{8\pi}\,+\,\sfrac{\pi}{12r_i^2}
\sum\limits_{i'=1}^N\Big(S_{ii'}\sfrac{r_{i'}}{\beta_{i'}}\Big)^{\hspace{-0.08cm}2}
\,\pm\,\sfrac{\pi}{12\beta_i^2}\,,
\label{noneq.K1}
\end{equation}
This results in the thermal conductance
\begin{equation}
{}^{\rm th\hspace{-0.02cm}}G_{ii'}\,
=\,-\beta_{i'}^2\sfrac{\partial}{\partial\beta_{i'}}\,
\omega_{\rm neq}\big(K_i^1(t,x)\big)\Big|_{\substack{\beta_j=\beta\\
V_j=V}}\,=\,\sfrac{\pi}{6\beta}\big((S_{ii'})^2\sfrac{r_{i'}^2}{r_i^2}
-\delta_{ii'}\big)\,.
\end{equation}

\nsection{Full counting statistics}
\label{sec:FCS}
\setcounter{equation}{0} 
\subsection{Charge transport}
\label{subsec:ch.transp}

Measuring transport of charges through the junction of quantum wires
requires specifying measurement protocol that may be not easy to
implement. Refs.\,\cite{LL,LLL} proposed an indirect measurement
of charge transferred through a quantum resistor and obtained
a closed formula for statistics of the results. The same charge transfer 
statistics could be obtained by considering a direct two-times measurement 
of the total charge accumulated in the system, provided the latter is finite. 
Following \cite{BD2}, we shall employ the second measurement protocol that is 
conceptually simpler although unpractical for large systems, keeping in mind 
that the charge transfer statistics obtained this way may be also accessed 
by a more practical indirect measurement protocol. 
\vskip 0.1cm

Consider first the system of disconnected wires 
of length $L$, each with Hamiltonian $H_i^0$ and charge operator $\,Q_i^0$.
Prepare the system in the product state 
$\,\omega^L_{0}=\mathop{\otimes}\limits_{i=1}^N\omega^{i,L}_{\beta_i,V_i}\,$ 
given by the
density matrix $\,\rho_0\equiv\mathop{\otimes}\limits_{i=1}^N
\rho^i_{\beta_i,V_i}$,
\,where $\,\rho^i_{\beta_i,V_i}=\frac{1}{Z^i_{\beta_i,V_i}}
\hspace{0.02cm}\ee^{-\beta_i(H^0_i-V_iQ_i^0)}$. \,At time zero, measure the 
total charge $Q_i^0$ in each wire. Then connect the wires 
instantaneously and let the system evolve. At time $t$, disconnect 
the wires and measure the total charge $Q_i(t)$ in each wire. By spectral 
decomposition,
\begin{equation}
 Q_i^0 = \sum_{q^0} q^0 P^0_{i,q^0}\,, \qquad  Q_i(t) = \sum_q q P_{i,q}(t)\,. 
\end{equation}
The probability that the first 
measurement gives the values of charges $(q_i^0)\equiv\bm{q}^0$ is equal
to $\,\tr\big(\mathop{\otimes}_{i=1}^N\rho^i_{\beta_i,V_i}P^0_{i,q_i^0}\big)$.
\,After the first measurement, the density matrix is reduced to
\qq
\rho_{0^+}\,\equiv\,\frac{\mathop{\otimes}\limits_{i=1}^NP^0_{i.q_i^0}\,
\rho^i_{\beta_i,V_i}P^0_{i.q_i^0}}
{\tr\big(\mathop{\otimes}_{i=1}^N\rho^i_{\beta_i,V_i}P^0_{i,q_i^0}\big)}\,,
\qqq 
The probability that the second measurement gives the values of charges 
$(q_i)\equiv\bm{q}$, 
\,is then equal to $\,\tr\big(\rho_{0^+}\prod\limits_{i=1}^NP_{i,q_i}(t)\big)$.
\,Altogether, the joint probability of the results $(\bm{q}^0,\bm{q})$ is
\begin{align}
 \mathbb P_t(\bm{q}^0,\bm{q}) &= \tr\big(\mathop{\otimes}_{i=1}^N
\rho^i_{\beta_i,V_i}P^0_{i,q_i^0}\big)\,\,\tr\,\rho_{0^+}\prod\limits_{i=1}^N
P_{i,q_i}(t)=\tr\,\Big(\mathop{\otimes}\limits_{i=1}^N\rho^i_{\beta_i,V_i}
P^0_{i.q_i^0}\Big)\prod\limits_{i=1}^N
P_{i,q_i}(t)\cr
&=\omega_{0}^L\Big(\prod\limits_i P_{i,q_i^0}\prod\limits_i
P_{i,q_i}(t)\Big),
\end{align}
where to obtain the second equality, we used the fact that 
$\,P_{i,q}^0\,$ commute with $\rho^i_{\beta_i,V_i}$.
The probability that the charges change by $\,\Delta q_i=q_i
-q_i^0\,$ is
\begin{equation}
 \mathbb P_t({\Delta\bm{q}})=\sum\limits_{(\bm{q}^0,
\bm{q})}\prod\limits_i\delta_
{\Delta q_i,q_i-q_i^0}\,\mathbb P_t(\bm{q}^0,\bm{q})\,.
\end{equation}
The latter probabilities may be encoded in their characteristic function
called the generating function of full counting statistics (FCS) for 
the electric charge transfers:
\begin{align}
{}^{{\rm el}\hspace{-0.04cm}}F^L_t(\bm\nu)&=
\sum\limits_{\Delta\bm{q}}\ee^{\ii\sum\limits_i\nu_i
\Delta q_i}\mathbb P_t(\Delta\bm{q})
\,=\sum\limits_{(\bm{q}^0,\bm{q})}\ee^{\ii\sum\limits_i
(q_i-q_i^0)}\,\mathbb P_t(\bm{q}^0,\bm{q})\cr
&=\,\omega^L_{0}
\bigg(\prod\limits_i\Big(\sum\limits_{q^0}
\ee^{-\ii\nu_iq^0}P^0_{i,q^0}\Big)\prod\limits_i\Big(\sum\limits_{q}
\ee^{\ii\nu_i q}P_{i,q}(t))\Big)\bigg)\cr
&=\omega^L_{0}
\Big(\ee^{-\ii\sum\limits_i\nu_iQ_i^0}\,
\ee^{\ii\sum\limits_i\nu_iQ_i(t)}\Big). 
\label{FCS}
\end{align}
For connected wires, the change of the wire charges in time is
\begin{align}
\Delta Q_i(t)&\equiv Q_i(t)-Q_i(0)= \int\limits_0^t\frac{_d}{^{ds}}Q_i(s)\,ds\,=\,
\int\limits_0^t \dd s\int\limits_{0}^{L}\partial_s
\big(J^\ell_i(s,x)+J^r_i(s,x)\big)\,\dd x\cr
&=\int\limits_0^t \dd s\int\limits_{0}^{L}
\big(\partial_xJ^\ell_i(s,x)-\partial_xJ^r_i(s,x)\big)\,\dd x
=-\int\limits_0^t
\big(J^\ell_i(s,0)-J^r_i(s,0)\big)\,\dd s\cr
&=-\int\limits_0^t
\big(J^\ell_i(0,s)-\sum_{i'} S_{ii'} J^\ell_{i'}(0,s)\big)\,\dd s\,.
\label{DQ}
\end{align}
After disconnecting the wires at time $\,t\leq L$, \,the latter
observables become the ones for unconnected wires given by the right hand
side of (\ref{DQ}). The crucial fact is that they are extensive in time 
but not in the wire length, unlike the total charges. Note the 
commutation: 
\begin{align}
\big[\Delta Q_i(t),Q^0_{i'}\big] & = 
-\int\limits_0^t\dd s\int\limits_{-L}^{L}\big[\big(J^\ell_i
(0,s)-\sum_{i''} S_{ii''} J^\ell_{i''}(0,s)\big),J^\ell_{i'}(0,x)\big]\,\dd x\cr
&=-(\delta_{ii'}-S_{ii'})\int\limits_0^t\dd s\int\limits_{-L}^{L}
\sum\limits_n\delta'(s-x+2nL)\,\dd x\,=\,0\,.
\label{commu}
\end{align}
Since the observable $\,Q_i(t)\,$ become equal to $\,Q_i^0+\Delta Q_i(t)\,$ 
after the disconnection of wires, the FCS generating function
(\ref{FCS}) may be rewritten due to (\ref{commu}) in the simpler form
\begin{align}
{}^{{\rm el}\hspace{-0.04cm}}F^L_t(\bm\nu)=
\omega^L_{0}\Big(\ee^{\ii\sum\limits_i\nu_i\Delta Q_i(t)}\Big)=
\prod\limits_{i=1}^N\omega^{i,L}_{\beta_i,V_i}
\Big(\ee^{-\ii\tilde\nu_i\int\limits_0^tJ^\ell_i(0,s)}\Big),
\label{FCSsf}
\end{align}
where we have set 
\qq
\tilde\nu_i\,\equiv\,\nu_i-\sum\limits_{i'}S_{i'i}\hspace{0.02cm}\nu_{i'}\,.
\label{nutilde}
\qqq
Due to the translation invariance of the state 
$\,\omega^{i,L}_{\beta_i,V_i}$,
\begin{equation}
\omega^{i,L}_{\beta_i,V_i}\Big(\ee^{-\ii\tilde\nu_i
\int\limits_0^tJ^\ell_i(0,s)\,\dd s}\Big)\hspace{-0.06cm}=
\omega^{i,L}_{\beta_i,V_i}\Big(\ee^{-\ii\tilde\nu_i\int\limits_{-t/2}^{t/2}
J^\ell_i(0,s)\,\dd s}\Big)\hspace{-0.06cm}=
\omega^{i,L}_{\beta_i,V_i}\Big(\ee^{-\ii\tilde\nu_i\int\limits_0^{t/2}
(J^\ell_i(0,s)+J^r(0,s))\,\dd s}\Big).
\label{8.8}
\end{equation}
In the limit $L\to\infty$,
\,the initial states $\,\omega^L_0$ tend to the product 
state $\,\omega_{\rm in}$ for semi-infinite wires considered in 
Sec.\,\ref{sec:NESS} so that
\begin{equation}
{}^{{\rm el}\hspace{-0.04cm}}F_t(\bm\nu)\equiv\lim\limits_{L\to\infty}\,
{}^{{\rm el}\hspace{-0.04cm}}F^L_t(\bm\nu)=
\omega_{\rm in}\Big(\ee^{\ii\sum\limits_i\nu_i\Delta Q_i(t)}\Big)
=\omega_{\rm neq}\Big(\ee^{\ii\sum\limits_i\nu_i\Delta Q_i(t)}\Big)
\end{equation}
with the last equality following from relations  (\ref{omomin}) and 
(\ref{DQ}). 
\vskip 0.1cm

We would like to study the large deviation form of the FCS generating 
function by calculating the rate function
\begin{equation}
{}^{{\rm el}\hspace{-0.05cm}}f(\bm\nu)=\lim\limits_{t\to\infty}\,t^{-1}\,
\ln{{}^{{\rm el}\hspace{-0.04cm}}F_t(\bm\nu)}\,.
\label{FCSL=infty}
\end{equation}
Refs. \cite{BD2,DHB} exposed a strategy for the calculation
of such rate functions for semi-infinite wires with a purely 
transmitting junction from its derivatives. Applying it to our case,
we note that such derivatives have the form
\begin{align}
 &\sfrac{\partial\,}{\partial \nu_j}\, {}^{{\rm el}\hspace{-0.05cm}}
f(\bm\nu)=\lim\limits_{t\to\infty}\,\frac{\ii}{t}\,
\frac{\omega_{\rm in}\Big(\ee^{\ii\sum\limits_i\nu_i\Delta Q_i(t)}
\Delta Q_j(t)\Big)}{\omega_{\rm in}\Big(\ee^{\ii\sum\limits_{i}
\nu_{i}\Delta Q_{i}(t)}\Big)}\cr
&=\,\lim\limits_{t\to\infty}\,\frac{1}{\ii t}\,
\frac{\omega_{\rm in}\Big(\ee^{-\ii\sum\limits_i
\tilde\nu_i\int\limits_0^tJ^\ell_i(0,s)}
\int\limits_0^t\big(J^\ell_j(0,s)-S_{jj'}J^\ell_{j'}(0,s)\big)\,ds\Big)}
{\omega_{\rm in}
\Big(\ee^{-\ii\sum\limits_i\tilde\nu_i\int\limits_0^tJ^\ell_i(0,s)}
\Big)}\cr
&=\,\lim\limits_{t\to\infty}\,\frac{1}{\ii t}\,
\frac{\omega_{\rm in}\Big(\ee^{-\ii\sum\limits_i\tilde\nu_i\int\limits_0^{t/2}
(J^\ell_i(0,s)+J^r(0,s))\,\dd s}
\int\limits_{-t/2}^{t/2}\big(J^\ell_j(0,s')-S_{jj'}J^\ell_{j'}(0,s')\big)\,
\dd s'\Big)}
{\omega_{\rm in}
\Big(\ee^{-\ii\sum\limits_i\tilde\nu_i\int\limits_0^{t/2}(J^\ell_i(0,s)+J^r(0,s))\,\dd s}\Big)}
\end{align}
with $\tilde\nu_i$ as above. 
\,It was argued in \cite{BD2,DHB}, \,following the approach set up in
\cite{BD1}, \,that
\begin{align}
\lim\limits_{t\to\infty}\,\frac{\omega_{\rm in}
\Big(\ee^{-\ii\sum\limits_i\tilde\nu_i\int\limits_{0}^{t/2}(J^\ell_i(0,s)+J^r(0,s))\,\dd s}
J^\ell_{j'}(0,s')\Big)}{\omega_{\rm in}\Big(\ee^{-\ii\sum\limits_i\tilde\nu_i
\int\limits_{0}^{t/2}(J^\ell_i(0,s)+J^r(0,s))\,\dd s}\Big)}\,
=\,\omega^{j'}_{\beta_{j'},V_{j'}-\ii\beta_{j'}^{-1}\tilde\nu_{j'}}\big(J^\ell_{j'}(0,s')\big)
\label{l1pt}
\end{align}
because for large $t$ the exponential factor becomes close to 
$\,\ee^{-i\sum\limits_i\tilde\nu_iQ^0_i}\,$ providing effectively 
the imaginary additions to potentials $\,V_i$. 
\,Since the one point function on the right hand side of (\ref{l1pt}) 
is independent of $s'$, \,this line of thought gives by the analytic 
continuation of \eqref{lim1pt} the identity
\begin{equation}
\sfrac{\partial\,}{\partial \nu_j}\, {}^{{\rm el}\hspace{-0.05cm}}
f(\bm\nu)\,=\,
-  \sfrac{\ii}{4\pi}\sum_{j'} (\delta_{jj'} - S_{jj'}) r_{j'}^2\big(V_{j'}
-\ii\beta_{j'}^{-1}\tilde\nu_{j'}\big)
\end{equation}
which, \,together with (\ref{nutilde}) and the relation $\,f(\bm0)=0$, \,implies 
that
\begin{align}
{}^{{\rm el}\hspace{-0.05cm}}f(\bm\nu)
\,=&\sum\limits_{i=1}^N\bigg(\sfrac{\big(V_i-\ii\beta_i^{-1}
(\nu_i-\sum\limits_{i'=1}^NS_{i'i}\nu_{i'})\big)^2\beta_ir_i^2}{8\pi}\,-\,
\sfrac{V_i^2\beta_ir_i^2}{8\pi}\bigg)\cr
\,=&-\sum\limits_{i=1}^N\bigg(
\sfrac{\big(\nu_i-\sum\limits_{i'=1}^NS_{i'i}\nu_{i'}
\big)^2r_i^2}{8\pi\beta_{i}}
+\ii\,\sfrac{V_i\big(\nu_i-\sum\limits_{i'=1}^NS_{i'i}\nu_{i'}
\big)r_i^2}{4\pi}\bigg).
\label{LD}
\end{align}
\vskip 0.1cm

The existence of the limit (\ref{FCSL=infty}) means that at long times 
the PDF of charge transfers takes the large-deviations form 
\begin{equation} 
\mathbb P_t(\Delta\bm{q})\ \sim\ 
\ee^{-t\,{}^{{\rm el}\hspace{-0.05cm}}I(\frac{1}{t}\Delta \bm{q})},
\label{LDF} 
\end{equation} 
where the rate function
\begin{equation} 
{}^{{\rm el}\hspace{-0.05cm}}I(\bm{\rho})\,=\,\mathop{\rm max}
\limits_{\bm\nu}\Big(\sum\limits_{i=1}^N\rho_i\nu_i\,-\,
{}^{{\rm el}\hspace{-0.05cm}}f(-\ii\bm\nu)\Big) 
\end{equation} 
is the Legendre transform of $\,{}^{{\rm el}\hspace{-0.05cm}}f(-i\bm\nu)$. 
\,For $\,{}^{{\rm el}\hspace{-0.05cm}}f(\bm\nu)\,$ 
given by (\ref{LD}), $\,{}^{{\rm el}\hspace{-0.05cm}}I(\bm{\rho})\,$ is a quadratic 
polynomial on the subspace where it is finite.  
\,In other words, the large deviations of charge transfers per unit time  
have the Gaussian distribution with mean 
\begin{equation} 
\big\langle\sfrac{\Delta q_{i}}{t}\big\rangle\,=\, 
\sum\limits_{i'=1}^N(S_{ii'}-\delta_{ii'})\sfrac{V_{i'}r_{i'}^2}{4\pi}\,,
\end{equation} 
equal to the mean current in the nonequilibrium state, see (\ref{noneq.J1}),
\,and covariance 
\begin{equation} 
\big\langle\sfrac{\Delta q_{i}}{t}\,\sfrac{\Delta q_{i'}}{t}\big\rangle 
-\big\langle\sfrac{\Delta q_{i}}{t}\big\rangle 
\big\langle\sfrac{\Delta q_{i'}}{t}\big\rangle\,=\, 
\sfrac{1}{t}\sum\limits_{i"=1}^N\sfrac{r_{i"}^2}{4\pi\beta_{i"}} 
(\delta_{i"i}-S_{ii"})(\delta_{i"i'}-S_{i'i"})\,. 
\end{equation} 

Note that the first of equalities (\ref{LD}) implies that the large-deviations
rate function for FCS of charge transfers is proportional to the difference 
of equilibrium free energies for different potentials:
\begin{equation}
{}^{{\rm el}\hspace{-0.05cm}}f(\bm\nu)\,
=\,\sfrac{1}{2}\sum\limits_{i=1}^N\beta_i\Big(f_i(\beta_i,V_i)
-f_i(\beta_i,V_i-\beta_i^{-1}\tilde\nu_i)\Big),
\label{turn}
\end{equation}
where $\,f_i(\beta,V)=-\frac{V^2r_i^2}{4\pi}-\frac{\pi}{6\beta^2}\,$ 
is the equilibrium free energy per unit length in a single decoupled 
semi-infinite wire with Neumann boundary conditions, \,see (\ref{freeen}).
\,Relation (\ref{turn}) implies in turn that
\begin{equation}
{}^{{\rm el}\hspace{-0.05cm}}f(\bm\nu)\,=\,\lim\limits_{t=2L\to\infty}\,
\frac{1}{t}\,\ln{}^{{\rm el}\hspace{-0.04cm}}F^L_t(\bm\nu)
\label{t=L}
\end{equation}
if we define ${}^{{\rm el}\hspace{-0.04cm}}F^L_t(\bm\nu)\,$ for $t>L$ by
the right hand side of (\ref{FCSsf}). Indeed, in that case
the $2^{\rm nd}$ equality in (\ref{8.8}) implies that
\begin{equation}
{}^{{\rm el}\hspace{-0.04cm}}F_{2L}^L(\bm\nu)\,
=\,\prod\limits_{i=1}^N\,\sfrac{Z^i_{\beta_i,V_i-\ii\beta_i^{-1}\tilde\nu_i}}
{Z^i_{\beta_i,V_i}}\,,
\end{equation}
where the partition functions on the right hand side pertain to the 
disconnected wires of length $L$. 
\,{\it A priori}, \,it is not clear that the same result for 
${}^{{\rm el}\hspace{-0.05cm}}f(\bm\nu)$ arises 
in the physically different limit that takes the thermodynamic limit 
$L\to\infty$ before sending $\,t\to\infty$. \,The 
calculation of \cite{BD2,DHB} amounts to the claim that both limits
are equal.

\subsection{Exact result for $\,{}^{\rm el}\hspace{-0.04cm}F^L_t(\bm\nu)$}
\label{subsec:exact.el}

The exactly soluble nature of the model considered here allows
to examine closer the distribution of charge transfers for
finite $L$ and $t$ and to see in more details how its large-deviations
form arises. A direct calculation performed in Appendix {A} gives the result
\begin{align}
{}^{{\rm el}\hspace{-0.04cm}}F^L_t(\bm\nu)_{\rm reg}\,=&\,
\exp\Big[-\sum\limits_{i=1}^N
\sfrac{\tilde\nu_i^2r_i^2}{8\pi^2}\Big(C_{\Lambda L}+\ln\sfrac{2\pi\,\theta_1(
\sfrac{i\beta_i}{2L};\sfrac{t}{2L})}
{\partial_z\theta_1(\sfrac{\ii\beta_i}{2L};0)}\Big)\Big]\cr
&\,\times\,\,\sfrac{\sum\limits_{\bm{k}\in{\mathbb Z}^N}\exp\Big[
\sum\limits_{i=1}^N\big(-\frac{\pi\beta_i}{Lr_i^2}k_i^2+\beta_iV_ik_i-\ii
\frac{\tilde\nu_it}{2L}k_i\big)\Big]}
{\sum\limits_{\bm{k}\in{\mathbb Z}^N}\exp\Big[
\sum\limits_{i=1}^N\big(-\frac{\pi\beta_i}{Lr_i^2}k_i^2+\beta_iV_ik_i
\big)\Big]},
\label{FCSex}
\end{align}  
where the subscript ``reg'' refers to a necessary ultraviolet regularization, 
that replaces the divergent constant 
$\,C_\infty=\sum\limits_{n>0}\frac{1}{n}\,$ by 
\begin{equation}
C_{\Lambda L}=\sum\limits_{n=1}^{\Lambda L}\sfrac{1}{n}=
\ln(\Lambda L)+C+\,O(\sfrac{1}{\Lambda L})
\label{CLambda}
\end{equation} 
with the ultraviolet cutoff $\Lambda$, 
\,see Appendix {A}.  \,Variables $\,\tilde\nu_i\,$ are as before, 
\,see (\ref{nutilde}), \,and 
\begin{equation}
\theta_1(\tau;z)=\sum\limits_{n\in\mathbb Z}
\ee^{\pi\ii\tau(n+\frac{1}{2})^2+2\pi i(n+\frac{1}{2})(z+\frac{1}{2})}
\end{equation}
is one of the Jacobi theta-functions. The first exponential factor
on the right hand side of (\ref{FCSex}) is the characteristic function
of a centered Gaussian distribution of charge transfers $\Delta\bm{q}$ 
with the covariance
\begin{equation}
\CC_{i'i''}\,=\,\sum\limits_{i=1}^N\sfrac{r_i^2}{4\pi^2}\big(\delta_{ii'}-S_{i'i}\big)
\big(\delta_{ii''}-S_{i''i}\big)\Big(C_{\Lambda L}+\ln\sfrac{2\pi\,\theta_1(
\sfrac{i\beta_i}{2L};\sfrac{t}{2L})}
{\partial_z\theta_1(\sfrac{\ii\beta_i}{2L};0)}\Big).
\end{equation}
The $\,t$-dependent expression under the logarithm is positive for $\,0<t<2L$.
\,Note that the ultraviolet divergent contribution to the covariance 
is independent of $\,t$. \,It describes the charge 
transfers that arise at the moments of the connection 
of wires at time $\,0\,$ or their
disconnection at time $\,t\,$ but do not contribute to the average charge
transfers realized during the long period of time when the wires are connected.
The second factor on the right hand side of (\ref{FCS}) 
is a characteristic function of the discrete distribution 
\begin{equation}
\sfrac{\sum\limits_{\bm{k}\in{\mathbb Z}^N}\exp\big[
\sum\limits_{i=1}^N\big(-\frac{\pi\beta_i}{Lr_i^2}k_i^2+\beta_iV_ik_i
\big)\big]\prod\limits_{i'=1}^N\delta\big(\Delta q_{i'}-\sum\limits_{i=1}^N(
\delta_{ii'}-S_{i'i})k_i\big)}
{\sum\limits_{\bm{k}\in{\mathbb Z}^N}\exp\big[
\sum\limits_{i=1}^N\big(-\frac{\pi\beta_i}{Lr_i^2}k_i^2+\beta_iV_ik_i
\big)\big]}
\end{equation}
of charge transfers $\Delta\bm q$. The two types of charge transfers
are realized independently and both correspond to the vanishing total 
charge transfer $\sum_{i}\Delta q_i$. As we shall see below,
they both contribute to the large deviations result (\ref{LD}).
\vskip 0.1cm

In order to study the behavior of the charge-transfer distribution for
large $L$ and large $t$, we shall rewrite (\ref{FCSex}) applying 
the Poisson resummation formula to the $\bm{k}$-sums and the
modular transformation
$\,\theta_1(\tau;z)=\ii(-\ii\tau)^{-1/2}\,\ee^{-\frac{\pi\ii z^2}{\tau}}\,
\theta_1(-\sfrac{1}{\tau};\sfrac{z}{\tau})\,$
to the Jacobi theta function. The resulting expression is
\begin{align}
{}^{{\rm el}\hspace{-0.04cm}}F^L_t(\bm\nu)_{\rm reg}\,
=&\,\exp\Big[-\sum\limits_{i=1}^N
\sfrac{\tilde\nu_i^2r_i^2}{8\pi^2}\Big(C_{\Lambda L}+\ln\sfrac{\pi\ii\beta_i\,
\ee^{-\frac{\pi t^2}{2L\beta_i}}\,
\theta_1(\sfrac{2iL}{\beta_i};-\sfrac{\ii t}{\beta_i})}
{L\,\partial_z\theta_1(\sfrac{2\ii L}{\beta_i};0)}
\Big)\Big]\cr
&\times\,\,\sfrac{\sum\limits_{\bm{k}\in{\mathbb Z}^N}\exp\Big[
\sum\limits_{i=1}^N\big(-\frac{2\pi r_i^2L}{\beta_i}k_i^2+
\ii V_ir_i^2Lk_i+\frac{\tilde\nu_ir_i^2t}{2\beta_i}k_i
-\frac{\ii\tilde\nu_iV_ir_i^2t}{4\pi}-\frac{\tilde\nu_i^2r_i^2t^2}
{16\pi\beta_iL}\big)\Big]}
{\sum\limits_{\bm{k}\in{\mathbb Z}^N}\exp\Big[
\sum\limits_{i=1}^N\big(-\frac{2\pi r_i^2L}{\beta_i}k_i^2+
\ii V_ir_i^2Lk_i\big)\Big]}\,.
\label{FCSex1}
\end{align}
Together with relation (\ref{CLambda}), it allows to extract the large 
$L$ behavior
\begin{align}
{}^{{\rm el}\hspace{-0.04cm}}F^L_t(\bm\nu)_{\rm reg}
=&\,\exp\Big[-\sum\limits_{i=1}^N\Big(\sfrac{\tilde\nu_i^2r_i^2}
{8\pi^2}\big(\ln(2\beta_i\tilde\nu_i)+\ln\sinh\sfrac{\pi t}{\beta_i}
\big)\Big]\,\exp\Big[-\sum\limits_{i=1}^N
\sfrac{\ii\tilde\nu_iV_ir_i^2t}{4\pi}\Big)\Big]\cr
&\,\times\,\,\Big(1\,+\,O(\sfrac{t^2}{L})\,+\,O(\ee^{-cL})\Big),
\label{FCSasymp}
\end{align}
where $c>0$ is some $\beta_i$- and $r_i$-dependent constant. 
The first exponential factor describes the leading behavior of the
contribution in the $1^{\rm st}$ line of (\ref{FCSex1}) and the second 
one that in the $2^{\rm nd}$ line. We infer that
\begin{equation}
{}^{{\rm el}\hspace{-0.04cm}}F_t(\bm\nu)_{\rm reg}\,\equiv\,
\lim\limits_{L\to\infty}\ 
{}^{{\rm el}\hspace{-0.04cm}}F^L_t(\bm\nu)_{\rm reg}\,
=\,\prod\limits_{i=1}^N\Big(\big(2\beta_i
\tilde\nu_i\sinh{\sfrac{\pi t}{\beta_i}}\big)^{\hspace{-0.03cm}
-\frac{\tilde\nu_i^2r_i^2}{8\pi^2}}\,\ee^{-\frac{\ii\tilde\nu_iV_ir_i^2t}{4\pi}}
\Big)
\end{equation}
and
\begin{equation}
{}^{{\rm el}\hspace{-0.05cm}}f(\bm\nu)_{\rm reg}\,\equiv\,
\lim\limits_{t\to\infty}\ \frac{1}{t}\ln{}^{{\rm el}\hspace{-0.04cm}}
F_t(\bm\nu)_{\rm reg}\,=\,
-\sum\limits_{i=1}^N\Big(\sfrac{\tilde\nu_i^2r_i^2}{8\pi\beta_i}+
\ii\sfrac{\tilde\nu_iV_ir_i^2}{4\pi}\Big),
\end{equation}
reproducing the large deviations result (\ref{LD}) up to the ultraviolet
regularization. Note that if follows from relation 
(\ref{FCSasymp}) that the same result is obtained for the limit
of $\,\frac{1}{t}\ln{}^{{\rm el}\hspace{-0.04cm}}
F^L_t(\bm\nu)_{\rm reg}\,$ obtained
by sending simultaneously $\Lambda,\ L$ and $t$ to infinity in such 
a way that the ratios $\frac{\ln\Lambda}{t}$ and $\frac{t}{L}$ tend to zero. 
This specifies more precisely the region where the distribution of charge 
transfers takes the Gaussian large deviation form (\ref{LDF}) described
previously. The above analysis does not cover, however, the
case  (\ref{t=L}) with $\,t=2L\to\infty\,$ which, although giving the same
limit, is somewhat special. In particular, no 
ultraviolet regularization of is required for $\,{}^{{\rm el}\hspace{-0.04cm}}
F_{2L}^L(\bm\nu)$.

\subsection{Heat transport}
\label{subsec:th.trans}

The protocol for the measurement of the thermal transfers is the same. 
It consists of preparing the system of wires of length $\,L\,$ in the 
initial product state 
$\,\omega^L_0=\mathop{\otimes}_{i=1}^N\omega^{i,L}_{\beta_i,V_i}$ and 
performing the measurements of the energies $\,H_i(0)=H_i^0\,$ 
and $\,H_i(t)\,$ in the disconnected wires at two times in between which 
the wires were connected. Denoting the results, respectively, 
$\,(e_i^0)\equiv\bm{e^0}\,$ and $\,(e_i)\equiv\bm e$, \,we encode 
the probability of the change of energies $\,\Delta e_i=e_i-e_i^0\,$ in 
the characteristic function 
\begin{equation}
{}^{{\rm th}\hspace{-0.04cm}}F^L_t(\bm\lambda) =  \sum_{\Delta\bm e} 
\ee^{\ii \sum_i \lambda_i \Delta e_i}\,\mathbb P_t
(\Delta\bm e)\,=\, 
\omega_0^L\Big(\ee^{-i\sum\limits_i\lambda_iH_i(0)}\,
\ee^{\,i\sum\limits_i\lambda_iH_i(t)}\Big),
\label{FCSenergy}
\end{equation}
the generating function of FCS for heat transfers.
The change of energy of the wires connected between times $0$ and $t$ is 
\begin{align}
&\Delta H_i(t)\,\equiv\,H_i(t) - H_i(0)  
= \int\limits_0^t \sfrac{\dd}{\dd s}H_i(s) \dd s = 
\int\limits_0^t \dd s
\int\limits_0^{L} \partial_s ( T_i^\ell(s,x) + T_i^r(s,x))\,\dd x \cr
&\hspace*{0.5cm}
=  \int\limits_0^t \dd s \int\limits_0^{L} (\partial_x T_i^\ell(s,x) 
- \partial_x T_i^r(s,x))\,\dd x  
= -  \int\limits_0^t  (T_i^\ell(0,s) - T_i^r(s,0))\,\dd s\,, 
\end{align}
compare to (\ref{DQ}). Moreover
\begin{align}
T_i^r(s,0)&  = T_i^r(0,-s)=\lim\limits_{\epsilon\to0}\Big(\sfrac{2\pi}{r_i^2}  
J_i^r(0,-s+\epsilon)\,J_i^r(0,-s)+\sfrac{1}{4\pi\epsilon^2}\Big)  \cr
& = \lim\limits_{\epsilon\to0}\Big(\sfrac{2\pi}{r_i^2} 
\sum_{i'i''} S_{ii'}S_{ii''} J_{i'}^\ell(0,s-\epsilon)\, J_{i''}^\ell(0,s)+\sfrac{1}
{4\pi\epsilon^2}\Big) \cr
& = \sfrac{1}{r_i^{2}}\sum_{i'} (S_{ii'})^2r_{i'}^2 T_{i'}^\ell(0,s) + \sfrac{2\pi}{r_i^2} 
\sum_{i'\neq i''} S_{ii'}
J_{i'}^\ell(0,s)\,S_{ii''} J_{i''}^\ell(0,s)
\end{align}
so that
\begin{align}
\hspace*{-0.1cm}\Delta H_i(t)= -\sfrac{1}{r_i^2}\int_0^t\hspace{-0.1cm}\Big( 
\sum_{i'}\big(\delta_{ii'}- (S_{ii'})^2\big)\,r_{i'}^2
T_{i'}^\ell(0,s)
- 2\pi\hspace{-0.08cm} \sum_{i'\neq i''} S_{ii'} J_{i'}^\ell(0,s) 
\,S_{ii''} J_{i''}^\ell(0,s)\Big)\,\dd s\,. 
\end{align}
Interpreting the latter operators as observables for disconnected
wires, we have the commutation relation
\begin{align}
\hspace{-0.07cm}\big[H^0_j,\Delta H_i(t)\big]&=\,\sfrac{2\pi\ii}{r_i^2}
\hspace{-0.08cm}\int\limits_0^t\hspace{-0.08cm}
\Big(\big(\delta_{ij}-(S_{ij})^2\big)\partial_s:(J^\ell_j(0,s))^2:
-2\sum\limits_{i'\not=j}S_{ii'}S_{ij}J^\ell_{i'}(0,s)\,\partial_s 
J^\ell_j(0,s)\Big)\dd s\cr
&=\,\sfrac{2\pi\ii}{r_i^2}\big(\delta_{ij}-(S_{ij})^2\big)\big(
:(J^\ell_j(0,t))^2:-:(J^\ell_j(0,0))^2:\big)\cr
&\quad\ -\,\sfrac{4\pi\ii}{r_i^2}
\sum\limits_{i'\not=j}S_{ii'}S_{ij}\int\limits_0^t
J^\ell_{i'}(0,s)\,\partial_s 
J^\ell_j(0,s)\,\dd s
\label{eal}
\end{align}
as a consequence of the identity
\begin{equation}
\big[H_j^0,J^\ell_i(t,x)\big]=-i\,\delta_{ij}\partial_t J^\ell_i(t,x)=
-i\,\delta_{ij}\partial_x J^\ell_i(t,x)\,.
\end{equation}
Note that $\,[H^0_j,\Delta H_i(t)]\not=0\,$ and the generating function 
(\ref{FCSenergy}) of FCS for heat transfers
\begin{equation}
{}^{{\rm th}\hspace{-0.04cm}}F_t^L(\bm\lambda)\,=\,\omega_0^L
\Big(\ee^{-i\sum\limits_i\lambda_i
H_i^0}\,\ee^{i\sum\limits_i\lambda_i\big(H_i^0+\Delta H_i(t)\big)}\Big)\,\not=\,
\omega_0^L\Big(\ee^{i\sum\limits_i\lambda_i\Delta H_i(t)}\Big).
\label{eFCS1}
\end{equation}
This difference occurs even for $t=2L$ when the first term on the right hand 
side of (\ref{eal}) vanishes, but not the second one. 
\vskip 0.1cm

We shall calculate explicitly
$\,{}^{{\rm th}\hspace{-0.04cm}}F^L_t(\bm\lambda)\,$ for $\,t=2L\,$ which is easier
than for general $\,t$. \,In this case,
\begin{equation}
H_i^0+\Delta H_i(2L)\,=\,\sfrac{\pi}{2L}
\sum\limits_{i',i''}O_{ii'}O_{ii''}
\Big(\tilde\alpha_{0i}\tilde\alpha_{0i'}\,+\,
2\sum\limits_{n>0}\tilde\alpha_{-2n,i}\tilde\alpha_{2n,i'}\Big)
\end{equation}
in terms of the modes, where $\,O\,$ is the orthogonal matrix related 
to matrix $\,S\,$ by (\ref{O}). One easily checks that the above observables
commute so that they may indeed be measured simultaneously
in the disconnected wires. Let
\begin{equation}
\CA\,=\,\sum\limits_{i}\lambda_i\big(H_i^0+\Delta H_i(2L)\big)\,=\,
\sfrac{\pi}{2L}
\sum\limits_{i,i'}(O\lambda O)_{ii'}\Big(\tilde\alpha_{0i}\tilde\alpha_{0i'}\,+\,
2\sum\limits_{n>0}\tilde\alpha_{-2n,i}\tilde\alpha_{2n,i'}\Big),
\label{itom}
\end{equation}
where $\lambda$ stands for the diagonal $N\times N$ matrix with entries
$\lambda_i$. The contributions of the zero modes 
and of the excited modes to the expectation 
\begin{equation}
\omega_0^L\Big(\ee^{-i\sum\limits_i\lambda_iH_i^0}\,
\ee^{\ii\sum\limits_i\lambda_i\big(H_i^0+\Delta H_i(2L)\big)}\Big)\,=\,
\frac{\tr_\CH\Big(\ee^{-\sum\limits_{i}\beta_i(H^0_i-V Q^0_i)}\,
\ee^{-\ii\sum\limits_i\lambda_iH_i^0}\,\ee^{\,\ii\CA}\Big)}
{\tr_\CH\Big(\ee^{-\sum\limits_{i}\beta_i(H^0_i-V Q^0_i)}\Big)}
\label{et=L}
\end{equation}
factorize. The first one has the form
\begin{align}
&\frac{\sum\limits_{{\bm k}\in{\mathbb Z}^N}\ee^{-\frac{\pi}{L}
(r^{-1}\bm k\,,\,\beta r^{-1}\bm k)\,+\,(\beta\bm V\,,\,\bm k)\,-\,\frac{\pi\ii}{L}
(r^{-1}\bm k\,,\,Cr^{-1}\bm k)}}
{\sum\limits_{{\bm k}\in{\mathbb Z}^N}\ee^{-\frac{\pi}{L}(r^{-1}\bm k\,,\,\beta r^{-1}\bm k)
\,+\,(\beta\bm V\,,\,\bm k)}}\,=\,
\det\big(I+i\beta^{-1}C\big)^{-1/2}\cr
&\times\ \frac{\sum\limits_{\bm k\in{\mathbb Z}^N}\ee^{-\pi L\,
(r\bm k\,,\,(\beta+\ii C)^{-1}r\bm k)
\,-\,\ii L\,(r\beta\bm V\,,\,(\beta+\ii C)^{-1}r\bm k)\,+\,\frac{L}{4\pi}
(r\beta\bm V\,,\,(\beta+\ii C)^{-1}r\beta\bm V)}}
{\sum\limits_{\bm k\in{\mathbb Z}^N}\ee^{-\pi L\,
(r\bm k\,,\,\beta^{-1}r\bm k)\,-\,\ii L\,(r\beta\bm V
\,,\,\beta^{-1}r\bm k)\,+\,\frac{L}{4\pi}
(r\beta\bm V\,,\,\beta^{-1}r\beta\bm V)}},
\label{12exp}
\end{align}
where $r$ and $\beta$ stand for the diagonal matrices with entries
$(r_i)$ and $(\beta_i)$, respectively, 
\begin{equation}
C=\lambda-O\lambda O
\label{C}
\end{equation} 
is a symmetric $N\times N$ matrix, and 
$\,\bm V\,$ denotes the vector with components $\,V_i$. 
\,The right hand side of (\ref{12exp}) was
obtained by the Poisson resummation. As for the contribution
of the excited modes, its calculation is given in Appendix {B}
and results in
\begin{equation}
\prod\limits_{n>0}\det\Big(I+
\big(I-\ee^{-\frac{\pi\ii}{L}n\lambda}\,
O\,\ee^{\frac{\pi\ii}{L}n\lambda}O\big)
\big(\ee^{\frac{\pi n}{L}\beta}-I\big)^{-1}\Big)^{-1}
\label{exccont}
\end{equation}
with a convergent infinite product. 
Gathering expressions (\ref{12exp}) and (\ref{exccont}), we obtain:
\begin{align}
{}^{{\rm th}\hspace{-0.04cm}}F^L_{2L}(\bm\lambda)\,=&
\,\det\big(I+i\beta^{-1}C\big)^{-1/2}\cr
&\times\ \frac{\sum\limits_{\bm k\in{\mathbb Z}^N}\ee^{-\pi L\,
(r\bm k\,,\,(\beta+\ii C)^{-1}r\bm k)
\,-\,\ii L\,(r\beta\bm V\,,\,(\beta+\ii C)^{-1}r\bm k)\,+\,\frac{L}{4\pi}
(r\beta\bm V\,,\,(\beta+\ii C)^{-1}r\beta\bm V)}}
{\sum\limits_{\bm k\in{\mathbb Z}^N}\ee^{-\pi L\,
(r\bm k\,,\,\beta^{-1}r\bm k)\,-\,\ii L\,(r\beta\bm V
\,,\,\beta^{-1}r\bm k)\,+\,\frac{L}{4\pi}
(r\beta\bm V\,,\,\beta^{-1}r\beta\bm V)}}\cr
&\times\ \prod\limits_{n>0}\det\Big(I+
\big(I-\ee^{-\frac{\pi\ii}{L}n\lambda}\,
O\,\ee^{\frac{\pi\ii}{L}n\lambda}O\big)
\big(\ee^{\frac{\pi n}{L}\beta}-I\big)^{-1}\Big)^{-1}.
\label{eFt=L}
\end{align}
In the limit $t=2L\to\infty$,
\begin{align}
\frac{1}{2L}\,\ln\,{}^{{\rm th}\hspace{-0.04cm}}F^L_{2L}(\bm\lambda)\,\ 
&\mathop{\longrightarrow}\limits_{L\to\infty}\ \,
\sfrac{1}{8\pi}\Big(\big(r\beta\bm V\,,\,(\beta+iC)^{-1}r\beta\bm V\big)
-\big(r\bm V\,,\,\beta r\bm V\big)\Big)\cr
&\hspace{1cm}-\,\int\limits_0^\infty
\ln\det\Big(I+\big(I-\ee^{-2\pi\ii\, x\lambda}\,O\,\ee^{2\pi\ii\,x\lambda}O\big)
\big(\ee^{2\pi x\beta}-I\big)^{-1}\Big)\,\dd x\cr
&\hspace{1cm}\equiv\,{}^{{\rm th}\hspace{-0.05cm}}f(\bm\lambda)
\label{eFt=L=inf}
\end{align}
for sufficiently small $|\lambda_i|$ so the zero-mode contribution with 
$\bm k=0$ dominates. 
\vskip 0.1cm

If we calculated the right hand side of (\ref{eFCS1}) for $\,t=2L\,$ 
instead of $\,{}^{{\rm th}\hspace{-0.04cm}}F^L_{2L}(\bm\lambda)$, \,the only change 
would be the replacement of matrix $\,\ee^{-\frac{2\pi\ii}{L}n\lambda}\,
O\,\ee^{-\frac{2\pi\ii}{L}n\lambda}O\,$ by 
$\,\ee^{-\frac{2\pi\ii}{L}n(\lambda-O\lambda O)}\,$
in the last line of (\ref{eFt=L}). In general, however, matrices 
$\,\lambda\,$ and $\,O\lambda O\,$ do not commute if $\,O\,$ has 
nondiagonal elements. Such a modification would also kill the symmetry 
(\ref{FRexc}) of the contribution (\ref{exccont}) showed in Appendix {B}. 
The difference of resulting expressions would persist also in 
the $\,L\to\infty\,$ limit of (\ref{eFt=L=inf}). 
\vskip 0.1cm

An explicit calculation of $\,{}^{{\rm th}\hspace{-0.04cm}}F^L_{t}(\bm\lambda)\,$
for $\,t\not=2L\,$ is also possible along the lines of Appendix {B}, using
the expansion of $\,H_i^0+\Delta H_i(t)\,$ in terms of the modes. 
We expect that the same large-deviations rate function (\ref{eFt=L=inf}) 
for energy transfers would result if we sent $L$ to infinity before $t$,
as suggested by the analysis of \cite{BD2}, but proving that basing on 
the exact formula for $\,{}^{{\rm th}\hspace{-0.04cm}}F^L_{t}(\bm\lambda)\,$
requires technical work that we postponed to the future.
\vskip 0.1cm

\subsection{FCS for charge and heat and fluctuation relations}
\label{subsec:ch.th.trans}

The characteristic function of joint measurements of charge and heat
transfers is defined as
\begin{align}
F^L_t(\bm\nu,\bm\lambda)\,=&\,
\sum\limits_{\Delta\bm q,\Delta\bm e}
\ee^{\,\ii\sum_i\big(\nu_i\Delta q_i+\lambda_i\Delta e_i}\,
\mathbb P_t\big(\Delta\bm q,\Delta\bm e\big)\cr
&=\,\omega^L_0\Big(\ee^{-\ii\sum\limits_i(\nu_i Q_i(0)+\lambda_iH_i(0))}
\,\ee^{\,\ii\sum\limits_i(\nu_i Q_i(t)+\lambda_iH_i(t))}\Big)
\end{align}
For $t=2L$, it can be easily computed since there is only a change in 
the contribution of the zero modes with respect to the calculation
of Subsec.\,(\ref{subsec:ch.th.trans}).
Indeed,
\begin{equation}
F^L_{2L}(\bm\nu,\bm\lambda)\,=
\,\omega_0^L\Big(\ee^{-\ii\sum\limits_i\big(\tilde\nu_i Q_i^0+\lambda_iH_i^0\big)}
\,\ee^{\,\ii\sum\limits_i\lambda_i\big(H_i^0+\Delta H_i(2L)\big)}\Big)
\end{equation}
so that the only effect is the change of $\,\bm V\,$ to 
$\,\bm V-i\beta^{-1}\tilde{\bm\nu}\,$ in the numerators 
of (\ref{12exp}). We infer that 
\begin{align}
&F^L_{2L}(\bm\nu,\bm\lambda)\,=\,
\,\det\big(I+i\beta^{-1}C\big)^{-1/2}\cr
&\times\ \frac{\sum\limits_{\bm k\in{\mathbb Z}^N}\ee^{-\pi L\,
(r\bm k\,,\,(\beta+\ii C)^{-1}r\bm k)
\,-\,\ii L\,(r(\beta\bm V-\ii\tilde{\bm\nu})\,,\,
(\beta+\ii C)^{-1}r\bm k)\,+\,\frac{L}{4\pi}
(r(\beta\bm V-\ii\tilde{\bm\nu})\,,\,(\beta+\ii C)^{-1}r
(\beta\bm V-\ii\tilde{\bm\nu}))}}
{\sum\limits_{\bm k\in{\mathbb Z}^N}\ee^{-\pi L\,
(r\bm k\,,\,\beta^{-1}r\bm k)\,-\,\ii L\,(r\beta\bm V
\,,\,\beta^{-1}r\bm k)\,+\,\frac{L}{4\pi}
(r\beta\bm V\,,\,\beta^{-1}r\beta\bm V)}}\cr
&\times\ \prod\limits_{n>0}\det\Big(I+
\big(I-\ee^{-\frac{\pi\ii}{L}n\lambda}\,
\ee^{\frac{\pi\ii}{L}n\,O\lambda O}\big)
\big(\ee^{\frac{\pi n}{L}\beta}-I\big)^{-1}\Big)^{-1}
\label{ceFt=L}
\end{align}
with
\begin{align}
\frac{1}{2L}\,\ln{F_{2L}^L(\bm\nu,\bm\lambda)}
\ \,&\mathop{\longrightarrow}\limits_{L\to\infty}\ \,
\sfrac{1}{8\pi}\Big(\big(r(\beta\bm V-\ii\tilde{\bm\nu})\,,\,(\beta+iC)^{-1}r
(\beta\bm V-\ii\tilde{\bm\nu})\big)-\big( r
\bm V\,,\,\beta r\bm V\big)\Big)\cr
&\hspace{1cm}-\,\int\limits_0^\infty
\ln\det\Big(I+\big(I-\ee^{-2\pi\ii\, x\lambda}\,\ee^{2\pi\ii\,x\,O\lambda O}\big)
\big(\ee^{2\pi x\beta}-I\big)^{-1}\Big)\,\dd x\cr
&\hspace{1cm}\equiv\ f(\bm\nu,\bm\lambda)\,.
\label{ceFt=L=inf}
\end{align}
The large-deviations rate function function 
(\ref{ceFt=L=inf}) of FCS for charge and heat transfers satisfies the 
fluctuation relation \cite{AGMT,BD3}
\begin{equation}
f(\bm\nu,\bm\lambda)\,=\,
f(-\bm\nu-\ii\beta\bm V,-\bm\lambda+\ii\bm\beta)
\label{FR}
\end{equation}
that reflects the time-reversal invariance of the dynamics.
The generating function (\ref{ceFt=L}) does not possess, 
however, the corresponding symmetry which arises only in the 
$t=2L\to\infty$ limit. Relation (\ref{FR}) is a consequence of the following 
matrix transformation properties under the change $\,(\bm\nu,\bm\lambda)
\longmapsto(-\bm\nu-\ii\beta\bm V,-\bm\lambda+\ii\bm\beta)$:
\begin{align}
&\hspace{-0.1cm}\ii\beta+\ii C=\beta+\ii\lambda-\ii O\lambda O\ \longmapsto\ 
\beta+\ii(-\lambda+\ii\beta)-\ii 
O(\-\lambda+\ii\beta)O=O\big(\beta+\ii C\big)O\,,\\
&\hspace{-0.1cm}r(\beta\bm V-\ii\tilde{\bm\nu})=r\beta\bm V-\ii(I-O)r\bm\nu
\ \longmapsto\ r\beta\bm V-\ii(I-O)r(-\bm\nu-\ii\beta\bm V)
=O\big(r(\beta\bm V-\ii\tilde{\bm\nu})\big)
\end{align}
and of the symmetry (\ref{FRexc}) showed in Appendix {B}. That the
same symmetry fails to hold for the generating function (\ref{ceFt=L}) 
follows from the fact that under the change $\,(\bm\nu,\bm\lambda)
\longmapsto(-\bm\nu-\ii\beta\bm V,-\bm\lambda+\ii\bm\beta)\,$
the sum over vectors $\bm k\in\mathbb Z^N$ in the numerator
of the middle line of (\ref{ceFt=L}) is transformed into the one
over vectors $r^{-1}Or\bm k$ that, in general, do not belong to
$\mathbb Z^N$.

\nsection{Comparison to Levitov-Lesovik formulae}
\label{sec:compLL}
\setcounter{equation}{0}

In \cite{LL}, L. S. Levitov and G. B. Lesovik obtained  a closed formula for 
the FCS of charge transfers between $N$ free fermionic 
systems, as those of Sec.\,\ref{subsec:qferm}. Such systems are assumed
to be initially in different equilibrium states and to interact subsequently 
during a period of time $\,t$. \,Their interaction is described by an 
$N\times N$ unitary mode-dependent matrix $\,\mathbb S_t(p)$ accounting 
for the scattering between the fermions of different systems, see also 
\cite{LLL}. The Levitov-Lesovik formula for the generating function
of charge FCS has the form of a product over the free fermionic modes 
of determinants:
\qq
\Phi_t(\bm\nu)\ =\,\prod\limits_{p}\det\big(I-f(p)+f(p)\,\ee^{-\ii s(p)\nu}\,
\mathbb S_t(p)^\dagger\ee^{\ii s(p)\nu}\,\mathbb S_t(p)\big)\,,
\qqq
where $\,s(p)\,$ is the sign function representing the 
charge of modes, $\,\nu\,$ is the diagonal $N\times N$ matrix of coefficients
$\,\nu_i$ and $\,f(p)\,$ that of Fermi functions 
$\,f_i(p)=(\ee^{\beta_i(\epsilon(p)-s(p)V_i)}+1)^{-1}$, with $\,\epsilon(p)\,$
representing the energy of modes. Upon taking the scattering matrix 
time and mode independent, $\,\mathbb S_t(p)=\mathbb S$, \,and the linear 
dispersion relation $\,\epsilon(p)=\frac{\pi}{L}|p|\,$
as in Sec.\,\ref{subsec:qferm}, \,and upon aligning the time and 
the size of the system by setting $\,t=2L$, \,the above generating
function leads in the rate function
\begin{align}
\phi(\bm\nu)\equiv
\lim\limits_{t=2L\to\infty}\,\frac{1}{t}\,\ln{\Phi_t(\bm\nu)}\,=&\ \sfrac{1}{2\pi}
\int\limits_0^\infty \ln\Big(\det\big(I-f^+(\epsilon)+f^+(\epsilon)\,
\ee^{-\ii\nu}\,\mathbb S^\dagger\ee^{\,\ii\nu}\,\mathbb S\big)\Big)\,\dd\epsilon\cr
+&\ \sfrac{1}{2\pi}
\int\limits_0^\infty \ln\Big(\det\big(I-f^-(\epsilon)+f^-(\epsilon)\,
\ee^{\,\ii\nu}\,\mathbb S^\dagger\ee^{-\ii\nu}\,\mathbb S\big)\Big)\,\dd\epsilon\,,
\label{LDLL}
\end{align}
where $\,f^\pm(\epsilon)\,$ are the diagonal matrices with entries
$\,f^\pm_{i}(\epsilon)=(\ee^{\beta_i(\epsilon\mp V_i)}+1)^{-1}$. \,Note that
this is a different expression than the rate function 
$\,{}^{{\rm el}\hspace{-0.05cm}}f(\bm\nu)\,$ of (\ref{LD}) obtained 
in Sec.\,\ref{subsec:ch.transp} which 
is quadratic in $\,\bm\nu\,$ and $\,\bm V$. \,For closer comparison, 
let us extract from (\ref{LDLL}) its leading quadratic contribution
describing the central-limit Gaussian distribution of charge transfers. 
In Appendix {C}, we show that
\begin{align}
\lim\limits_{\theta\to\infty}\,\theta^{2}\,\phi(\theta^{-1}\bm\nu)\big|_{\bm\beta,
\,\theta^{-1}\bm V}\,=
&-\sfrac{\ii}{2\pi}\sum\limits_iV_i\big(\nu_i-
\sum\limits_{i'}|\mathbb S_{i'i}|^2\nu_{i'}\big)+
\sum\limits_i\sfrac{\nu_i}{2\pi\beta_i}
\sum\limits_{i'}|\mathbb S_{i'i}|^2\nu_{i'}\cr
&-\sum\limits_i
\sfrac{\nu_i^2}{4\pi\beta_i}
-\sum\limits_i\sfrac{1}{4\pi\beta_i}
\sum\limits_{i',i''}|\mathbb S_{i'i}|^2\nu_{i'}|\mathbb S_{i''i}|^2\nu_{i''}
\cr
&+\sum\limits_i\sfrac{\ln{2}}{2\pi\beta_i}\sum\limits_{i',i''}
|\mathbb S_{i'i}|^2\nu_{i'}|\mathbb S_{i''i}|^2\nu_{i''}
-\sum\limits_i\sfrac{\ln{2}}{2\pi\beta_i}\sum\limits_{i'}|\mathbb S_{i'i}|^2
\nu_{i'}^2\cr
&+\,\sfrac{1}{2\pi}\sum\limits_{i\not=i'}g(\beta_i,\beta_{i'})\sum\limits_{j,j'}
\mathbb S^\dagger_{ij}\nu_j\hspace{0.03cm}
\mathbb S_{ji'}\mathbb S^\dagger_{i'j'}\nu_{j'}\mathbb S_{j'i}\,,
\label{LLRF}
\end{align}
where $\,g(\beta_i,\beta_{i'})\,$ is given by the integral formula (\ref{fbb}).
The first two lines on the right hand side reproduce the rate function 
(\ref{LD}) for the compactification radii squared $\,r_i^2=2\,$ that 
correspond to free fermions if we set $\,S_{ii'}=|\mathbb S_{ii'}|^2$. 
The last two lines represent terms not present
in the rate function (\ref{LD}). Of course, in spite of similarities,
the coupling between the free fermions realized by the junction of wires 
with matrix $\,S\,$ describing the scattering of the currents at the junction 
is different than that assumed in the Levitov-Lesovik approach, so 
there is no {\it a priori} reason for the two systems to lead to the 
same charge transport statistics. Note also that for arbitrary unitary 
matrix $\,(\mathbb S_{ii'})\,$ the matrix $\,(|\mathbb S_{ii'}|^2)\,$ 
is not necessarily orthogonal. 
\vskip 0.1cm

In the particular case when all temperatures are equal $\beta_i=\beta\,$ for
$\,i=1,\dots,N$, \,the last line of (\ref{LLRF}) reduces to
\begin{align}
&\sfrac{2\ln{2}-1}{4\pi\beta}\Big(\sum\limits_{i\not,i'}\sum\limits_{j.j'}
\mathbb S^\dagger_{ij}\nu_j\hspace{0.03cm}
\mathbb S_{ji'}\mathbb S^\dagger_{i'j'}\nu_{j'}\mathbb S_{j'i}\,-\sum\limits_i
\sum\limits_{j,j'}
\mathbb S^\dagger_{ij}\nu_j\hspace{0.03cm}
\mathbb S_{ji}\mathbb S^\dagger_{ij'}\nu_{j'}\mathbb S_{j'i}\Big)\cr
&=\,
\sfrac{2\ln{2}-1}{4\pi\beta}\sum\limits_i\Big(\nu_i^2-\sum\limits_{j,j'}
|\mathbb S_{ji}|^2\nu_j\,|\mathbb S_{j'i}|^2\nu_{j'}\Big),
\end{align} 
if we use relation (\ref{fbbe}) and the unitarity of matrix $\,\mathbb S$, 
\,and expression (\ref{LLRF}) reduces to
\qq
-\sfrac{1}{2\pi}\sum\limits_i(\ii V_i+\beta^{-1}\nu_i)\big(\nu_i-
\sum\limits_{i'}|\mathbb S_{i'i}|^2\nu_{i'}\big)
\label{LLRFb}
\qqq
On the other hand, the rate function (\ref{LD}) becomes in this case equal to
\qq
{}^{{\rm el}\hspace{-0.05cm}}f(\bm\nu)\,
=\,-\sfrac{1}{2\pi}\sum\limits_i(\ii V_i+\beta^{-1}\nu_i)\big(\nu_i-
\sum\limits_{i'}S_{i'i}\nu_{i'}\big)
\qqq
upon using the orthogonality of matrix $\,S=O$. \,It follows that 
for equal temperatures,
\qq
{}^{{\rm el}\hspace{-0.05cm}}f(\bm\nu)\,=\,\lim\limits_{\theta\to\infty}\,\theta^2\,
\phi(\theta^{-1}\bm\nu)\big|_{\bm\beta,\,\theta^{-1}\bm V}
\qqq
if we identify $\,|\mathbb S_{ij}|^2=S_{ij}$, \,assuming that the latter
identification leads to a matrix $\,S\,$ with the desired properties.
In that case, the fluctuations of charge transfers induced by different 
electric potentials at the same ambient temperature agree in the two setups 
on the level of the Gaussian central limit contributions. One should
remark, however, that the scaling limit (\ref{LLRF}) removes from the 
Levitov-Lesovik rate function (\ref{LDLL}) the term linear in $\,\bm V\,$ 
and quadratic in $\,\bm\nu\,$ that is responsible for the zero-temperature 
shot noise given by the Khlus-Lesovik-B\"{u}ttiker formula \cite{BB}.    
\vskip 0.1cm

There is another relation of the FCS statistics that we have obtained
for the junction of wires and the Levitov-Lesovik type formulae, this time 
for the energy transfers. Indeed, the contribution (\ref{exccont})
of the excited modes to the generating function 
$\,{}^{{\rm th}\hspace{-0.04cm}}F^L_{2L}(\bm\lambda)\,$ of energy FCS coincides with 
the version of the Levitov-Lesovik formula for $N$ free bosons with 
the dispersion relation $\,\epsilon_n=\frac{\pi n}{L}\,$ and the interaction 
described by the scattering matrix $\mathbb S=O$. \,The bosonic version 
of the Levitov-Lesovik formula was obtained in \cite{Klich}. Its proof 
in that reference provides a more direct way to calculate the excited 
modes contribution to $\,{}^{{\rm th}\hspace{-0.04cm}}F^L_{2L}(\bm\lambda)\,$ than 
the one followed in Appendix {B}. Unlike the proof of \cite{Klich}, however, 
our calculation may be extended to the case of 
$\,{}^{{\rm th}\hspace{-0.04cm}}F^L_{t}(\bm\lambda)\,$ for general $\,t\,$ in which
case matrices $\,(A_{ni,n'i'})\,$ in formula (\ref{CAA}) do not vanish.

\nsection{Examples}
\label{sec:exampl}
\setcounter{equation}{0}
\subsection{Case $\,N=2$}
\label{subsec:N=2}

In the case of two wires, the dimension $M$ of the brane should be $1$
for an interesting junction, since $M=2$ leads to a disconnected junction 
with $S=I$ and $M=0$ gives $S=-I$, which does not conserve the total charge.  
For $M=1$, let $\kappa = (a,b)$ and the compactification
radii $r_1^2$  and $r_2^2$. The injectivity of $\kappa$ requires
$a \wedge b =1$, and the conservation of charge in ensured for $a=b=1$. 
Forgetting this last requirement for a while, the $S$-matrix takes
the form
\begin{equation}
 S = \frac{1}{r_1^2 a^2+ r_2^2 b^2}\begin{pmatrix}
      {r_1^2 a^2- r_2^2 b^2} & {2 r_1^2 ab} \\ {2 r_2^2 ab} 
      & {r_2^2 b^2- r_1^2 a^2}
     \end{pmatrix}.  
\end{equation}
Two simple but interesting cases arise here. The first one will 
require the charge conservation ($a=b=1$) but will keep general radii 
of compactification for each wire, and
the second one will relax the charge conservation for the equal radii
$r_1 = r_2 = r$. \,In the second case, we shall consider only the heat 
transport. 

\begin{figure}[hbt]
\psfrag{title}{\hspace{-1cm} $\scriptstyle (2 \pi)/(r_1^2t)\, \langle\Delta
q_1\rangle $}
\psfrag{datadata}{\hspace{-0.05cm}\tiny$\scriptstyle \Delta\hspace{-0.05cm}
V =1$}
\psfrag{data2}{\hspace{-0.05cm}\tiny$\scriptstyle \Delta\hspace{-0.05cm}V
=-\sfrac{1}{2}$}
\psfrag{data3}{\hspace{-0.05cm}\tiny$\scriptstyle \Delta\hspace{-0.05cm}V =2$}
\psfrag{rho}{$\scriptstyle \rho$}
\hspace{0.5cm}\includegraphics[scale=0.5]{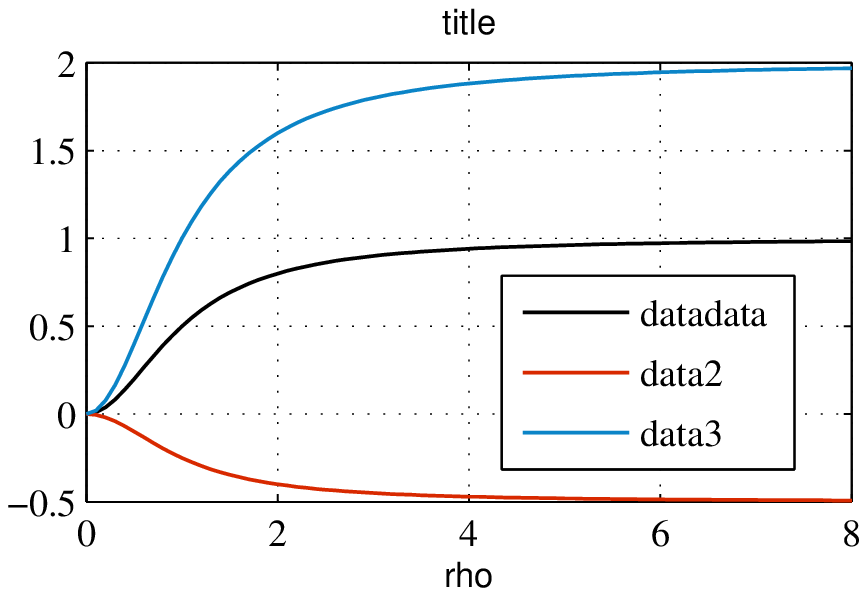} 
\psfrag{title}{\hspace{-1.5cm} $\scriptstyle (\pi t) / (r_1^2)\, C_{11} \text{ at
} \beta_1=1$}
\psfrag{datadata}{$\scriptstyle \beta_2 =1$}
\psfrag{data2}{$\scriptstyle \beta_2 =3$}
\psfrag{data3}{$\scriptstyle \beta_2 =10$}
\includegraphics[scale=0.5]{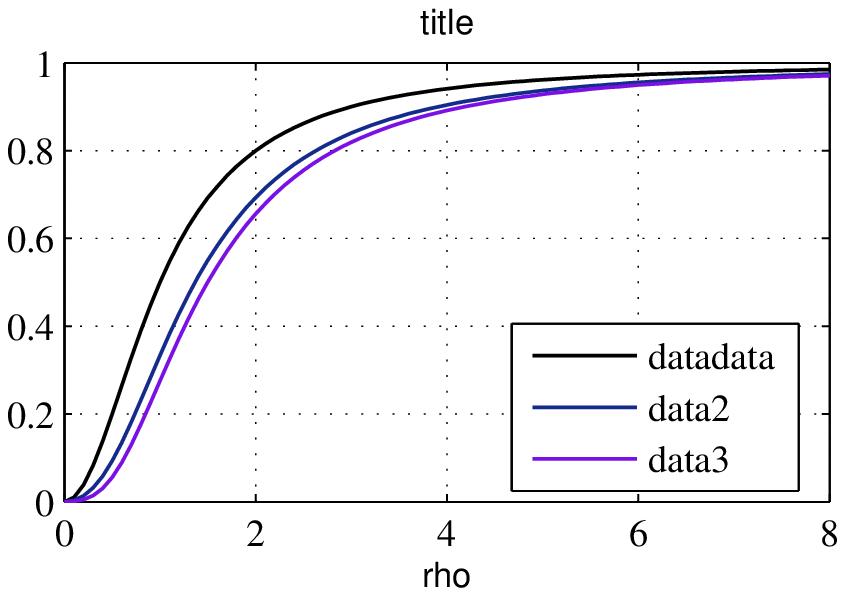} 
\psfrag{title}{\hspace{-1.5cm} $\scriptstyle (\pi t) / (r_1^2)\, C_{11} \text{ at
} \beta_2=1$}
\psfrag{datadata}{$\scriptstyle \beta_1 =1$}
\psfrag{data2}{$\scriptstyle \beta_1 =3$}
\psfrag{data3}{$\scriptstyle \beta_1 =10$}
\includegraphics[scale=0.5]{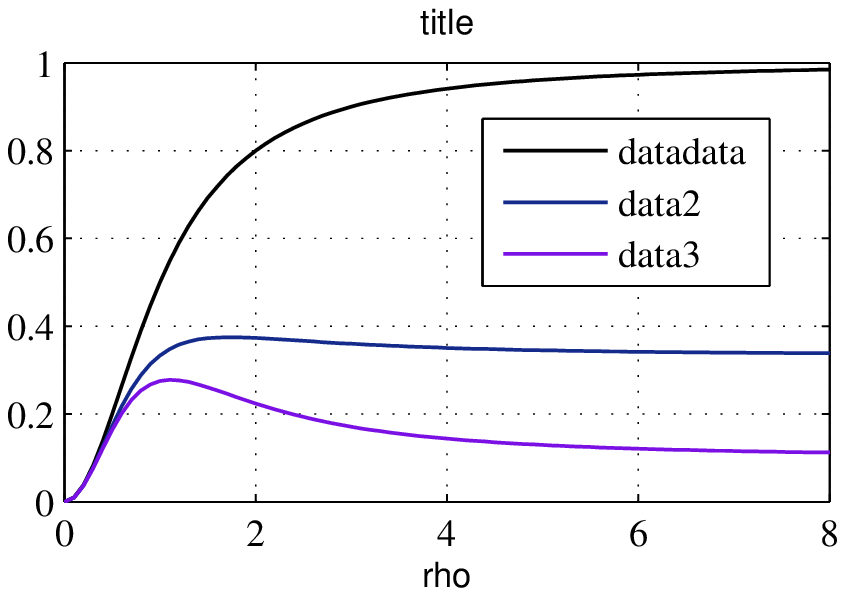} 
\vskip -0.25cm
\caption{Mean and covariance for different values of $\Delta V= V_2-V_1$,
$\beta_1$ and $\beta_2$. \label{fig:electric}}
\vskip 0.25cm
\end{figure}

\paragraph{General radii, \,charge conserved.}

\ Here
\begin{equation}\label{S_1}
 S = \frac{1}{r_1^2 + r_2^2}\begin{pmatrix}
      r_1^2 - r_2^2  & 2 r_1^2 \\ 2 r_2^2  & r_2^2 - r_1^2 
     \end{pmatrix},\qquad O = \frac{1}{r_1^2 + r_2^2}\begin{pmatrix}
      r_1^2 - r_2^2  & 2 r_1r_2 \\ 2 r_1 r_2  & r_2^2 - r_1^2 
     \end{pmatrix}, 
\end{equation}
see (\ref{ON2}).
Note that in the particular case $\,r_1 = r_2$,
\begin{equation}
S = O = \begin{pmatrix}
      0  & 1\\ 1  & 0 
     \end{pmatrix} 
\end{equation}
corresponding to the fully transmitting junction. 
For general radii, one obtains from Eqs.\,(\ref{noneq.J1}) and 
(\ref{noneq.K1}) for the mean electric and thermal currents 
in the non equilibrium stationary state the expressions
\begin{align}
&\hspace{-0.05cm}\omega_{\rm neq}(J_1^1(t,x)) = -\omega_{\rm neq}(J_2^1(t,x)) =
\frac{r_1^2r_2^2}{r_1^2+r_2^2} \frac{V_2-V_1}{2 \pi}\\
&\hspace{-0.05cm}\omega_{\rm neq}(K_1^1(t,x)) =  -\omega_{\rm neq}(K_2^1(t,x))=
\frac{r_1^2 r_2^2}{(r_1^2 +r_2^2)^2}\Big(\frac{V_2-V_1}{2\pi}(r_1^2 V_1
+ r_2^2 V_2) + \frac{\pi}{3} \Big(
\frac{1}{\beta_2^2}- \frac{1}{\beta_1^2} \Big)\hspace{-0.05cm}\Big)
\end{align}
implying for the electric and thermal conductance the formulae
\begin{equation}
{}^{\rm el}G = \frac{1}{2\pi}\frac{r_1^2r_2^2}{r_1^2+r_2^2} 
\begin{pmatrix}-1 & 1 \\ 1 & -1
\end{pmatrix}, \qquad  G^{\rm th} = \frac{2\pi}{3 \beta}\frac{r_1^2r_2^2}{(r_1^2+r_2^2)^2}
\begin{pmatrix}-1 & 1 \\ 1 & -1 \end{pmatrix}.
\end{equation}
Hence, in mean, (with our convention) the electric current flows through 
the junction from the wire at higher potential to the one at the lower 
one and, when the potentials are equal, the energy current flows from 
the wire at higher temperature to the one at lower temperature, although
the latter direction may be reversed by putting the lower-temperature wire
in sufficiently high electric potential. 
The large deviation rate function associated to charge only is
\begin{equation}
{}^{\rm el}\hspace{-0.05cm}f(\nu) = - \frac{1}{2\pi} \frac{r_1^2 r_2^2}{(r_1^2 + r_2^2)^2} \Big(
\frac{r_2^2}{\beta_1} + \frac{r_1^2}{\beta_2} \Big) \nu^2 + \frac{\ii}{2 \pi}
\frac{r_1^2 r_2^2}{r_1^2 + r_2^2} (V_2-V_1) \nu\,,
\end{equation}
where $\,\nu= \nu_1 - \nu_2$, \,see Eq.\,(\ref{LD}).
In the special case $\,r_1 = r_2 =r\,$ of a fully transmitting junction
\begin{equation}
{}^{\rm el}\hspace{-0.05cm}f(\nu) 
= - \frac{r^2}{8\pi} \Big( \frac{1}{\beta_1} + \frac{1}{\beta_2}
\Big) \nu^2 + \frac{\ii r^2}{4 \pi} (V_2-V_1) \nu
\end{equation}
which is compatible\footnote{Ref.\,\cite{BD2}
uses a different normalization of the $U(1)$-charges so $\,\nu_i\,$ and 
$\,V_i\,$ there are rescaled by $\,\frac{r}{2\pi}\,$ relative to the ones
used here.} with Eq.\,(86) of \cite{BD2}. 
The quadratic dependence of $\,{}^{\rm el}\hspace{-0.05cm}f(\nu)\,$ 
on $\,\nu\,$ implies that for large time the charge transfers per unit 
time become Gaussian random variables with mean and covariance equal to
\begin{equation}
 \left\langle \sfrac{\Delta q_1}{t} \right\rangle =  - \left\langle 
\sfrac{\Delta q_2}{t}
\right\rangle = \frac{r_1^2 r_2^2}{r_1^2 + r_2^2} \frac{V_2-V_1}{2\pi}\,,
\qquad
\CC = \frac{1}{\pi t} \frac{r_1^2 r_2^2}{(r_1^2 + r_2^2)^2} \Big(
\frac{r_2^2}{\beta_1} + \frac{r_1^2}{\beta_2} \Big)\begin{pmatrix}1 & -1 \\ -1
& 1 \end{pmatrix}.
\end{equation}
To illustrate the latter formulae, we trace in Fig.\,\ref{fig:electric} the
dependence of $\,\frac{2\pi}{r_1^2t}\langle\Delta q_1\rangle\,$ and of
$\,\frac{\pi t}{r_1^2}C_{11}\,$ on $\,\rho=\frac{r_2}{r_1}\,$
for few values of potential difference and temperatures. 

\vskip 0.1cm
 
\begin{figure}[h]
\centering
\psfrag{title}{$\scriptstyle\hspace{-0.9cm}f(\lambda)
\equiv\, ^{\text{th}}\hspace{-0.05cm}
f(-\ii \lambda)$}
\psfrag{-1}{$\scriptstyle-\beta_1$}
\psfrag{5}{$\scriptstyle\beta_2$}
\psfrag{data1}{$\scriptstyle \rho =1$}
\psfrag{data2}{$\scriptstyle \rho =2$}
\psfrag{data3}{$\scriptstyle \rho =3$}
\includegraphics[scale=0.57]{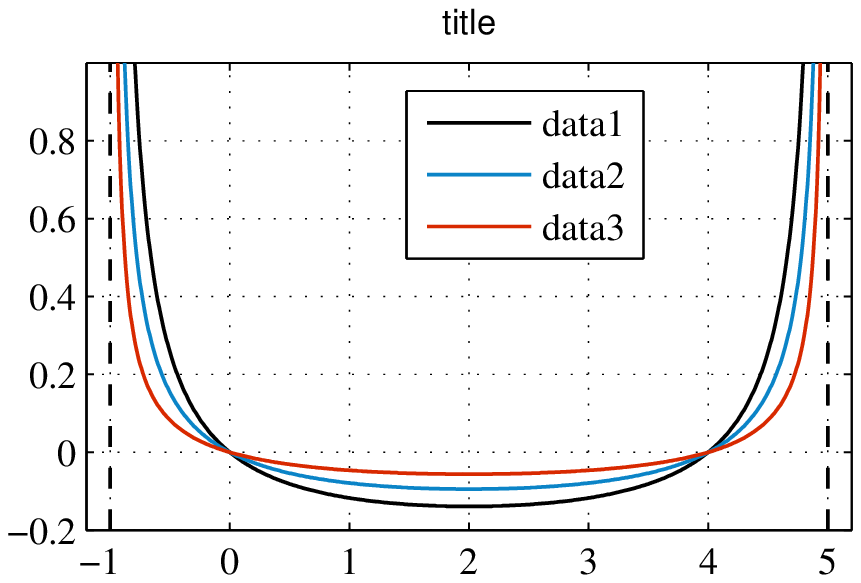} 
\hspace{1cm}
\psfrag{5}{$\scriptstyle 5$}
\psfrag{title}{\hspace{-0.15cm}$\scriptstyle I(x)$}
\psfrag{data1}{$\scriptstyle \rho =1$}
\psfrag{data2}{$\scriptstyle \rho =2$}
\psfrag{data3}{$\scriptstyle \rho =3$}
\psfrag{-1}{$-1$}
\includegraphics[scale=0.57]{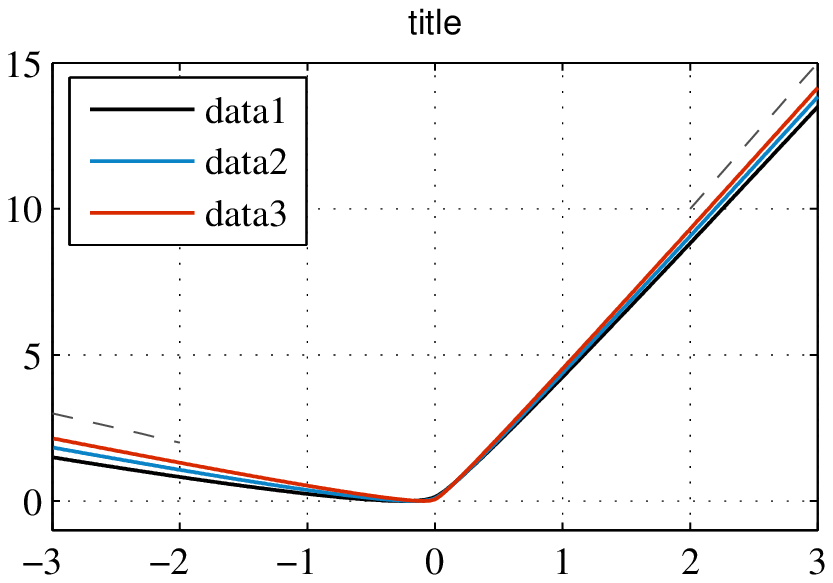}
\caption{Large deviation rate function $f(\lambda)$ and its Legendre transform
$\,I(x)\,$ for different $\rho = r_2/r_1$ at $\beta_1 = 1$, $\beta_2=5$, 
$V_1=V_2=0$). }
\label{fig:example_rho}
\end{figure}

\begin{figure}[h]
\hspace{0.5cm}
\includegraphics[scale=0.48]{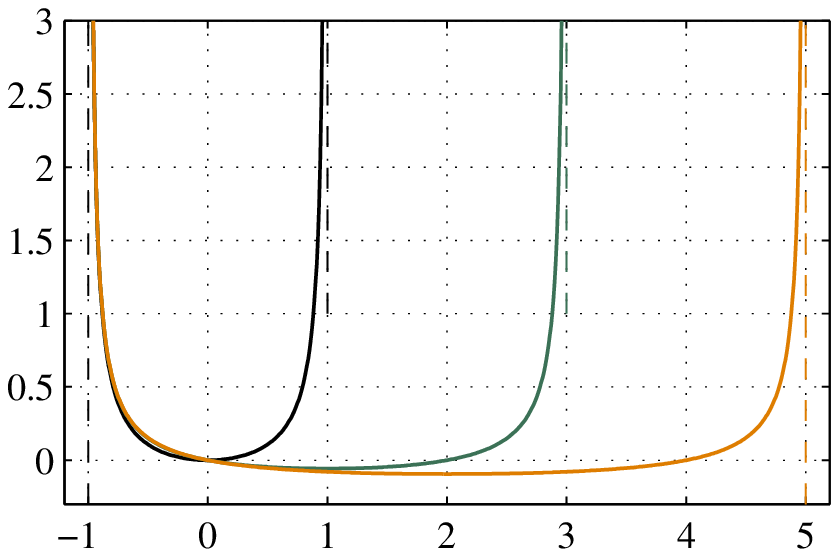} 
\hspace{-0.1cm}
\psfrag{data1}{\hspace{-0.06cm}\tiny$\scriptstyle \beta_2 =1$}
\psfrag{data2}{\hspace{-0.06cm}\tiny$\scriptstyle \beta_2 =3$}
\psfrag{data3}{\hspace{-0.06cm}\tiny$\scriptstyle \beta_2 =5$}
\includegraphics[scale=0.586]{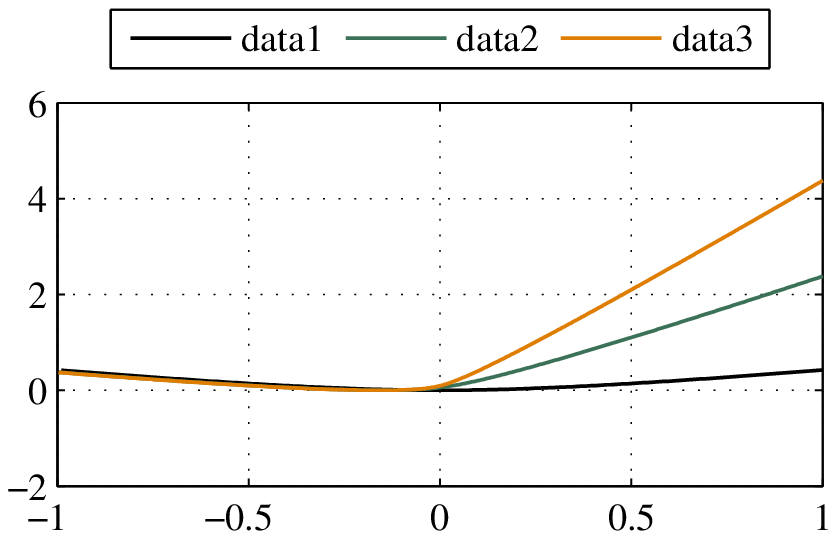}
\hspace{-0.1cm}
\includegraphics[scale=0.47]{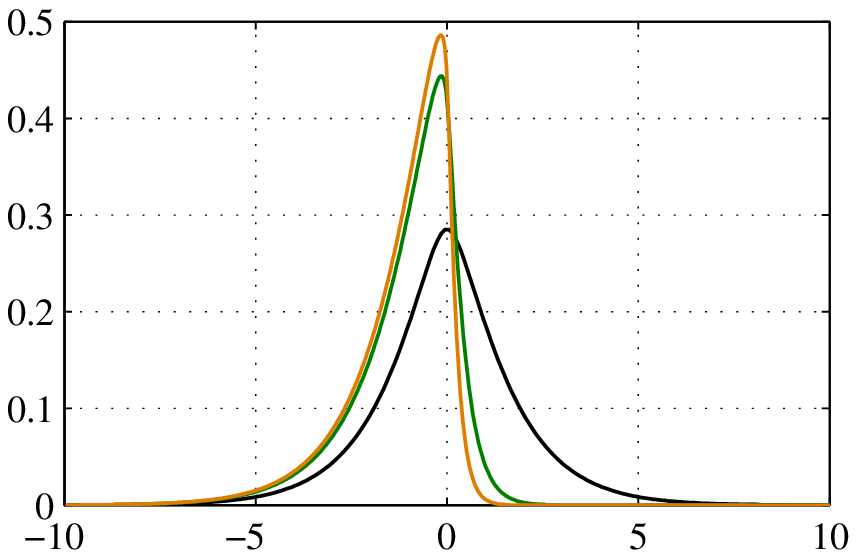}
\vspace{-0.1cm}
\caption{Large deviations rate function $f(\lambda)$, \,its Legendre
transform $I(x)$, \,and probability density $\propto \exp[-I(x)]$ for different 
$\beta_2\,$ ($N=2$, $\beta_1 = 1$, 
$r_2=2r_1$, and $V_1=V_2=0$)}
\label{fig:example_beta}
\vskip -0.4cm
\end{figure}

The large deviation rate function associated to energy only is
\begin{align}
&{}^{\rm th}\hspace{-0.05cm}f(\lambda) \,=\,
-\frac{r_1^2r_2^2}{2\pi} \frac{(r_1^2 \beta_1 V_1^2 + r_2^2 \beta_2 V_2^2)
\lambda^2 + \ii \beta_1 \beta_2 (V_1-V_2) (r_1^2 V_1 + r_2^2 V_2 )\lambda
}{4r_1^2 r_2^2 \lambda^2 + 4 \ii r_1^2 r_2^2 (\beta_2-\beta_1) \lambda + \beta_1
\beta_2 (r_1^2 + r_2^2)^2} \cr
& - \int_0^\infty \ln \sfrac{\big( 1+ \ee^{-2\pi x (\beta_1+\beta_2)}\big)
- (O_{11})^2 \big( \ee^{-2\pi x \beta_1} + \ee^{-2\pi x \beta_2}\big) -
(O_{12})^2 \big( \ee^{-2\pi x (\beta_1+\ii \lambda)}+\ee^{-2\pi x (\beta_2-\ii
\lambda)} \big)}{\big(1-\ee^{-2\pi x \beta_1}\big)\big(1-\ee^{-2\pi x
\beta_2}\big)}\,\dd x\qquad 
\label{thrf}
\end{align}
for $\,\lambda \equiv \lambda_1-\lambda_2$, \,see Eq.\,(\ref{eFt=L=inf}).
For a fully transmitting junction with $\,r_1 = r_2 =r$,  
\,the integral becomes computable, resulting 
in the expression
\begin{align}\label{ex_f_heat_R}
{}^{\rm th}\hspace{-0.05cm}f(\lambda) = & - \frac{r^2}{8\pi} 
\frac{(\beta_1V_1^2\hspace{-0.05cm}+\hspace{-0.05cm}\beta_2V_2^2) \lambda^2 
+ \ii \beta_1 \beta_2 (V_1^2\hspace{-0.05cm} -\hspace{-0.05cm}V_2^2) 
\lambda}{(\beta_1+
\ii \lambda)(\beta_2 - \ii \lambda)}
+ \frac{\pi}{12}\Big(\hspace{-0.04cm}\frac{1}{\beta_1+ \ii \lambda} 
\hspace{-0.05cm}-\hspace{-0.05cm}\frac{1}{\beta_1}
\hspace{-0.05cm}+\hspace{-0.05cm}
\frac{1}{\beta_2- \ii \lambda}\hspace{-0.05cm} -\hspace{-0.05cm}
\frac{1}{\beta_2}\hspace{-0.05cm}\Big)
\end{align}
which agrees with Eq.\,(90) of \cite{BD2} taken at $\nu = 0$. 
\,Let us look more closely at the analytic continuation 
$\,{}^{\rm th}\hspace{-0.05cm}f(-i\lambda)\equiv f(\lambda)\,$ of the rate 
function (\ref{thrf}) for $\,V_1=V_2=0$. $\,f(\lambda)\,$ is finite 
for $\,-\beta_1<\lambda<\beta_2\,$ and symmetric around 
$\,\frac{1}{2}(\beta_2-\beta_1)$. \,Outside that interval, $\,f(\lambda)\,$
diverges to $\,+\infty$. \,Fig.\,\ref{fig:example_rho} presents the graph
of $\,f(\lambda)\,$ and of its Legendre transform
\begin{equation}
 I(x) =  \max_{\lambda \in ]-\beta_1,\beta_2[} \{ \lambda x - f(\lambda) \}
\end{equation}
for $\,\beta_1=1,\ \beta_2=5\,$ and $\,\rho=\frac{r_2}{r_1}=1,2,3$. 
\,The change with increasing $\,\rho\,$ is clearly visible. 
\noindent In the limit
$\,\rho\to\infty$, function $\,f(\lambda)\,$ vanishes in 
the interval $\,]-\beta_1,\beta_2[\,$ and stays infinite outside of it. 
The large deviations rate function $\,I(x)\,$ is that of the probability 
distribution of the energy change in the first wire per unit time 
$\,\Delta H_1(t)/t$. $I(x)\,$ has linear asymptotes with the 
slopes $\,-\beta_1\,$ and $\,\beta_2\,$ on the left and on the right,
respectively, indicating the exponential decay of the distribution function
of $\,\Delta H_1(t)/t\,$ arising at long times, with the rate linearly 
growing with time. 
\begin{figure}[h]
\centering
\psfrag{data1}{$\scriptstyle \rho =1$}
\psfrag{data2}{$\scriptstyle \rho =2$}
\psfrag{data3}{$\scriptstyle \rho =3$}
\includegraphics[scale=0.58]{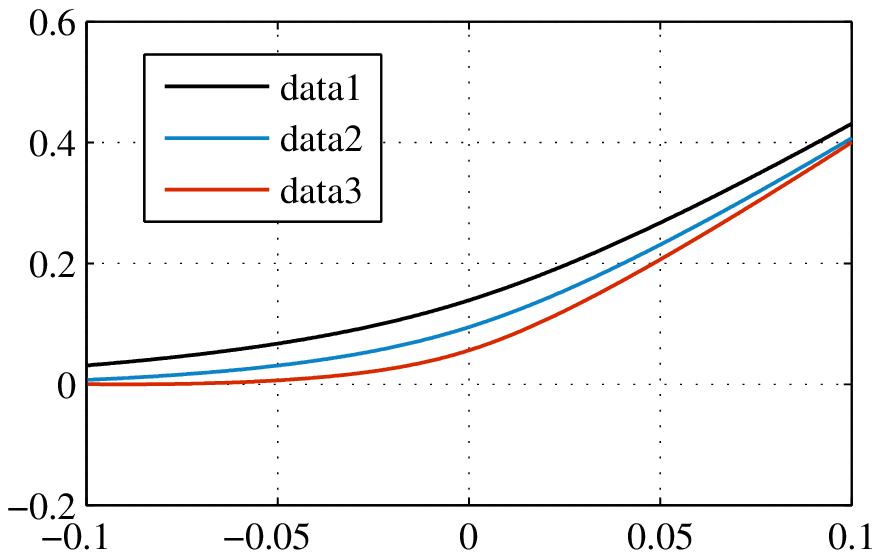}
\hspace{1cm}
\psfrag{title}{\hspace{-6.5cm}$\scriptstyle I(x)$\hspace{4.7cm}$\scriptstyle
\exp[-I(x)]\ {\rm normalised}$}
\psfrag{data1}{$\scriptstyle \rho =1$}
\psfrag{data2}{$\scriptstyle \rho =2$}
\psfrag{data3}{$\scriptstyle \rho =3$}
\includegraphics[scale=0.6]{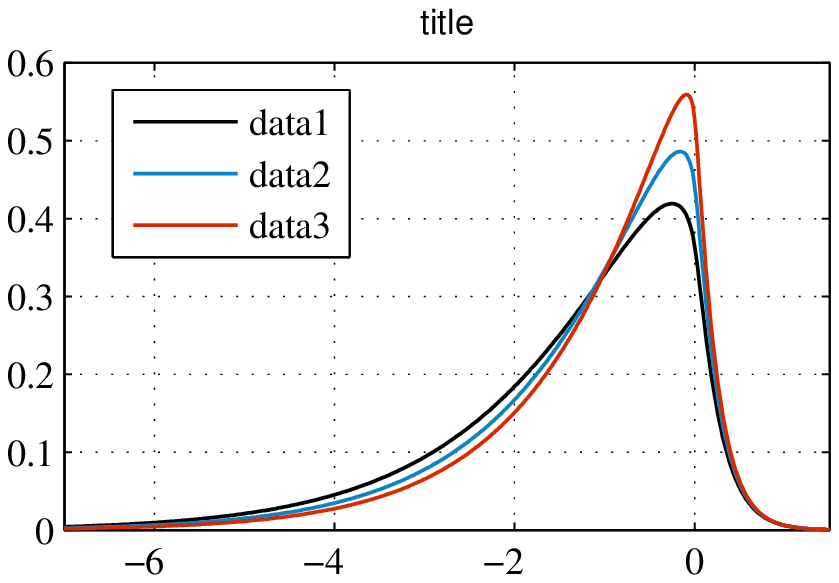}
\caption{Influence of $\,\rho\,$ on the rate function (zoom), and on the 
corresponding normalized density probability}
\label{fig:example_rho_zoom}
\end{figure}
\noindent Fig.\,\ref{fig:example_rho_zoom} zooms on the central region 
of $\,I(x)\,$ around $\,x=0\,$ and, for illustrative purpose, presents 
the graphs of normalized distribution functions $\,\propto\exp[-I(x)]$.    
\begin{figure}[t]
\centering
\psfrag{title}{\hspace{-0.4cm}\small Mean/$t$}
\psfrag{rho}{$\rho$}
\includegraphics[scale=0.57]{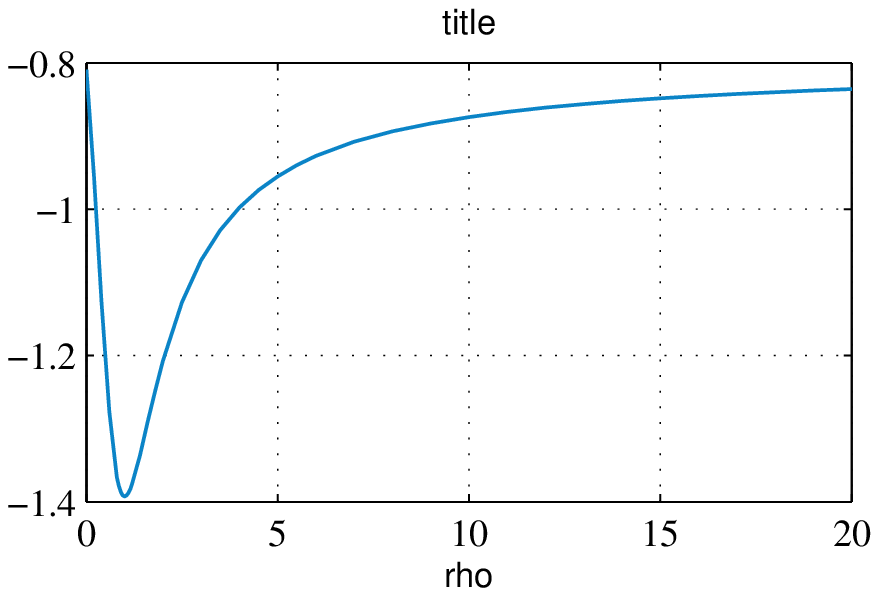}
\hspace{1cm}
\psfrag{title}{\hspace{-0.7cm}\small Variance/$t$}
\psfrag{rho}{$\rho$}
\includegraphics[scale=0.58]{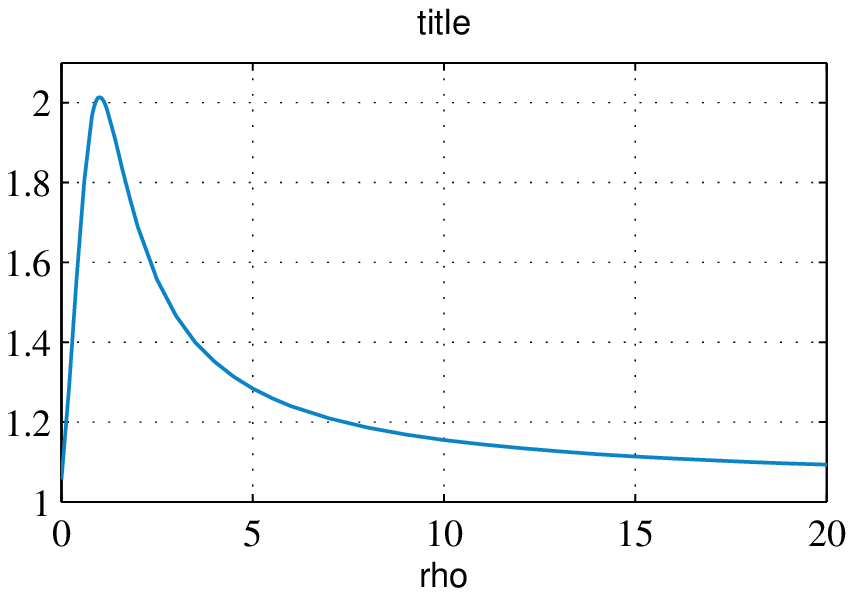}
\vskip -0.3cm
\caption{Influence of $\,\rho\,$ on the mean and 
the variance of $\Delta H_1(t)$ \label{fig:mean_cov1}}
\vskip -0.3cm
\end{figure}
\noindent The influence of $\,\rho\,$ on $\,f'(0)\,$ and $\,f''(0)\,$ 
representing the long-time mean  
of $\,\Delta H_1(t)\,$ and of its variance, both divided 
by $\,t$, \,is depicted in Fig.\,\ref{fig:mean_cov1}.
The mean and the variance per unit time represent, respectively, 
the mean heat current and the thermal noise in the first wire.  
The increase of $\,\rho\geq 1\,$ increases the absolute value of the 
current and decreases the noise. Both exhibit the 
$\,\rho\mapsto1/\rho\,$ symmetry implying that they are
least sensitive to the change of $\,\rho\,$ around $\,\rho=1$.
\,The influence of the temperature on the rate function
$\,f(\lambda)$, \,its Legendre transform $\,I(x)$, \,and on 
the probability distribution $\,\propto\exp[-I(x)]\,$ is illustrated 
on Fig.\,\ref{fig:example_beta}. 
The asymmetry of the curves increases when the temperature
of the second wire is lowered below that of the first wire.

\paragraph{Same radii, \,charge not conserved.} \ The interest 
in this case is due to the fact that it corresponds to a reflecting 
and transmitting junction for wires of the same type.
Indeed, for $\,r_1 = r_2 =r\,$ but $\,a \neq b$, 
\vskip -0.2cm
\begin{equation}
 S = O = \sfrac{1}{1+ \alpha^2}\begin{pmatrix}
      1-\alpha^2 & 2\alpha \\ 2\alpha & \alpha^2-1
     \end{pmatrix} \qquad \text{ for } \alpha = \sfrac{b}{a}\,.
\end{equation}
Since charge is not conserved when $\,\alpha \neq 1$, 
\.we focus on the energy transport only, and set $\,V_1 = V_2 =0$. \,Then
\vskip -0.5cm
\begin{align}
  \omega_{\rm neq}(K_1^1(t,x)) =&\,-\omega_{\rm neq}(K_2^1(t,x)) = 
\sfrac{\pi}{12} (S_{12})^2 \Big( \frac{1}{\beta_2^2}- \frac{1}{\beta_1^2}
\Big),\\
{}^{{\rm th}\hspace{-0.02cm}}G=&\,\frac{2\pi}{3}\Big(\frac{\alpha}{1+\alpha^2}\Big)^2
\begin{pmatrix}-1&1\cr 1&-1\end{pmatrix} 
\end{align}
\begin{figure}[b!]
\vskip -0.2cm
\includegraphics[scale=0.5]{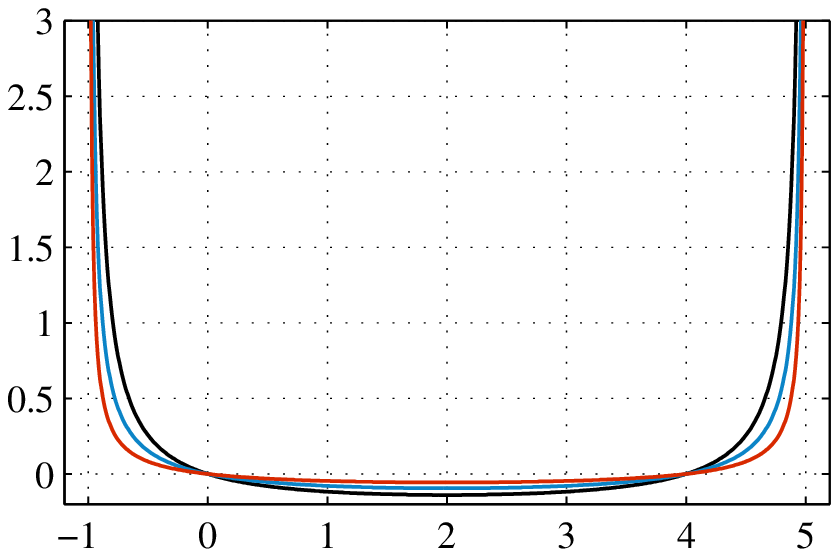} 
\hspace{-0.2cm}
\psfrag{data1}{\hspace{-0.06cm}$\scriptstyle \alpha =1$}
\psfrag{data2}{\hspace{-0.06cm}$\scriptstyle \alpha =2$}
\psfrag{data3}{\hspace{-0.06cm}$\scriptstyle \alpha =3$}
\includegraphics[scale=0.6]{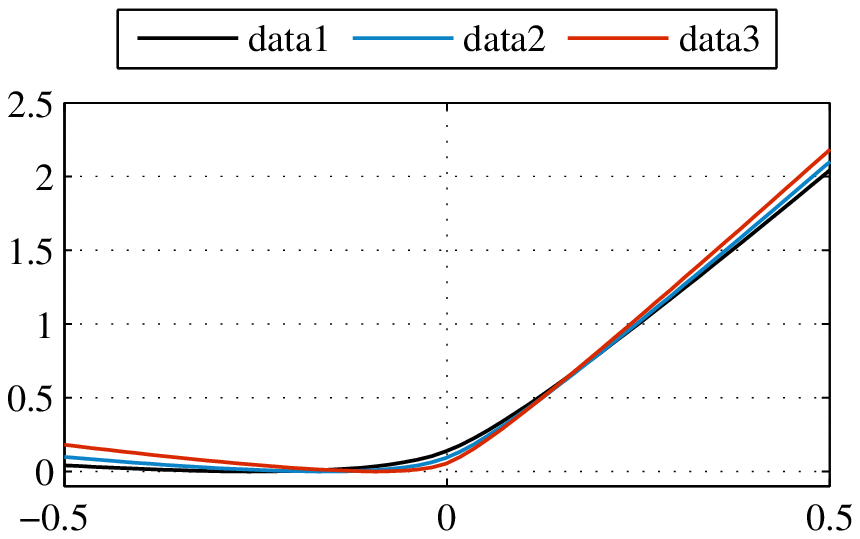}
\hspace{-0.1cm}
\includegraphics[scale=0.5]{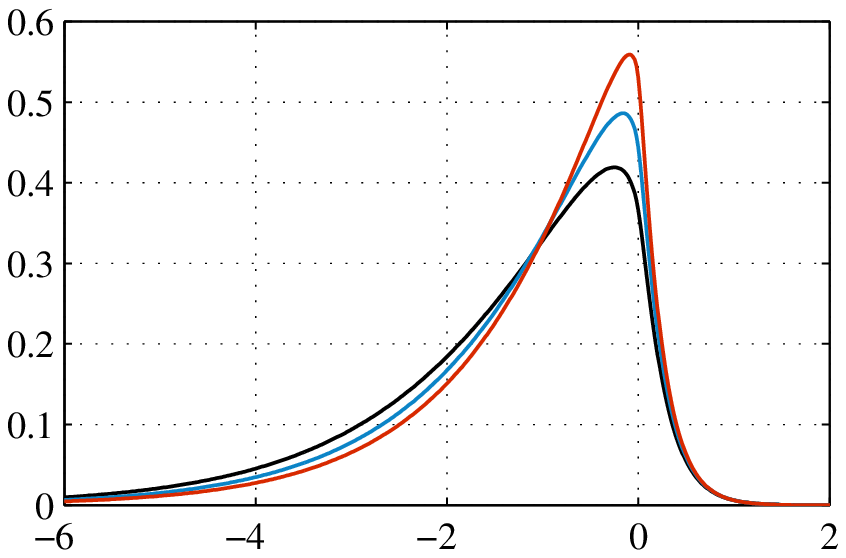}
\vskip -0.2cm
\caption{Large deviations rate function $f(\lambda)$, \,its Legendre
transform $I(x)$, \,and probability density $\propto \exp[-I(x)]$ 
for different $\alpha\,$ ($N=2$, $\beta_1 = 1$, $\beta_2=5$, 
$r_2=r_1$, and $V_1=V_2=0$)}
\label{fig:example_alpha}
\vskip -0.1cm
\end{figure}
\noindent and the large deviation function for 
energy transfer is, \,with $\,\lambda =
\lambda_1 - \lambda_2$,
\begin{align}
{}^{\rm th}\hspace{-0.05cm}f(\lambda)
=  - \int_0^\infty 
\hspace{-0.1cm}\ln \sfrac{1-\ee^{-2 \pi  x \beta_1
}(1-S_{12}^2(1-\ee^{-2 \pi \ii x \lambda }))-\ee^{-2 \pi  x \beta_2
}(1-S_{12}^2(1-\ee^{2 \pi \ii x \lambda }))+\ee^{-2 \pi  x
(\beta_1+\beta_2)}}{(1-\ee^{-2 \pi  x \beta_1 })(1-\ee^{-2 \pi  x \beta_2 })}\,
\dd x\,,
\end{align}
which is illustrated in Fig.\,\ref{fig:example_alpha}. 
The above expression simplifies for $\,\alpha =1\,$ when $\,S_{12}=1$, 
\,reducing again to the relation \eqref{ex_f_heat_R} for fully 
transmitting junction.

\subsection{Case $\,M=1$} 
\label{subsec:M=1}

The case $\,M=1\,$ with the total charge conservation corresponds 
to the brane $\,\CB\cong U(1)\,$ diagonally embedded into $\,U(1)^N$ 
so that
\begin{equation}
 \kappa = (1, \ldots , 1)
\end{equation}
leading to the $S$-matrix
\begin{equation}
 S_{ij} = - \rho_i \delta_{ij} + \tau_{i} (1-\delta_{ij})
\end{equation}
where 
\begin{equation}
 \rho_i = \sfrac{- r_i^2+\sum_{k\neq i} r_k^2 }{\sum_k r_k^2}, \qquad \tau_i =
\sfrac{2 r_j^2}{\sum_k r_k^2}
\end{equation}
are the reflection and transmission coefficients. For the equal radii
$\,r_i=r$,
\begin{equation}
\rho_i= \rho = \sfrac{N-2}{N}, \qquad \tau_i=\tau = \sfrac{2}{N}
\end{equation}
leading to a simple nontrivial $S$-matrix 
\begin{equation}\label{S_M=1}
 S_{ij} = - \rho \delta_{ij} + \tau(1-\delta_{ij})
\end{equation}
that was already considered in \cite{NFLL}, see also \cite{BMS}. 

\begin{figure}[htb]
\centering
\psfrag{title}{$\scriptstyle\hspace{-1.55cm}f(\lambda_{12},\lambda_{12})
\ \text{for}\ (\beta_1,\beta_2,\beta_3)=(1,1,1)$
}
\psfrag{lambda12}{$\scriptstyle\hspace{0.3cm}\lambda_{12}$}
\psfrag{lambda13}{$\scriptstyle\hspace{1.2cm}\lambda_{13}$}
\hspace{-0.3cm}\includegraphics[scale=0.655]{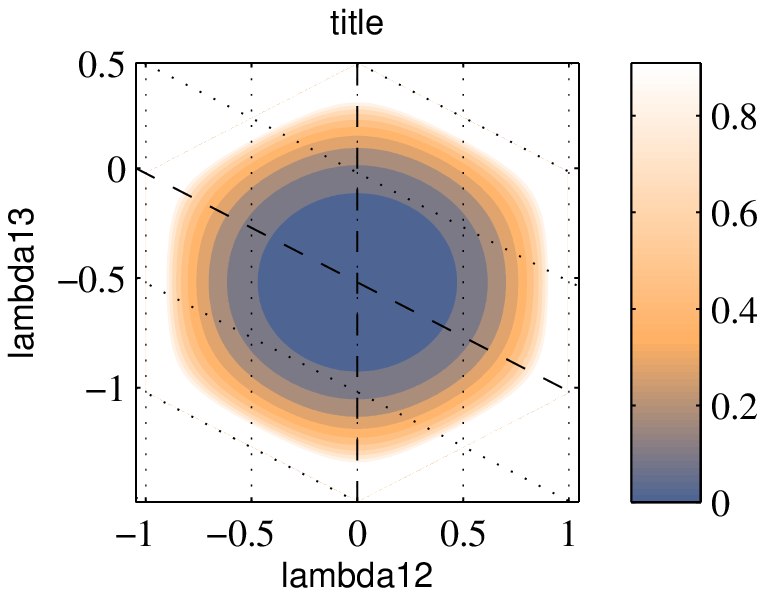}
\hspace{1cm}
\psfrag{title}{$\scriptstyle\hspace{-1.58cm}\text{and \,for\ }\ 
(\beta_1,\beta_2,\beta_3)=(1,2,3)$}
\psfrag{lambda13}{$\scriptstyle\hspace{0.9cm}\lambda_{13}$}
\psfrag{lambda12}{$\scriptstyle\hspace{0.3cm}\lambda_{12}$}
\includegraphics[scale=0.65]{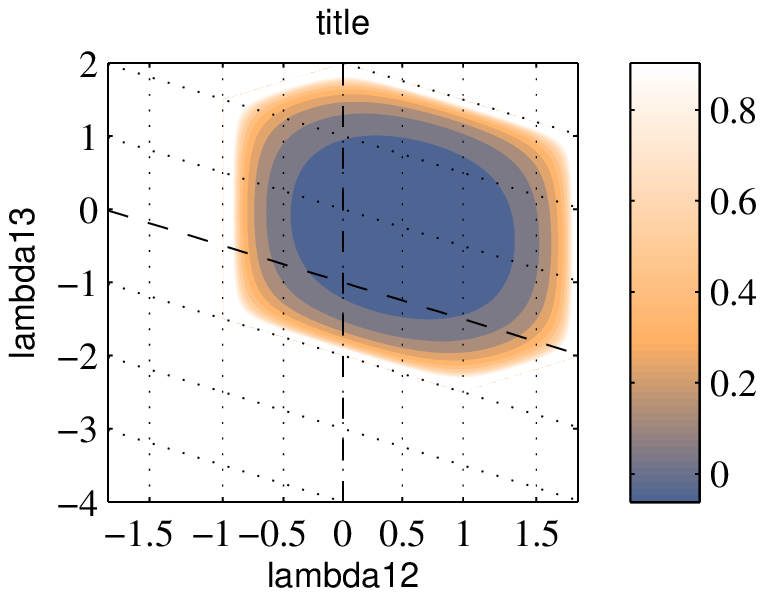}
\vskip -0.2cm 
\caption{Rate function $\,f(\lambda_{12},\lambda_{13})\,$
for \,3 \,wires\label{fig:f3wires}}
\end{figure}

\begin{figure}[h]
\centering
\psfrag{title}{$\scriptstyle\hspace{-1.55cm}I(x_{12},x_{12})
\ \text{for}\ (\beta_1,\beta_2,\beta_3)=(1,1,1)$
}
\psfrag{x12}{$\scriptstyle\hspace{0.1cm}x_{12}$}
\psfrag{x13}{$\scriptstyle\hspace{1cm}x_{13}$}
\hspace{-0.3cm}\includegraphics[scale=0.655]{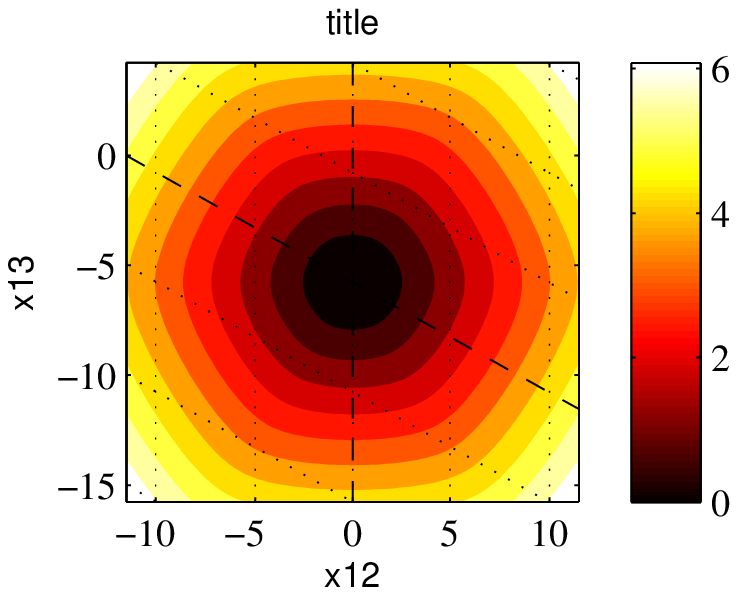}
\hspace{1cm}
\psfrag{title}{$\scriptstyle\hspace{-1.1cm}\text{and \,for}\ 
(\beta_1,\beta_2,\beta_3)=(1,2,3)$}
\psfrag{x13}{$\scriptstyle\hspace{1.1cm}x_{13}$}
\psfrag{x12}{$\scriptstyle\hspace{0.1cm}x_{12}$}
\includegraphics[scale=0.65]{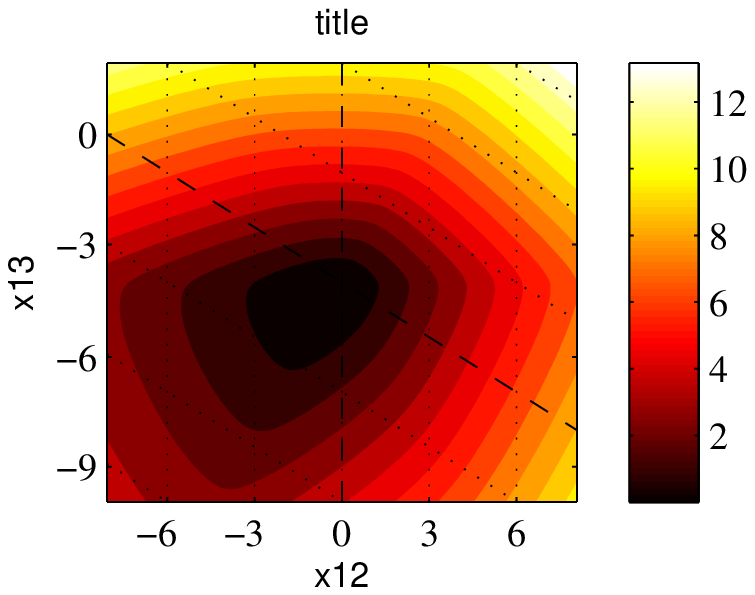}
\vskip -0.12cm 
\caption{Legendre transform $\,I(x_{12},x_{13})\,$
for \,3 \,wires\label{fig:I3wires}}
\end{figure}

\paragraph{Application to 3 wires} \ We consider the simplest 
case with the same radii of compactification. 
Here we have $\,\rho= 1/3\,$ and $\,\tau = 2/3\,$ and the $S$-matrix is
\begin{equation}
 S = \frac{1}{3}\begin{pmatrix}
      -1 & 2 & 2 \\ 2 & -1 & 2 \\ 2 & 2 & -1
     \end{pmatrix}
\end{equation}
The charge and energy currents are
\begin{align}
 & \omega_{neq}(J_1^L) = \sfrac{r^2\tau}{4\pi} (V_2 + V_3 - V_1) \cr
& \omega_{neq}(K_1^L) =  \sfrac{r^2\tau^2}{8\pi}(V_2 + V_3 - 2
V_1)(V_1+V_2+V_3) + \sfrac{\pi\tau^2}{12}\big(\sfrac{1}{\beta_2^2} +
\sfrac{1}{\beta_3^2} -\sfrac{2}{\beta_1^2} \big) 
\end{align}
and similarly on the other wires, cyclicly permuting the indices.
For the electric and thermal conductance, this gives:
\begin{equation}
{}^{{\rm el}\hspace{-0.02cm}}G\,=\,\sfrac{r^2\tau}{4\pi}\begin{pmatrix}
-1&1&1\\1&-1&1\\1&1&-1\end{pmatrix},\qquad
{}^{{\rm th}\hspace{-0.02cm}}G\,=\,\sfrac{\pi\tau^2}{6}\begin{pmatrix}
-2&1&1\\1&-2&1\\1&1&-2\end{pmatrix}.
\end{equation} 
The large deviation function for the FCS of charge transfers is 
\begin{equation}
{}^{\rm el}\hspace{-0.05cm}f(\bm\nu) = \sfrac{r^2\tau^2}{8\pi} 
\Big( \sfrac{(2 \nu_1 - \nu_2 -
\nu_3)^2}{\beta_1} \ +\ {\rm cycl.}\Big)+ \ii
\sfrac{r^2\tau}{4\pi}\Big( V_1 (2 \nu_1 - \nu_2 - \nu_3) \ 
+\ {\rm cycl.}\,\Big),
\end{equation}
where ``${\rm cycl.}$'' stands for terms  obtained by cyclic permutation 
of the indices.
The large deviation function for the FCS of energy transfers for 
$\,V_i = 0\,$ is:
\begin{equation}
{}^{\rm th}\hspace{-0.05cm}f(\bm\lambda) 
= - \int_0^\infty \ln \sfrac{N(x,\bm\lambda)}{D(x)}\, \dd x\qquad
\end{equation}
where
\begin{align}
N(x,\bm\lambda) = &\, 1 - \ee^{-2\pi x (\beta_1+\beta_2+\beta_3)} + \rho^2 \big( \ee^{-2\pi
x(\beta_1+\beta_2)} \ + \ {\rm cycl.}\,\big) - \rho^2 \big(
\ee^{-2\pi x \beta_1} \ + \ {\rm cycl.}\,\big)\cr
& + \tau^2 \Big( \ee^{-2\pi x(\beta_1+\beta_2)}\big(\ee^{-2\pi \ii
x(\lambda_2-\lambda_3)} + \ee^{-2\pi \ii x(\lambda_1-\lambda_3)}\big) 
\ + \ {\rm cycl.}\,\Big) \cr
& - \tau^2 \Big(\ee^{-2\pi x\beta_1}\big(\ee^{-2\pi \ii x(\lambda_1-\lambda_2)}
+ \ee^{-2\pi \ii x(\lambda_1-\lambda_3)}\big)\ + \ {\rm cycl.}\,\Big),\cr
 D(x) = &\,(1- \ee^{-2\pi x\beta_1}) (1- \ee^{-2\pi x\beta_2}) (1- \ee^{-2\pi
x\beta_3})\,. 
\end{align}
Upon the analytic continuation, $\,{}^{\rm th}\hspace{-0.05cm}f(-\ii\bm\lambda)
\equiv f(\lambda_{12},\lambda_{13})\,$ for 
$\,\lambda_{12}=\lambda_1-\lambda_2\,$ and 
$\,\lambda_{13}=\lambda_1-\lambda_3\,$
which is finite only in the region
\begin{equation}
 - \beta_1 < \lambda_{12} < \beta_2\,, \qquad  - \beta_1 < \lambda_{13} < 
\beta_3\,,\qquad- \beta_2 < \lambda_{13} - \lambda_{12} < \beta_3\,. 
\label{3w_limlambda}
\end{equation}
Function $\,f\,$ is plotted 
in Fig.\,\ref{fig:f3wires} \,for $\,(\beta_1,\beta_2,\beta_3)$$=(1,1,1)$
(the equilibrium case) and $\,(\beta_1,\beta_2,\beta_3)=(1,2,3)$
in the coordinate system with axes at $120^\circ$ so that the 
counter-clockwise rotation of the graph by $120^\circ$ corresponds 
to the cyclic permutation $(\lambda_1,\lambda_2,\lambda_3)
\mapsto(\lambda_3,\lambda_1,\lambda_2)$. \,In equilibrium, $\,f\,$ is 
symmetric under such a transformation but out of equilibrium, the 
above $\,\bm Z_3\,$ symmetry is broken to a degree that 
may be used as a measure of distance from equilibrium.
The Legendre transform of $\,{}^{\rm th}\hspace{-0.04cm}f(-\ii\bm\lambda)\,$ 
is infinite
out of the plane $\,x_1+x_2+x_3=0\,$ and on that plane, it may be
regarded as a function 
\qq
I(x_{12},x_{13})=\mathop{\rm max}\limits_{\lambda_{12},\lambda_{13}}\,\big\{
\frac{_1}{^3}(2x_{12}\lambda_{12}-x_{12}\lambda_{13}-x_{13}\lambda_{12}
+2x_{13}\lambda_{13})-f(\lambda_{12},\lambda_{13})\big\}\,.
\qqq
Fig.\,\ref{fig:I3wires} presents the plot of $\,I(x_{12},x_{13})\,$
for the equilibrium and nonequilibrium choice of temperatures. 
The level lines of $\,I\,$ are equally 
spaced in various direction far from the origin, indicating the asymptotic 
linear increase of the function. \,The similar breaking 
of $\,\bm Z_3\,$ symmetry as for $\,f\,$ may be observed. 
\begin{figure}[ht]
\centering
\psfrag{title}{$\scriptstyle\hspace{-2.4cm}\text{normalized}\ 
\exp[-I(x_{12},x_{12})]
\ \text{for}\ (\beta_1,\beta_2,\beta_3)=(1,1,1)$
}
\psfrag{x12}{$\scriptstyle\hspace{0.1cm}x_{12}$}
\psfrag{x13}{$\scriptstyle\hspace{1cm}x_{13}$}
\includegraphics[scale=0.655]{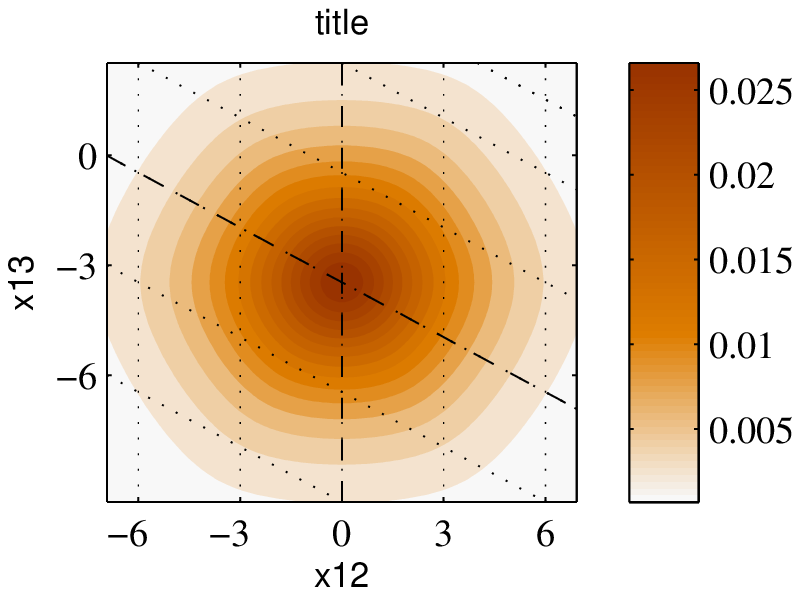}
\hspace{0.6cm}
\psfrag{title}{$\scriptstyle\hspace{-1.1cm}\text{and \,for}
\ (\beta_1,\beta_2,\beta_3)=(1,2,3)$}
\psfrag{x13}{$\scriptstyle\hspace{1cm}x_{13}$}
\psfrag{x12}{$\scriptstyle\hspace{0.1cm}x_{12}$}
\includegraphics[scale=0.65]{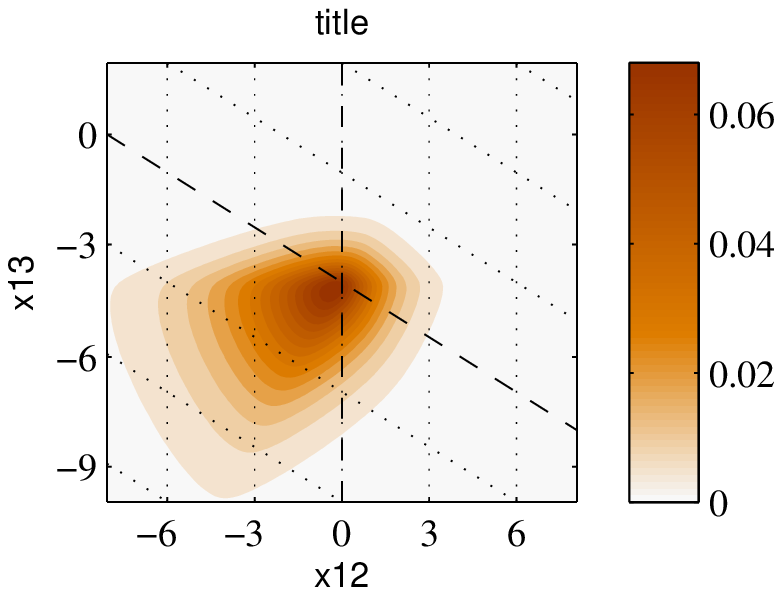}
\vskip -0.15cm 
\caption{Probability density $\,\propto\exp[-I(x_{12},x_{13})]\,$ 
for \,3 \,wires\label{fig:expI3wires}}
\end{figure}
Finally, 
Fig.\,\ref{fig:expI3wires} plots for illustration the probability 
densities $\,\propto\exp[-I(x_{12},x_{13})]$.
\,Note that most mass of the distribution is in the negative quadrant
indicating the heat transfer from the hotter $1^{\rm st}$ and the $2^{\rm nd}$
wires to the colder $3^{\rm rd}$ one.

\nsection{Conclusions} 
\label{sec:concl}
\vskip 0.15cm

We have studied a model of a junction of $\,N\,$ quantum wires. 
The Luttinger liquids in the bulk of the wires were represented by 
a toroidal compactification of the $N$-component massless free 
bosonic field, with the junction modeled by a simple boundary condition 
restricting the values of the compactified field 
to a brane $\,\CB\,$ forming a subgroup isomorphic to $\,U(1)^M\,$ 
of the target torus $\,U(1)^N$. 
\,The brane $\,\CB\,$ was assumed to be invariant under the
diagonal multiplication by phases in order to assure the global
$\,U(1)\,$ invariance and the conservation of the total electric 
charge. We constructed the theory on the classical and
the quantum level and showed that the boundary condition at 
the junction leads to a linear relation between the right-moving 
and the left-moving components of the electric currents in
different wires. Such a relation describes the scattering by 
the junction of charges carrying the current. 
The equilibrium state of the system of connected wires 
of length $\,L\,$ kept at inverse temperature $\,\beta\,$ and in 
electric potential $\,V\,$ was discussed in the functional-integral 
language and in the open-string and closed-string operator formalism, 
the latter being well suited to describe the thermodynamic
limit $\,L\to\infty$. \,We obtained the exact solution 
for the equilibrium current correlation functions both for wires 
of finite length $\,L\,$ and for $\,L=\infty$.
\,In the latter case, the resulting theory provides a special 
case of the system studied in \cite{MSLL} and we adapted
from that paper the construction of a stationary nonequilibrium 
state (NESS) in which the wires are kept at different temperatures 
and electric potentials. Following the lines of \cite{BD1,BD2},
it was shown that such a state is attained at long times if we 
prepare disjoint semi-infinite wires in equilibrium states at
different temperatures and potentials and then connect them
by the junction and let the dynamics operate. This is a particular 
realization of the scenario for construction of quantum nonequilibrium 
states proposed in \cite{Ruelle}. By considering the constructed 
NESS close to equilibrium, we extracted formulae for the electric 
and thermal conductance of the junction. 
\vskip 0.1cm

The main result of this paper has been the calculation of the full
counting statistics (FCS) for charge and heat transfers through 
the junction and the analysis 
of its large deviations asymptotics at long transfer times in 
the presence of both transmission and reflection of the conserved 
charges. This was done first for charge, then for heat, and finally, 
jointly for both. We confirmed by an exact calculation that the large 
deviations regime of the charge FCS for a junction of semi-infinite 
wires may be obtained from a large class of limiting procedures 
sending to infinity the length of the wires as well as the transfer 
time. The computation of FCS for heat transfers was explicitly 
done only aligning the length of the wires and the time of the transfer, 
although we developed tools for performing the general calculation. 
We expect that the result for the large deviations regime of the
FCS for heat transfers obtained in the explicitly treated case
applies as well to the situation when the length of the wires
is sent to infinity faster than the evolution time. The expressions 
obtained for the large deviations rate functions of FCS were compared 
with the ones given by the Levitov-Lesovik formulae. For the charge 
transfers, we showed that our results for the junctions under 
consideration differ from the Levitov-Lesovik formula for free fermions, 
although, for the vanishing Luttinger couplings, some similarity could 
be observed in the quadratic part of the rate functions that describes 
the central-limit asymptotics. For the energy transfers through 
the junction, the part of the FCS that was contributed by the excited
bosonic Fock-space modes appeared to coincide with the bosonic 
Levitov-Lesovik-type formula for FCS obtained in \cite{Klich}.  
\vskip 0.1cm

The simple class of conformal boundary defects considered in 
the present paper have been chosen for illustrative purpose rather 
than from phenomenological considerations. The latter might require 
introducing a larger family of boundary defects. The simplest extension 
of the class considered here would include conformal boundary defects 
with displaced branes $\,\bm g_0\CB\,$ for $\,\bm g_0\in U(1)^N\,$ or/and 
the ones with added Wilson lines (w.r.t.\,\,\,a constant gauge field 
on $\,\CB$). Such boundary defects could be dealt with 
by the same technique, leading to a richer class of $S$-matrices 
describing current scattering. More complicated 
conformal boundary defects would require more powerful 
boundary CFT techniques for calculation. For the case of vanishing 
Luttinger couplings, one could use the nonabelian bosonization 
\cite{WittenNA,AleksSchom} that comes with a class of conformal 
boundary defects with nonabelian symmetries. Some partial results
in this direction have been already obtained \cite{GT}. The other 
physically relevant question, not disjoint from the previous one, is 
the stability of the boundary defects in the renormalization group 
sense. This problem was addressed for some simple cases of junctions
in \cite{NFLL,OCA0,OCA}. It can be also studied with 
the boundary CFT techniques.
We postpone a discussion of the above questions to the future research.


\nappendix{A}
\label{app:A}
\vskip 0.5cm

\noindent Here we shall calculate directly the quantity
\begin{equation}
\omega^{L}_{\beta,V}\Big(\ee^{-\ii\nu\int\limits_0^tJ^\ell
(0,s)\,\dd s}\Big)
\end{equation}
for one wire of length $L> t$ with the Neumann boundary 
conditions. In this case,
\begin{equation}
J^\ell(0,s)\,=\,\sfrac{r}{4L}\sum\limits_{n\in\mathbb Z}\tilde\alpha_{2n}
\,\ee^{-\frac{\pi\ii ns}{L}}
\end{equation} 
for $\,\tilde\alpha_{2n}=r\alpha_{2n}\,$ and
\begin{equation}
\int\limits_0^tJ^\ell(0,s)\,\dd s\,=\,\sfrac{ir}{4\pi}\sum\limits_{n\not=0}
\sfrac{1}{n}\tilde\alpha_{2n}\big(\ee^{-\frac{\pi\ii nt}{L}}-1\big)
+\sfrac{rt}{4L}
\tilde\alpha_0\,=\,\sfrac{r^2}{4\pi}\big(\varphi^\ell(0,t)
-\varphi^\ell(0,0)\big),
\end{equation}
where the chiral field $\,\varphi^\ell\,$ is given by (\ref{phiell}).
We shall reorder the exponential of that operator writing
\begin{align}
\ee^{-\ii\nu\int\limits_0^tJ^\ell(0,s)\,\dd s}\,=&\,\,
\ee^{-\ii\frac{\nu r t}{4L}\tilde\alpha_0}\,\,
\ee^{\sfrac{\nu r}{4\pi}
\sum\limits_{n<0}\frac{1}{n}\tilde\alpha_{2n}
\big(\ee^{-\frac{\pi\ii nt}{L}}-1\big)}\,
\ee^{\sfrac{\nu r}{4\pi}
\sum\limits_{n>0}\frac{1}{n}\tilde\alpha_{2n}
\big(\ee^{-\frac{\pi\ii nt}{L}}-1\big)}\cr
&\times\,\,
\ee^{-\frac{\nu^2r^2}{16\pi^2}\sum\limits_{n>0}\frac{1}{n}\big(2-
\ee^{\,\frac{\pi\ii nt}{L}}-\ee^{-\frac{\pi\ii nt}{L}}\big)}\cr
=&\,\,
\ee^{-\ii\frac{\nu rt}{4L}\tilde\alpha_0}\,\,
\ee^{\sfrac{\nu r}{4\pi}
\sum\limits_{n<0}\frac{1}{n}\tilde\alpha_{2n}
\big(\ee^{-\frac{\pi\ii nt}{L}}-1\big)}\,
\ee^{\sfrac{\nu r}{4\pi}
\sum\limits_{n>0}\frac{1}{n}\tilde\alpha_{2n}
\big(\ee^{-\frac{\pi\ii nt}{L}}-1\big)}\cr
&\times\,\,
\big(4\sin^2(\sfrac{\pi t}{2L})\big)^{-\frac{\nu^2r^2}{16\pi^2}}
\,\ee^{-\frac{\nu^2 r^2}{8\pi^2}\sum\limits_{n>0}\frac{1}{n}}.
\label{A4}
\end{align}
Note that the last factor, that may be interpreted as providing
the Wick ordering of the left-moving vertex operators
$\,\ee^{-\ii\frac{\nu r^2}{4\pi}\varphi^\ell(0,t)}\,$ and
$\,\ee^{\,\ii\frac{\nu r^2}{4\pi}\varphi^\ell(0,0)}$,
\,is ultraviolet singular. We shall replace it by its regularized version 
\begin{equation}
\ee^{-\sfrac{\nu^2r^2}{8\pi^2}\sum\limits_{n<\Lambda L}\frac{1}{n}}\equiv\,
\ee^{-\sfrac{\nu^2r^2}{8\pi^2}C_{\Lambda L}}
\label{UVcutoff}
\end{equation}
where $\Lambda$ is the ultraviolet cutoff.
This leads to the definition:
\begin{align}
\Big(\ee^{-\ii\nu\int\limits_0^tJ^\ell(0,s)\,\dd s}\Big)_{\hspace{-0.05cm}{\rm reg}}
=&\,\,\,
\ee^{-\ii\frac{\nu r t}{4L}\tilde\alpha_0}\,\,
\ee^{\sfrac{\nu r}{4\pi}
\sum\limits_{n<0}\frac{1}{n}\tilde\alpha_{2n}
\big(\ee^{-\frac{\pi\ii nt}{L}}-1\big)}\,
\ee^{\sfrac{\nu r}{4\pi}
\sum\limits_{n>0}\frac{1}{n}\tilde\alpha_{2n}
\big(\ee^{-\frac{\pi\ii nt}{L}}-1\big)}\cr
&\times\,\,
\big(4\sin^2(\sfrac{\pi t}{2L})\big)^{\hspace{-0.05cm}-\frac{\nu^2r^2}{16\pi^2}}\,
\,\ee^{-\sfrac{\nu^2r^2}{8\pi^2}\sum\limits_{n<\Lambda L}\frac{1}{n}}\,.
\label{A5}
\end{align}
The commutation relation
\begin{equation}
\Big[\ee^{\pm\sfrac{\nu r}{4\pi}
\sum\limits_{n>0}\frac{1}{n}\tilde\alpha_{2n}
\big(\ee^{-\frac{\pi\ii nt}{L}}-1\big)},
\tilde\alpha_{-2m}\Big]\,=\,\pm\sfrac{\nu r}{2\pi}
\big(\ee^{-\frac{\pi\ii mt}{L}}-1\big)\,\ee^{\pm\sfrac{\nu r}{4\pi}
\sum\limits_{n>0}\frac{1}{n}
\tilde\alpha_{2n}\big(\ee^{-\frac{\pi\ii nt}{L}}-1\big)},
\end{equation}
implies that
\begin{equation}
\ee^{\pm\sfrac{\nu r}{4\pi}
\sum\limits_{n>0}\frac{1}{n}\tilde\alpha_{2n}\big(
\ee^{-\frac{\pi\ii nt}{L}}-1\big)}
(\tilde\alpha_{-2m})^{p}|0\rangle\,=\,
\sum\limits_{k=0}^{p}(\pm1)^k\big(\substack{{{p}}\\{k}}\big)
\big(\sfrac{\nu r}{2\pi}
\big(\ee^{-\frac{\pi\ii mt}{L}}-1\big)\big)^k\,
(\tilde\alpha_{-2m})^{p-k}|0\rangle\,.
\end{equation}
Hence
\begin{align}
&\frac{\big\langle0\big|\,(\tilde\alpha_{2m})^{p}\,\ee^{\sfrac{\nu r}{4\pi}
\sum\limits_{n<0}\frac{1}{n}\tilde\alpha_{2n}\big(
\ee^{-\frac{\pi\ii nt}{L}}-1\big)}
\,\ee^{\sfrac{\nu r}{4\pi}
\sum\limits_{n>0}\frac{1}{n}\tilde\alpha_{2n}\big(
\ee^{-\frac{\pi\ii nt}{L}}-1\big)}
(\tilde\alpha_{-2m})^{p}\big|0\big\rangle}{p!(2m)^{p}}\cr
&=\,\sfrac{1}{p!(2m)^{p}}
\,\sum\limits_{k=0}^{p}(-1)^k\big(\substack{{p}\\{k}}\big)^{\hspace{-0.05cm}2}
\big(\sfrac{\nu^2r^2}{4\pi^2}\big|\ee^{-\frac{\pi\ii mt}{L}}-1\big|^2\big)^{k}
(p-k)!(2m)^{p-k}\cr
&=\,\sum\limits_{k=0}^{p}(-1)^k\big(\substack{{p}\\{k}}\big)\sfrac{1}{k!(2m)^{k}}
\big(\sfrac{\nu^2r^2}{\pi^2}
\sin^2({\sfrac{\pi mt}{2L}})\big)^{k}.
\end{align}
The orthonormal basis of the Fock space $\CF_e$ is given by the vectors
\begin{equation}
\prod\limits_{m=1}^\infty\sfrac{(\tilde\alpha_{-2m})^{p_m}}
{\sqrt{p_m!(2m)^{p_m}}}\,\big|0\big\rangle
\end{equation}
with all but a finite number of $\,p_m=0,1,\dots\,$ equal to zero.
Such vectors are eigenvectors of the Hamiltonian $\,H^0\,$ with
eigenvalues $\,\frac{\pi}{2L}\big(\sum\limits_{m=1}^\infty 2mp_m\,
-\frac{1}{8}\big)\,$ and are annihilated by $\,Q^0$.
\,Hence
\begin{align}
&\tr_{\CF_e}\,\ee^{-\beta(H^0-V Q^0)}\,\ee^{\sfrac{\nu 
r}{4\pi}
\sum\limits_{n<0}\frac{1}{n}\tilde\alpha_{2n}\big(
\ee^{-\frac{\pi\ii nt}{L}}-1\big)}
\,\ee^{\sfrac{\nu r}{4\pi}
\sum\limits_{n>0}\frac{1}{n}\tilde\alpha_{2n}\big(
\ee^{-\frac{\pi\ii nt}{L}}-1\big)}
\cr
&=\,\ee^{\frac{\pi\beta}{16L}}\prod\limits_{m=1}^\infty
\Big(\sum\limits_{p=1}^\infty
\ee^{-\frac{\pi\beta}{L}pm}\sum\limits_{k=0}^p(-1)^k
\big(\substack{{p}\\{k}}\big)\sfrac{1}{k!(2m)^{k}}
\big(\sfrac{\nu^2r^2}{\pi^2}\sin^2({\sfrac{\pi mt}{2L}})\big)^{k}\Big)\cr
&=\,\ee^{\frac{\pi\beta}{16L}}\prod\limits_{m=1}^\infty
\Big(\sum\limits_{k=0}^{\infty}
\sfrac{(-1)^k}{(k!)^2(2m)^k}\big(\sfrac{\nu^2r^2}{\pi^2}
\sin^2(\sfrac{\pi mt}{2L})\big)^{k}
\sum\limits_{p=k}^{\infty}
p(p-1)\cdots (p-k+1)\,\ee^{-\frac{\pi\beta}{L}pm}\Big).\qquad
\end{align}
Since by a straightforward calculation
\begin{align}
&\sum\limits_{p=k}^\infty p(p-1)\cdots(p-k+1)z^p\,
=\,\sfrac{k!z^k}{(1-z)^{k+1}}
\end{align}
for $\,|z|<1$, \,we infer that
\begin{align}
&\tr_{\CF_e}\,\ee^{-\beta(H^0-V Q^0)}\,\ee^{\sfrac{\nu 
r}{4\pi}
\sum\limits_{n<0}\frac{1}{n}\tilde\alpha_{2n}\big(
\ee^{-\frac{\pi\ii nt}{L}}-1\big)}
\,\ee^{\sfrac{\nu r}{4\pi}
\sum\limits_{n>0}\frac{1}{n}\tilde\alpha_{2n}\big(
\ee^{-\frac{\pi\ii nt}{L}}-1\big)}
\cr
&=\,\ee^{\frac{\pi\beta}{16L}}\prod\limits_{m=1}^\infty\Big(
\sum\limits_{k=0}^\infty\sfrac{(-1)^k}{k!(2m)^k}\big(\sfrac{\nu^2r^2}{\pi^2}
\sin^2(\sfrac{\pi mt}{2L})\big)^{k}
\sfrac{\ee^{-\frac{\pi\beta}{L}mk}}{(1-
\ee^{-\frac{\pi\beta}{L}m})^{k+1}}\Big)\cr
&=\,\ee^{\frac{\pi\beta}{16L}}\prod\limits_{m=1}^\infty\Big(
\exp\Big[-\sfrac{\nu^2r^2}{2\pi^2m}
\sin^2(\sfrac{\pi mt}{2L})\,
\sfrac{\ee^{-\frac{\pi\beta}{L}m}}{1-
\ee^{-\frac{\pi\beta}{L}m}}\Big]\Big)\prod\limits_{m=1}^\infty
\sfrac{1}{1-\ee^{-\frac{\pi\beta}{L}m}}\,.
\end{align}
The trace over the zero mode space $\,\CH_0\,$ spanned by the orthonormal
vectors $\,|k\rangle\,$ with $\,k\in\mathbb Z\,$ such that 
$\,\tilde\alpha_0|k\rangle=2r^{-1}|k\rangle\,$ is
\begin{align}
\tr_{\CH_0}\,\ee^{-\beta(H^0-V Q^0)}\,\ee^{-\ii\nu\frac{rt}
{4L}\tilde\alpha_0}=\sum\limits_{q\in\mathbb Z}\ee^{-\frac{\pi\beta}
{Lr^2}k^2+\beta V k-\ii\nu\frac{t}{2L}k}\,.
\end{align}
Multiplying those two traces and the factor of the last line
of (\ref{A5}) and dividing them by the partition function, we obtain the
expression
\begin{align}
\omega^{L}_{\beta,V}\Big(\Big(\ee^{-\ii\nu\int\limits_0^tJ^\ell
(0,s)\,\dd s}\Big)_{\hspace{-0.05cm}\rm reg}\Big)\,=&\,\,
\big(4\sin^2(\sfrac{\pi t}{2L})\big)^{\hspace{-0.05cm}
-\frac{\nu^2 r^2}{16\pi^2}}\ 
\ee^{-\sfrac{\nu^2r^2}{8\pi^2}\sum\limits_{n<\Lambda L}\frac{1}{n}}
\ \sfrac{\sum\limits_{k\in\mathbb Z}\ee^{-\frac{\pi\beta}
{Lr^2}k^2+\beta V k-\ii\nu\frac{t}{2L}k}}
{\sum\limits_{k\in\mathbb Z}\ee^{-\frac{\pi\beta}
{Lr^2}k^2+\beta V k}}
\cr
&\times\,\,\prod\limits_{m=1}^\infty\Big(
\exp\Big[-\sfrac{\nu^2r^2}{2\pi^2m}
\sin^2(\sfrac{\pi mt}{2L})\,
\sfrac{\ee^{-\frac{\pi\beta}{L}m}}{1-
\ee^{-\frac{\pi\beta}{L}m}}\Big]\Big).
\label{ssum}
\end{align}
By Eq.\,16.30.1 of \cite{AS}, 
\begin{equation}
4\sum\limits_{m=1}^\infty\sfrac{1}{m}\sin^2(\sfrac{\pi mt}{2L})\,
\sfrac{\ee^{-\frac{\pi\beta}{L}m}}{1-\ee^{-\frac{\pi\beta}{L}m}}
=\ln\sfrac{\pi\,\theta_1(\frac{\ii\beta}{2L};\frac{t}{2L})}
{\sin(\frac{\pi t}{2L})\,\partial_z\theta_1(\frac{\ii\beta}{2L};0)}\,,
\end{equation}
where we use the definition 
\begin{equation}
\theta_1(\tau;z)=\sum\limits_{n\in\mathbb Z}
\ee^{\pi\ii\tau(n+\frac{1}{2})^2+2\pi i(n+\frac{1}{2})(z+\frac{1}{2})}
\end{equation}
for the first of the Jacobi theta-functions. Hence (\ref{ssum}) may be
rewritten in the form
\begin{align}
\omega^{L}_{\beta,V}\Big(\Big(\ee^{-\ii\nu\int\limits_0^tJ^\ell
(0,s)\,\dd s}\Big)_{\rm reg}\Big)\,=&\,\,\exp\Big[-\sfrac{\nu^2r^2}{8\pi^2}
\big(\sum\limits_{n<\Lambda L}\sfrac{1}{n}\,+\,\ln\sfrac{2\pi\,
\theta_1(\frac{\ii\beta}{2L};\frac{t}{2L})}
{\partial_z\theta_1(\frac{i\beta}{2L};0)}\Big)\Big]\cr
&\times\,\,
\sfrac{\sum\limits_{k\in\mathbb Z}\ee^{-\frac{\pi\beta}
{Lr^2}k^2+\beta V k-\ii\nu\frac{t}{2L}k}}
{\sum\limits_{k\in\mathbb Z}\ee^{-\frac{\pi\beta}
{Lr^2}k^2+\beta V k}}
\label{ssum1}
\end{align}
The substitution of the above relation to the (ultraviolet regularized 
version of) (\ref{FCSsf}) results in the identity (\ref{FCSex}).

\nappendix{B}
\label{app:B}
\vskip 0.5cm

\noindent Let us consider a quadratic selfadjoint operator
\begin{equation}
\CA\,=\,\sum\limits_{i,i'=1}^N\sum\limits_{n,n'=1}^\infty
\big(\sfrac{1}{2}A_{ni,n'i'}\,a^\dagger_{ni}a^\dagger_{n'i'}+B_{ni,n'i'}\,a^\dagger_{ni}
a_{n'i'}+\sfrac{1}{2}\overline{A_{ni,n'i'}}\,a_{ni}a_{n'i'}\big)
\label{CAA}
\end{equation}
acting in the Fock space $\CF$ of the vacuum representation of the canonical 
commutation relations (CCR)
\begin{equation}
[a_{ni},a_{n'i'}]=0=[a^\dagger_{ni},a^\dagger_{n'i'}]\,,\qquad[a_{ni},a^\dagger_{n'i'}]=
\delta_{ii'}\delta_{n,-n'}
\end{equation}
built on the normalized vacuum state $|0\rangle$ annihilated by operators 
$a_{ni}$. We assume that $A_{ni,n'i'}=A_{n'i',ni}$ and $B_{ni,n'i'}=
\overline{B_{n'i',ni}}$. \,Note the commutation relations
\begin{equation}
\bigg[\CA\,,\bigg(\begin{matrix}a^\dagger_{ni}\cr a_{ni}\end{matrix}\bigg)\bigg]
=\sum_{i'}\sum_{n'>0}\bigg(\begin{matrix}\overline{B_{in,i'n'}}&
\overline{A_{in,i'n'}}\cr
-A_{in,i'n'}&-B_{in,i'n'}\end{matrix}\bigg)\bigg(\begin{matrix}a^\dagger_{n'i'}\cr 
a_{n'i'}\end{matrix}\bigg).
\end{equation}
The exponentiation of the above relation gives
\begin{equation}
\ee^{\hspace{0.03cm}\ii\CA}\bigg(\begin{matrix}a^\dagger_{ni}\cr a_{ni}\end{matrix}
\bigg)\ee^{-\ii\CA}
=\sum_{i'}\sum_{n'>0}\bigg(\begin{matrix}\overline{P_{in,i'n'}}
&\overline{Q_{in,i'n'}}\cr
Q_{in,i'n'}&P_{in,i'n'}\end{matrix}\bigg)\bigg(\begin{matrix}a^\dagger_{n'i'}\cr 
a_{n'i'}\end{matrix}\bigg), 
\end{equation}
where, in the language of infinite matrices,
\begin{equation}
\bigg(\begin{matrix}\overline{P}
&\overline{Q}\cr
Q&P\end{matrix}\bigg)\,=\,\exp\bigg(\begin{matrix}\overline{B}&\overline{A}\cr
-A&-B\end{matrix}\bigg).
\end{equation}
One of the results of the theory of the Bogoliubov transformations associated
to the quadratic Hamiltonians $\,\CA\,$ is the formula
\begin{equation}
\big\langle0\big|\ee^{i\CA}\big|0\rangle\,=\,\det\big(\ee^{iB}P\big)^{-1/2}
\label{derger}
\end{equation}
holding under conditions that guarantee the finiteness of the left and
right hand sides which will be satisfied in the cases of interest for us,
\,see e.g. \cite{DerGer}. 
\vskip 0.1cm

We shall need a generalization of that formula to the expectations
of operator $\,\ee^{i\CA}\,$ in certain mixed states. Consider such a state 
$\,\omega_\rho$ corresponding to the density matrix 
\begin{equation}
\rho\,=\,\sfrac{1}{Z}
\,\ee^{-\sum\limits_{i=1}^N\sum\limits_{n=1}^\infty \epsilon_{ni}\,a^\dagger_{ni}a_{ni}}\,,
\end{equation} 
with $\,\epsilon_{in}>0\,$ diverging sufficiently fast when $\,n\to\infty\,$ 
and the normalization factor
\begin{equation}
Z\,=\,\prod\limits_{i=1}^N\prod\limits_{n=1}^\infty
(1-\ee^{-\epsilon_{ni}})^{-1}.
\end{equation}
We would like to calculate the expectation value 
$\,\omega_\rho\big(\ee^{i\CA}\big)$. \,To this end, we may use the Araki-Woods 
representation \cite{AW} of the CCR acting in the double Fock space 
$\,\CF\otimes\CF\,$ and given  by the formula
\begin{equation}
\bigg(\begin{matrix}a_{ni}^\dagger\cr a_{ni}\end{matrix}\bigg)\,\longmapsto\,
\bigg(\begin{matrix}\hat a_{ni}^\dagger\cr\hat a_{ni}\end{matrix}\bigg)
=\bigg(\begin{matrix}\sqrt{1+b_{ni}}\,
(a^1_{ni})^\dagger+\sqrt{b_{ni}}\,a^2_{ni}\cr 
\sqrt{1+b_{ni}}\,
a^1_{ni}+\sqrt{b_{ni}}\,(a^2_{ni})^\dagger
\end{matrix}\bigg),
\end{equation} 
where 
\begin{equation}
b_{ni}=\frac{1}{\ee^{\epsilon_{ni}}-1}\,,\qquad a^1_{ni}=a_{ni}\otimes I\,,
\qquad a^2_{ni}=I\otimes a_{ni}\,.
\end{equation}
The Araki-Woods representation has the property that the matrix element
on the vacuum $|0,0\rangle=|0\rangle\otimes|0\rangle$ of a CCR 
observables taken in that representation reproduces their $\omega_\rho$
expectations. In particular
\begin{equation}
\omega_\rho\big(\ee^{\hspace{0.03cm}\ii\CA}\big)\,=\,
\tr_{_\CF}\Big(\rho\,\ee^{\hspace{0.03cm}\ii\CA}\Big)
\,=\,\big\langle0,0\big|\ee^{i\hat\CA}\big|0,0\big\rangle\,,
\end{equation}
where
\begin{align}
&\hat\CA\,=\,\sum\limits_{i=1}^N\sum\limits_{n=1}^\infty
\big(\sfrac{1}{2}A_{ni,n'i'}\,\hat a^\dagger_{ni}\hat a^\dagger_{n'i'}
+B_{ni,n'i'}\hat a^\dagger_{ni}\,
\hat a_{ni}+\sfrac{1}{2}\overline{A_{ni,n'i'}}\,\hat a_{ni}\hat a_{n'i'}\big)\cr
&=\,\sum\limits_{\sigma,\sigma'=1}^2\sum\limits_{i=1}^N\sum\limits_{n=1}^\infty
\big(\sfrac{1}{2}\hat A^{\sigma,\sigma'}_{ni,n'i'}(\hat a^\sigma_{ni})^\dagger
({\hat a}^{\sigma'}_{n'i'})^\dagger
+\hat B^{\sigma,\sigma'}_{ni,n'i'}({\hat a}^\sigma_{ni})^\dagger
{\hat a}^{\sigma'}_{n'i'}+\sfrac{1}{2}\overline{\hat A^{\sigma,\sigma'}_{ni,n'i'}}
{\hat a}^\sigma_{ni}{\hat a}^{\sigma'}_{n'i'}\big)\cr
&\hspace{0.3cm}+\sum\limits_{i=1}^N\sum\limits_{n=1}^NB_{ni,ni}b_{ni}\,,
\end{align}
with
\begin{align}
&\bigg(\begin{matrix}\hat A^{1,1}_{ni,n'i'}&\hat A^{1,2}_{ni,n'i'}\cr
\hat A^{2,1}_{ni,n'i'}&\hat A^{2,2}_{ni,n'i'}\end{matrix}\bigg)\,=\,
\bigg(\begin{matrix}A_{ni,n'i'}\sqrt{(1+b_{ni})
(1+b_{n'i'})}&B_{ni,n'i'}\sqrt{(1+b_{ni})
b_{n'i'}}\cr\overline{B_{ni,n'i'}}
\sqrt{b_{ni}(1+b_{n'i'})}&
\overline{A_{ni,n'i'}}\sqrt{b_{ni}b_{n'i'}}
\end{matrix}\bigg),\cr
&\bigg(\begin{matrix}\hat B^{1,1}_{ni,n'i'}&\hat B^{1,2}_{ni,n'i'}\cr
\hat B^{2,1}_{ni,n'i'}&\hat B^{2,2}_{ni,n'i'}\end{matrix}\bigg)\,=\,
\bigg(\begin{matrix}B_{ni,n'i'}\sqrt{(1+b_{ni})(1+b_{n'i'})}&
A_{ni,n'1'}\sqrt{(1+b_{ni})b_{n'i'}}
\cr\overline{A_{ni,n'1'}}\sqrt{b_{ni}(1+b_{n'i'})}
&\overline{B_{ni,n'i'}}\sqrt{b_{ni}b_{n'i'}}\end{matrix}\bigg).
\end{align}
Denoting 
\begin{equation}
\exp\bigg[\ii\bigg(\begin{matrix}\overline{\hat B}&\overline{\hat A}\cr
-\hat A&-\hat B\end{matrix}\bigg)\bigg]\,\equiv\,
\bigg(\begin{matrix}\overline{\hat P}&\overline{\hat Q}\cr
\hat Q&\hat P\end{matrix}\bigg),
\end{equation}
we infer from the previous result (\ref{derger}) that
\begin{equation}
\omega_\rho\big(\ee^{\hspace{0.03cm}\ii\CA}\big)\,=\,\ee^{i\sum\limits_{i,n}B_{in,in}
b_{ni}}\det\big(\ee^{\ii\hat B}\hat P\big)^{-1/2}\,.
\label{fpr}
\end{equation}
\vskip 0.1cm

The above calculation can be applied to the operator
$\CA$ corresponding to the excited modes part of 
$\,\sum\limits_i\lambda_i\big(H_i^0+\Delta H_i(t)\big)\,$ for 
any time $\,t$, \,but we shall limit ourselves to the simpler instance 
when $\,t=2L$. \,In that case
\begin{equation}
\CA\,=\,\sfrac{\pi}{L}\sum\limits_{i,i'}(O\lambda O)_{ii'}
\sum\limits_{n=1}^\infty\tilde \alpha_{(-2n)i}\tilde\alpha_{(2n)i'}\,,
\end{equation}
see (\ref{itom}), \,with the identification
\begin{equation}
\sfrac{\tilde\alpha_{(2n)i}}{(2n)^{1/2}}\equiv a_{ni}\,,\qquad
\sfrac{\tilde\alpha_{(-2n)i}}{(2n)^{1/2}}\equiv a^\dagger_{ni}\,,
\end{equation}
so that
\begin{equation}
A_{ni,n'i'}=0\,,\qquad B_{ni,n'i'}=\sfrac{2\pi}{L}(O\lambda O)_{ii'}n
\,\delta_{nn'}\,.
\end{equation}
For the density matrix, we shall take
\begin{equation}
\rho\,=\,\sfrac{1}{Z_{\beta}}\,
\ee^{-\frac{\pi}{2L}\sum\limits_i\beta_i\sum\limits_{n>0}\tilde\alpha_{(-2n)i}\tilde
\alpha_{(2n)i}}
\end{equation}
so that the state $\,\omega_\rho\,$ coincides with $\,\omega_0^L\,$ on the 
algebra generated by the excited modes. The normalization
\begin{equation}
Z_\beta\,=\,\prod\limits_{i,n>0}(1-\ee^{-\frac{\pi n\beta_i}{L}})^{-1}\,.
\label{Zbet}
\end{equation}
In this case $b_{ni}=
(\ee^{\frac{\pi}{L}n\beta_i}-1)^{-1}$. \,It follows that
\begin{align}
&\bigg(\begin{matrix}\hat A^{1,1}_{ni,n'i'}&\hat A^{1,2}_{ni,n'i'}\cr
\hat A^{2,1}_{ni,n'i'}&\hat A^{2,2}_{ni,n'i'}\end{matrix}\bigg)=
\sfrac{\pi}{L}n\,\delta_{nn'}\bigg(\begin{matrix}0&(O\lambda O)_{ii'}
\sqrt{(1+b_{ni})b_{ni'}}\cr
(O\lambda O)_{ii'}\sqrt{b_{ni}(1+b_{ni'})}&
0\end{matrix}\bigg),\cr
&\bigg(\begin{matrix}\hat B^{1,1}_{ni,n'i'}&\hat B^{1,2}_{ni,n'i'}\cr
\hat B^{2,1}_{ni,n'i'}&\hat B^{2,2}_{ni,n'i'}\end{matrix}\bigg)=
\sfrac{\pi}{L}n\,\delta_{nn'}\bigg(\begin{matrix}(O\lambda O)_{ii'}
\sqrt{(1+b_{ni})(1+b_{ni'})}&0\cr
0&(O\lambda O)_{ii'}\sqrt{b_{ni}b_{ni'}}
\end{matrix}\bigg).
\end{align}
A little of straightforward algebra shows that
\begin{align}
&\bigg(\begin{matrix}\hat P^{1,1}_{ni,n'i'}&\hat P^{1,2}_{ni,n'i'}\cr
\hat P^{2,1}_{ni,n'i'}&\hat P^{2,2}_{ni,n'i'}\end{matrix}\bigg)=\delta_{nn'}
\delta_{ii'}\bigg(\begin{matrix}1&0\cr0&1\end{matrix}\bigg)\cr
&+\delta_{nn'}\bigg(\begin{matrix}
\big(O\,\ee^{-\frac{\pi\ii}{L}n\lambda}O-I\big)_{ii'}
\sqrt{(1+b_{ni})(1+b_{ni'})}&0\cr
0&-\big(O\,\ee^{\frac{\pi\ii}{L}n\lambda}O-I\big)_{ii'}\sqrt{b_{ni}b_{ni'}}
\end{matrix}\bigg),\quad\cr
&\bigg(\begin{matrix}\hat Q^{1,1}_{ni,n'i'}&\hat Q^{1,2}_{ni,n'i'}\cr
\hat Q^{2,1}_{ni,n'i'}&\hat Q^{2,2}_{ni,n'i'}\end{matrix}\bigg)\cr
&=\delta_{nn'}\bigg(\begin{matrix}0&
\big(O\,\ee^{-\frac{\pi\ii}{L}n\lambda}O-I\big)_{ii'}
\sqrt{(1+b_{ni})b_{ni'}}\cr
-\big(O\,\ee^{\frac{\pi\ii}{L}n\lambda}O-I\big)_{ii'}\sqrt{b_{ni}(1+b_{ni'})}&0
\end{matrix}\bigg),\quad
\end{align}
Hence, denoting by $b_n$ the diagonal $N\times N$ matrix with entries
$b_{ni}$, we obtain
\begin{align}
\det\big(\hat P_{n,n}\big)^{-{1/2}}\,&=\,\bigg(\det\Big(I+
\big(O\,\ee^{-\frac{\pi\ii}{L}n\lambda}O-I\big)(I+b_n)\Big)
\det\Big(I-\big(O\,\ee^{\frac{\pi\ii}{L}n\lambda}O-I\big)b_n\Big)\bigg)^{-1/2}\cr
&=\,\bigg(\det\Big(O\,\ee^{-\frac{\pi\ii}{L}n\lambda}O\Big)
\det\Big(I+\big(I-O\,\ee^{\frac{\pi\ii}{L}n\lambda}O\big)b_n\Big)^2\bigg)^{-1/2}\cr
&=\,\ee^{\frac{\pi\ii}{2L}n\sum\limits_i\lambda_i}\,
\det\Big(I+\big(I-O\,\ee^{\frac{\pi\ii}{L}n\lambda}O\big)b_n\Big)^{-1}.
\end{align}
On the other hand,
\begin{equation}
\ee^{\ii\sum\limits_iB_{ni,ni} b_{ni}}\,\det\big(\ee^{i\hat
B_{n,n}}\big)^{-1/2}=
\ee^{\frac{\pi\ii}{L}n\sum\limits_i(O\lambda O)_{ii}
b_{ni}-\frac{\pi\ii}{2L}n\sum\limits_i(O\lambda O)_{ii}(1+2b_{ni})}=
\ee^{-\frac{\pi\ii}{2L}n\sum\limits_i\lambda_i}.
\end{equation}
We infer then from (\ref{fpr}) that
\begin{equation}
\tr_{_\CF}\Big(\rho\,\ee^{i\CA}\Big)\,=\,\prod\limits_{n=1}^\infty
\det\Big(I+\big(I-O\,\ee^{\frac{\pi\ii}{L}n\lambda}O\big)
\big(\ee^{\frac{\pi}{L}n\beta}-I\big)^{-1}\Big)^{-1}.
\label{todivide}
\end{equation}
What we have to compute, however, is the contribution of the excited modes 
to (\ref{et=L}) which is equal to   
\begin{equation}
\sfrac{1}{Z_\beta}\,
\tr\Big(\ee^{-\frac{\pi}{2L}\sum\limits_i(\beta_i+\ii\lambda_i)
\sum\limits_{n>0}\alpha_{(-2n)i}\tilde\alpha_{(2n)i}}\,\ee^{i\CA}\Big). 
\end{equation}
This is clearly obtained by multiplying (\ref{todivide}) by
$\,Z_\beta$, \,shifting $\,\beta\,$ to $\,\beta+i\lambda\,$
in the result, and re-dividing it by $\,Z_\beta$, \,which gives 
\begin{align}
&\prod\limits_{n=1}^\infty\bigg(
\sfrac{\det\big(I-\ee^{-\frac{\pi}{L}n\beta}\big)}
{\det\big(I-\ee^{-\frac{\pi}{L}n(\beta+i\lambda)}\big)}
\det\Big(I+\big(I-O\,\ee^{\frac{\pi\ii}{L}n\lambda}O\big)\big(
\ee^{\frac{\pi}{L}n
(\beta+i\lambda)}-I\big)^{-1}\Big)^{-1}\bigg)\cr
&=\,\prod\limits_{n=1}^\infty\det\Big(I+\big(I-
\ee^{-\frac{\pi\ii}{L}n\lambda}
O\,\ee^{\frac{\pi\ii}{L}n\lambda}O\big)
(\ee^{\frac{\pi}{L}n\beta}-I)^{-1}\Big)^{-1},
\end{align}
i.e. the result (\ref{exccont}).
Note that the last expression is invariant under the change of $\,\lambda\,$
to $\,-\lambda+i\beta$. \,Indeed,
\begin{align}
&\det\Big(I+
\big(I-\ee^{-\frac{\pi\ii}{L}n(-\lambda+\ii\beta)}\,
O\,\ee^{\frac{\pi\ii}{L}n(-\lambda+\ii\beta)}O\big)
\big(\ee^{\frac{\pi n}{L}\beta}-I\big)^{-1}\Big)^{-1}\cr
&=\sfrac{\det\Big(\ee^{\frac{\pi}{L}n\beta}\Big)^{-1}\det\Big(I\,-\,
\ee^{\frac{\pi\ii}{L}n\lambda}\,O\,\ee^{\frac{\pi
i}{L}n(-\lambda+\ii\beta)}O\Big)^{-1}}
{\det\Big(\ee^{\frac{\pi}{L}n\beta}-1\Big)^{-1}}=
\sfrac{\det\Big(\ee^{\frac{\pi}{L}n\beta}\Big)^{-1}
\det\Big(I\,-\,
O\,\ee^{\frac{\pi\ii}{L}n\lambda}O\,\ee^{\frac{\pi i}{L}n\,
(-\lambda+\ii\beta)}\Big)^{-1}}
{\det\Big(\ee^{\frac{\pi}{L}n\beta}-1\Big)^{-1}}\cr
&=\sfrac{
\det\Big(\ee^{\frac{\pi}{L}n\beta}\,-\,
O\,\ee^{\frac{\pi\ii}{L}n\lambda}O\,\ee^{-\frac{\pi i}{L}n\lambda}\Big)^{-1}}
{\det\Big(\ee^{\frac{\pi}{L}n\beta}-1\Big)^{-1}}
=\sfrac{
\det\Big(\ee^{\frac{\pi}{L}n\beta}\,-\,
\ee^{-\frac{\pi i}{L}n\lambda}\,O\,\ee^{\frac{\pi\ii}{L}n\lambda}O\Big)^{-1}}
{\det\Big(\ee^{\frac{\pi}{L}n\beta}-1\Big)^{-1}}
\cr
&=\det\Big(I+\big(I-\ee^{-\frac{\pi\ii}{L}n\lambda}
O\,\ee^{\frac{\pi\ii}{L}n\lambda}O\big)(\ee^{\frac{\pi}{L}n
\beta}-I)^{-1}\Big)^{-1}.
\label{FRexc}
\end{align}

\nappendix{C}
\label{app:C}
\vskip 0.5cm

\noindent We compute here the scaling limit
\begin{equation}\label{scaling}
\lim_{\theta \rightarrow \infty}
\theta^2\,\phi(\theta^{-1}\bm\nu)\big|_{\bm\beta,
\,\theta^{-1}\bm V}\,, 
\end{equation}
where the rate function $\,\phi(\theta^{-1}\bm\nu)\,$ is given by
relation (\ref{LDLL}). The logarithms of determinants on the right hand 
side of (\ref{LDLL}) will be computed by expanding:
\begin{equation}
 \ln\big(\det(1+A_\pm)\big) = \sum_{n=1}^{\infty}(-1)^{n-1}\sfrac{1}{n} 
\tr(A_\pm^n)\,,
\end{equation}
for $N\times N$ matrices
\qq
A_\pm\,=\,f^\pm(\epsilon)\big|_{\bm\beta,\,\theta^{-1}\bm
V}\big(\ee^{\mp\ii\theta^{-1}\nu}
\mathbb S^\dagger\ee^{\pm\ii\theta^{-1}\nu}\mathbb S-I\big)\,=\,
f^\pm(\epsilon)\big|_{\bm\beta,\,\theta^{-1}\bm
V}\big(\pm\ii\theta^{-1}B+\theta^{-2}D
+O(\theta^{-3})\big)
\qqq
with
\qq
B=\mathbb S^\dagger\nu\,\mathbb S-\nu\,,\qquad D=\nu\,\mathbb S^\dagger
\nu\,\mathbb S-\sfrac{1}{2}\mathbb S^\dagger\nu^2\mathbb S-\sfrac{1}{2}\nu^2\,.
\label{BD}
\qqq
Only the first two terms of that expansion will contribute to the scaling
limit (\ref{scaling}).
The integration over $\,\epsilon\,$ is reduced to the terms
\begin{equation}
\int_0^\infty (1+\ee^{\beta_i(\epsilon\mp\theta^{-1}V_i)})^{-1}\,\dd\epsilon  
= \sfrac{1}{\beta_i}\ln \big(1+\ee^{\pm\beta_i\theta^{-1}V_i} \big) 
= \sfrac{1}{\beta_i}\big(\ln 2 \pm \sfrac{1}{2}\beta_i\theta^{-1} V_i  
+ O(\theta^{-2})\big)
\end{equation}
so that
\begin{equation}
\int_0^\infty\tr(A_\pm)\,\dd\epsilon = \sum_i\Big(\pm\sfrac{\ii \ln 2}
{\theta \beta_i} B_{ii} + \sfrac{\ln 2}{\theta^2 \beta_i} D_{ii} 
+ \sfrac{1}{2} \sfrac{\ii V_i}{\theta^2} B_{ii}\Big)
+ O(\theta^{-3})\,.
\end{equation}
Thus
\qq
\int_0^\infty\big(\tr(A_+)+\tr(A_-)\big)\,\dd\epsilon =\sum\limits_i\Big( 
\sfrac{2\ln 2}{\theta^2 \beta_i} D_{ii} + \sfrac{\ii V_i}{\theta^2} B_{ii}\Big) 
+ O(\theta^{-3})\,.
\qqq
Using formulae (\ref{BD}), we finally get:
\begin{align}
&\lim\limits_{\theta\to\infty}\,\theta^2
\int_0^\infty\big(\tr(A_+)+\tr(A_-)\big)\,\dd\epsilon\cr
&\hspace{3cm}=\,
2\ln{2}\sum\limits_i\beta_i^{-1}\Big(\nu_i\sum\limits_{i'}
|\mathbb S_{i'i}|^2\nu_{i'}
-\sfrac{1}{2}\sum\limits_{i'}|S_{i'i}|^2\nu_{i'}^2-\sfrac{1}{2}\nu_i^2\Big)\cr
&\hspace{3.5cm}+\ii\sum\limits_iV_i\Big(\nu_i-\sum\limits_{i'}
|\mathbb S_{i'i}|^2\nu_{i'}\Big).
\label{cnt1}
\end{align}
The other non-vanishing terms in the scaling limit (\ref{scaling}) come from
\qq
\int\limits_0^\infty\tr(A_\pm^2)\,\dd\epsilon\,=\,-\,\theta^{-2}\sum\limits_{i.i'}
B_{ii'}B_{i'i}g(\beta_i,\beta_{i'})\,,
\qqq
where
\qq
g(\beta_{i},\beta_{i'})\,=\,\int\limits_0^\infty\sfrac{1}{(1+\ee^{
\beta_i\epsilon})
(1+\ee^{\beta_{i'}\epsilon})}\,\dd\epsilon\,.
\label{fbb}
\qqq
The distinction between the contributions for different signs
disappears at this order so that
\qq
\lim\limits_{\theta\to\infty}\,\theta^2
\int_0^\infty\big(-\sfrac{1}{2}\tr(A_+^2)-\sfrac{1}{2}\tr(A_-^2)\big)\,
\dd\epsilon\,=\,\sum\limits_{i.i'}
B_{ii'}B_{i'i}g(\beta_i,\beta_{i'})\,.
\qqq
For $\,i=i'$,
\qq
g(\beta_i,\beta_i)\,=\,\sfrac{2\ln{2}-1}{2\beta_i}
\label{fbbe}
\qqq
and
\qq
B_{ii}^2\,=\,\sum\limits_{j,j'}|\mathbb S_{ji}|^2\nu_j|\mathbb
S_{j'i}|^2\nu_{j'}
-2\nu_i\sum\limits_j|\mathbb S_{ji}|^2\nu_j+\nu_i^2\,.
\qqq
On the other hand, for $\,i\not=i'$,
\qq
B_{ii'}B_{i'i}\,=\,\sum\limits_{j,j'}\mathbb S^\dagger_{ij}\nu_j\mathbb S_{ji'}
\mathbb S^\dagger_{i'j'}\nu_{j'}\mathbb S_{j'i}\,.
\qqq
Hence
\begin{align}
&\lim\limits_{\theta\to\infty}\,\theta^2
\int_0^\infty\big(-\sfrac{1}{2}\tr(A_+^2)-\sfrac{1}{2}\tr(A_-^2)\big)\,
\dd\epsilon\cr
&\hspace{3cm}=\,\sum\limits_i\sfrac{2\ln{2}-1}{2\beta_i}\Big(
\sum\limits_{j,j'}|\mathbb S_{ji}|^2\nu_j|\mathbb S_{j'i}|^2\nu_{j'}
-2\nu_i\sum\limits_j|\mathbb S_{ji}|^2\nu_j+\nu_i^2\Big)\cr
\quad
&\hspace{3.4cm}+\sum\limits_{i\not=i'}\sum\limits_{j,j'}
\mathbb S^\dagger_{ij}\nu_j\mathbb S_{ji'}
\mathbb S^\dagger_{i'j'}\nu_{j'}\mathbb S_{j'i}\,g(\beta_i,\beta_{i'})\,.
\label{cnt2}
\end{align}
The addition of contributions (\ref{cnt1}) and (\ref{cnt2}) results
in the identity (\ref{LLRF}).


\begin{thebibliography}{bib}

\bibitem{Affleck}
I. Affleck: \textit{Conformal Field Theory Approach to the Kondo Effect},
Acta Phys. Polon. B {\bf 26} (1995), 1869-1932

\bibitem{AGMT}
D. Andrieux, P. Gaspard, T. Monnai, S. Tasaki: \textit{The fluctuation theorem 
for currents in open quantum systems}, New J. Phys. {\bf 11} (2009), 043014

\bibitem{AW}
H. Araki, E. J. Woods: \textit{Representations of the canonical 
commutation relations describing a nonrelativistic infinite free Bose gas}, 
J. Math. Phys. {\bf 4} (1963), 637-662

\bibitem{AJPP}
W. Aschbacher, V. Jak$\check{\rm s}$ic, Y. Pautrat, C.-A. Pillet: 
\textit{Transport properties of quasi-free fermions}, J. Math. Phys.
{\bf 48} (2007), 032101 

\bibitem{AS}
\textit{Handbook of Mathematical Functions: 
with Formulas, Graphs, and Mathematical Tables}, editors
M. Abramowitz and I. A. Stegun, Dover 1972

\bibitem{AleksSchom}
A. Alekseev, V. Schomerus, \textit{D-branes in the WZW model}, 
Phys. Rev. D {\bf 60} (1999), 061901

\bibitem{BdeBDO}
C. Bachas, J. de Boer, R. Dijkgraaf, H. Ooguri: \textit{Permeable conformal 
walls and holography},  JHEP 0206:027 (2002) 

\bibitem{BDLS}
M.J. Bhaseen, B. Doyon, A. Lucas, K. Schalm: \textit{Far from equilibrium 
energy flow in quantum critical systems}, arXiv:1311.3655 [hep-th]

\bibitem{BMS}
B.Bellazzini, M. Mintchev, P. Sorba, \textit{Bosonization and Scale Invariance
on Quantum Wires}, J. Phys. A \textbf{40} (2007), 2485-2508 

\bibitem{BD1}
D. Bernard, B. Doyon:
{\it Energy flow in non-equilibrium conformal field theory},
J. Phys. A: Math. Theor. 45 (2012), 362001 

\bibitem{BD2}
D. Bernard, B. Doyon:
{\it Non-equilibrium steady-states in conformal field theory}, Ann. H. 
Poincar\'e, online first, arXiv:1302.3125 [math-ph], 

\bibitem{BD3}
D. Bernard, B. Doyon: \textit{Time-reversal symmetry and fluctuation 
relations in non-equilibrium quantum steady states},
J. Phys. A: Math. Theor. {\bf 46} (2013), 372001 

\bibitem{BDV}
D. Bernard, B. Doyon, J. Viti: \textit{Non-equilibrium conformal 
field theories with impurities}, arXiv:1411.0470 [math-ph]

\bibitem{BB}
Ya. M. Blanter, M. B\"{u}ttiker: \textit{Shot noise in mesoscopic conductors},
Phys. Rep. {\bf 336} (2000), 1-166

\bibitem{BJP}
L. Bruneau, V. Jak$\check{\rm s}$ic, C.-A. Pillet: 
\textit{Landauer-B\"{u}ttiker formula and Schr\"{o}dinger conjecture},
Commun. Math. Phys. {\bf 319} (2013), 501-513 

\bibitem{Datta}
S. Datta: \textit{Electronic Transport in Mesoscopic Systems}, 
Cambridge University Press 1995

\bibitem{DerGer}
J. Derezi\'{n}ski, Ch. G\'{e}rard, \textit{Mathematics of Quantization 
and Quantum Fields}, Cambridge University Press 2013

\bibitem{D}
B. Doyon: \textit{Lower bounds for ballistic current and noise in 
non-equilibrium quantum steady states}, arXiv:1410.0292 [cond-mat.str-el]

\bibitem{DHB}
B. Doyon, M. Hoogeveen, D. Bernard:
{\it Energy flow and fluctuations in non-equilibrium conformal field 
theory on star graphs},
J. Stat. Mech. (2014) P03002

\bibitem{FLS}
P. Fendley, A. W. W. Ludwig, H. Saleur: \textit{Exact nonequilibrium transport 
through point contacts in quantum wires and fractional quantum Hall devices},
Physical Review B {\bf 52} (1995), 8934-8950

\bibitem{Gaberdiel}
M. Gaberdiel, \textit{Boundary conformal field theory and D-branes} Lectures
notes (2003) http://www.phys.ethz.ch/$\sim$mrg/lectures2.pdf 

\bibitem{G}
T. Giamarchi: \textit{Quantum Physics in One Dimension},
Oxford University Press 2003

\bibitem{GT}
K. Gaw\c{e}dzki, C. Tauber: \textit{unpublished}

\bibitem{GGM1}
D. B. Gutman, Y. Gefen, A. D. Mirlin: \textit{Bosonization out of equilibrium},
Europhys. Lett. {\bf 90} (2010), 37003 

\bibitem{GGM2}
D. B. Gutman, Y. Gefen, A. D. Mirlin: \textit{Full counting statistics 
of Luttinger liquid conductor}, Phys. Rev. Lett. {\bf 105} (2010), 256802

\bibitem{Ishii}
H. Ishii \textit{et al.}: \textit{Direct observation of 
Tomonaga-Luttinger-liquid state in carbon nanotubes at low temperatures}, 
Nature {\bf 426} (2003), 540-544

\bibitem{Kane}
C. L. Kane: \textit{Lectures on Bosonization}, Boulder Summer School lectures 
on electrons in one dimension, \#1+2, www.physics.upenn.edu/$\sim$kane/

\bibitem{Klich}
I. Klich: \textit{Full Counting Statistics: An elementary derivation 
of Levitov's formula}, in: \textit{Quantum Noise in Mesoscopic Physics},
ed. Yu. V. Nazarov, Springer 2003, pp. 397-402 

\bibitem{LL}
L. S. Levitov, G. B. Lesovik: \textit{Charge distribution in quantum shot 
noise}, JETP Lett. {\bf 58} (1993), 230-235

\bibitem{LLL}
L. S. Levitov, H. W. Lee, and G. B. Lesovik: \textit{Electron counting 
statistics and coherent states of electric current},  
J. Math. Phys. {\bf 37} (1996), 4845-4866 

\bibitem{Mintchev}
M. Mintchev: \textit{Non-equilibrium steady states of
quantum systems on star graphs}, J. Phys. A: Math. Theor. {\bf44} (2011), 
415201 

\bibitem{MSLL}
M. Mintchev, P. Sorba: \textit{Luttinger Liquid in Non-equilibrium Steady State}
J. Phys. A: Math. Theor. \textbf{46} (2013), 095006  

\bibitem{NDBM}
S. Ngo Dinh, D. A. Bagrets, A. D. Mirlin: \textit{Nonequilibrium functional 
bosonization of quantum wire networks}, Ann. Phys. {\bf 327} (2012), 2794-2852

\bibitem{OCA0}
M. Oshikawa, C. Chamon, I. Affleck: \textit{Junctions of three quantum 
wires and the dissipative Hofstadter model}, Phys. Rev. Lett. {\bf 91}
(2003), 206403

\bibitem{OCA}
M. Oshikawa, C. Chamon, I. Affleck: 
{\it Junctions of three quantum wires}, J. Stat. Mech. 0602:P02008 (2006)

\bibitem{NFLL}
C. Nayak, M. F. A. Fisher, A. W. W. Ludwig, H. H. Lin:
\textit{Resonant multilead point-contact tunneling}, Phys. Rev. B {\bf 59}
(1999), 15694-15704 

\bibitem{OA}
M. Oshikawa and I. Affleck: {\it Boundary conformal field theory approach 
to the critical two-dimensional Ising model with a defect line}, Nucl. Phys. 
B {\bf 495} (1997), 533-582
    
\bibitem{Polchinski}
J. Polchinski: \textit{String Theory Vol. I: An Introduction 
to the Bosonic String}, Cambridge University Press 2005

\bibitem{RHFCA}
A. Rahmani, C.-Y. Hou, A. Feiguin, C. Chamon, I. Affleck:
{\it How to find conductance tensors of quantum multi-wire junctions 
through static calculations: application to an interacting Y junction},
Phys. Rev. Lett. {\bf 105} (2010), 226803 

\bibitem{RHFOCA}
A. Rahmani, C.-Y. Hou, A. Feiguin, M. Oshikawa, C. Chamon, I. Affleck:
{\it General method for calculating the universal conductance of strongly 
correlated junctions of multiple quantum wires}, Phys. Rev. Rev. B 
{\bf 85} (2012), 045120

\bibitem{Ruelle}
D. Ruelle: \textit{Natural nonequilibrium states in quantum statistical
mechanics}, J. Stat. Phys. {\bf 98} (2000), 57-75

\bibitem{SassKram}
M. Sassetti, B. Kramer: \textit{Quantum wires as Luttinger liquids: theory},
in: \textit{Advances in Solid State Physics} Vol. 40, ed. B. Kramer,   
Springer 2000, pp 117-132

\bibitem{Senechal}
D. S\'en\'echal: \textit{An introduction to bosonization}, 
arXiv:cond-mat/9908262v1

\bibitem{Sims}
C. C. Sims: \textit{Computation with Finitely Presented Groups}, Cambridge 
University Press 1994

\bibitem{Stone}
\textit{Bosonization}, editor M. Stone, World Scientific, Singapore 1994

\bibitem{Voit}
J. Voit: \textit{One-dimensional Fermi liquids}, Rep. Prog.
Phys. {\bf 58} (1995), 977-1116

\bibitem{WW}
E.T. Whittaker, G.N. Watson, \textit{A Course of Modern Analysis, Fourth 
Edition}, Cambridge University Press  1990

\bibitem{WittenNA}
E. Witten: \textit{Non-abelian bosonization in two dimensions}, 
Commun. Math. Phys. {\bf 92} (1984), 455-472 

\bibitem{WA}
E. Wong, I. Affleck:
{\it Tunneling in quantum wires: a boundary Conformal Field Theory approach},
Nucl. Phys. B {\bf 417} (1994), 403-438




\end{thebibliography}
\end{document}